\documentclass[preprint,
nofootinbib,
longbibliography,
amsmath,amssymb,
aps,
prd
]{revtex4-2}

\usepackage[utf8]{inputenc}
\usepackage{amsmath}
\usepackage{amsfonts,dsfont}
\usepackage{mathrsfs}
\usepackage{cancel} 
\usepackage{amssymb,ulem}
\usepackage{amsmath}
\usepackage{romannum}
\usepackage{subfigure}
\usepackage{adjustbox}
\usepackage{dcolumn}
\usepackage{graphicx}
\usepackage{bm}
\usepackage{enumerate}

\usepackage[colorlinks=true,linkcolor=blue,urlcolor=blue,filecolor=black,
citecolor=red,pdfstartview=FitV,pdftitle={},pdfsubject={},pdfkeywords={},
pdfpagemode=None,bookmarksopen=true]{hyperref}
\usepackage[capitalise]{cleveref}
\crefrangeformat{table}{#3#1#4-#5#2#6}
\crefrangeformat{equation}{(#3#1#4)-(#5#2#6)}

\usepackage{array}
\usepackage{diagbox}

\usepackage{xcolor}
\usepackage{float}
\usepackage{enumitem}  
\usepackage{bbold}
\usepackage{tikz}
\usepackage{multirow}
\usepackage{caption}
\usepackage{supertabular,adjustbox}
\newcolumntype{L}[1]{>{\raggedright\let\newline\\\arraybackslash\hspace{0pt}}m{#1}}
\newcolumntype{C}[1]{>{\centering\let\newline\\\arraybackslash\hspace{0pt}}m{#1}}
\newcolumntype{R}[1]{>{\raggedleft\let\newline\\\arraybackslash\hspace{0pt}}m{#1}}
\usepackage[figuresright]{rotating}

\newcommand{\dd}{{\text{d}}}

\begin{document}
	\preprint{APS/PRD}
	\title{\boldmath
		Tensor-current contributions to B Anomalies}
	
	\author{Qiaoyi Wen}
	\author{Fanrong Xu}
	\email{fanrongxu@jnu.edu.cn}
	
	\affiliation{Department of Physics and Siyuan Laboratory, Jinan University,
		Guangzhou 510632, P.R. China}
	
\begin{abstract}

 Tensor-current operators, potentially generated by scalar leptoquarks in grand unified theories (GUTs), 
are among the plausible new physics (NP) candidates suggested by the anomalies observed in $B$-meson decays. 
As experimental data continue to accumulate, exploring this possibility remains timely and well motivated. 
In this work, we present a systematic analysis of representative tensor-current Wilson coefficients 
($C_T, C_{T5}$) in $b \to s \ell^+ \ell^-$ transitions. 
By incorporating contributions from the high-$q^2$ region, our framework fully exploits the available experimental data across the entire $q^2$ range. 
Within this setup, five distinct lepton-flavor-universal (LFU) scenarios are proposed and tested through global fits. 
Our results show that it is difficult to resolve the tension between experimental measurements and theoretical predictions using only $C_T$ and $C_{T5}$. 
Meanwhile, the significance of $\Delta C_9$ remains essentially unchanged, even in the presence of tensor contributions. 
In one representative scenario (S-III), we obtain 
$[C_9,C_{10}, C_T, C_{T5}] \simeq [-1.05,\,0.22,\,0.02,\,0.01]$, 
with a reduced chi-squared statistic $\tilde{\chi}^2 \equiv \chi^2_{\rm min}/{\rm d.o.f.} = 708.7/486 = 1.46$. 
Furthermore, we derive a stringent 95\% C.L. constraint on tensor operators,
$$
	F(x,y)\Big|^{\text{S-I}}_{x=\Delta C_T,\,y=\Delta C_{T5}} 
	= x^2 + 0.063\,xy + 0.989\,y^2 + 0.034\,x + 0.043\,y \leq 0.003,
$$
which provides one of the strongest bounds to date on $C_T$ and $C_{T5}$.

\end{abstract}
\maketitle
\clearpage
\clearpage
\newpage
\pagenumbering{arabic}

\section{Introduction}
\label{sec: Intro}

Much like a splendid symphony yet to reach its finale, the anomalies in the 
$b \to s \ell^+ \ell^-$ continue to persist. 
Recent analyses by LHCb, carried out through various approaches, suggest a deviation of the Wilson coefficient $C_9$ from its Standard Model (SM) prediction~\cite{LHCb:2023gel, LHCb:2023gpo, 
LHCb:2024onj}. 
This observed trend, that potential new physics (NP) effects may be encoded in $C_9$, is well supported by global theoretical fits~\cite{Wen:2023pfq, Hurth:2023jwr, Alguero:2023jeh} based on a broad range of available experimental data. 
Meanwhile, the possibility of NP as an underlying cause remains an intriguing and open question.

Leptoquarks, which often arise from grand unified theories (GUTs), are among the leading candidates for addressing the $B$-anomaly problems. 
While much attention has been given to vector leptoquarks, scalar leptoquarks offer a distinct and compelling perspective for resolving these anomalies~\cite{Lee:2021jdr, Angelescu:2021lln, Becirevic:2018afm, Goto:2023qch}. 
A particularly interesting feature of scalar leptoquark models is their ability to generate new tensor-current operators at low energies. 
Given this, it is worthwhile to further explore whether current and future experimental data can offer meaningful insights into models with scalar leptoquarks and their connection to GUT frameworks~\cite{Becirevic:2018afm, Goto:2023qch}.

Motivated by the above considerations, this work presents a model-independent analysis based on a complete set of effective operators.  
In addition to incorporating tensor-current operators, following the approach in~\cite{Wen:2023pfq}, we also update the set of form factors (FFs) for the $B \to K^{(*)} \ell^+ \ell^-$ transitions.  
The previous FFs were only valid in the low-$q^2$ region, whereas the current version extends across the full $q^2$ range.  
This advancement enables the inclusion of more experimental data from the high-$q^2$ region, thereby improving the statistical power of the analysis.

To explore the impact of tensor-current operators, we consider five distinct benchmark scenarios.  
In particular, the full scenario involves a high-dimensional parameter space of up to 32 dimensions.  
Our analysis incorporates data across the entire $q^2$ spectrum for exclusive processes such as $B \to V \ell^+ \ell^-$ and $B \to P \ell^+ \ell^-$, as well as inclusive processes like $B \to X_s \ell^+ \ell^-$ and the rare decays $B_{s(d)} \to \mu^+ \mu^-$.  
Taking into account correlations among the fitted parameters, we find that the reduced $\chi^2$ values across different scenarios lie in the range of $1.4 \sim 2.1$, indicating the reliability of our fits.  
As a result, the allowed ranges for the tensor-current operators are extracted, along with an intriguing relation satisfied by the corresponding Wilson coefficients (WCs).

The remainder of the paper is organized as follows.  
In \cref{sec:frame}, we present the theoretical framework of the analysis, including the effective Hamiltonian formalism, the five working scenarios under consideration, the relevant observables for all decay channels, and the fitting strategies.  
The numerical analysis is given in \cref{sec: num}, with input parameters summarized in \cref{sub: num_input} and the results presented in \cref{sub: num_res}.  
Our conclusions are provided in \ref{sec:con}.  
Additional details regarding theoretical formulas, input parameters, experimental data, and supplementary material can be found in \cref{app:Theo}, \ref{app:input}, \ref{app:exp_input}, and \ref{app:sup_materials}, respectively.

\section{The Working Frame}
\label{sec:frame}

\subsection{Effective Hamiltonian}
\label{sub: eff_Ham}

The most generic effective Hamiltonian 
to describe $b \to s \ell^+ \ell^-$ processes are given as
\begin{equation}
	\begin{aligned}
		\mathscr{H}=-\frac{4G_F}{\sqrt{2}}V_{tb}V^*_{ts}\frac{e^2}{16\pi^2}\sum_i\left(C_i \mathcal{O}_i+C'_i\mathcal{O}'_i\right)+\text{H.c.},
	\end{aligned}
\end{equation}
in which high energy information is encoded in Wilson coefficients (WCs) $C_i$, while low energy
effective operators $\mathcal{O}_i$ are 
\begin{equation}
	\begin{aligned}
		&\mathcal{O}_7=\frac{m_b}{e}(\bar{s}\sigma_{\mu\nu}P_R b)F^{\mu\nu},
		&\mathcal{O}'_7=\frac{m_b}{e}(\bar{s}\sigma_{\mu\nu}P_L b)F^{\mu\nu},\\
		&\mathcal{O}_8=\frac{g_sm_b}{e^2}(\bar{s}\sigma_{\mu\nu}T^aP_R b)G^{a\,\mu\nu},
		&\mathcal{O}'_8=\frac{g_sm_b}{e^2}(\bar{s}\sigma_{\mu\nu}T^aP_L b)G^{a\,\mu\nu},\\	
		&\mathcal{O}_9=(\bar{s}\gamma_\mu P_L b)(\bar{\ell}\gamma^\mu \ell),
		&\mathcal{O}'_9=(\bar{s}\gamma_\mu P_R b)(\bar{\ell}\gamma^\mu \ell),\nonumber\\
		&\mathcal{O}_{10}=(\bar{s}\gamma_\mu P_L b)(\bar{\ell}\gamma^\mu\gamma_5 \ell),
		&\mathcal{O}'_{10}=(\bar{s}\gamma_\mu P_R b)(\bar{\ell}\gamma^\mu \gamma_5\ell),\nonumber\\
		&\mathcal{O}_{S}=m_b(\bar{s}P_R b)(\bar{\ell}  \ell),
		&\mathcal{O}'_{S}=m_b(\bar{s}P_L b)(\bar{\ell}  \ell),\;\nonumber\\
		&\mathcal{O}_{P}=m_b(\bar{s}P_R b)(\bar{\ell} \gamma_5 \ell),
		&\mathcal{O}'_{P}=m_b(\bar{s}P_L b)(\bar{\ell} \gamma_5 \ell),\nonumber	\\
		&\mathcal{O}_{T}=(\bar{s}\sigma^{\mu\nu} b)(\bar{\ell} \sigma_{\mu\nu} \ell),
		&\mathcal{O}_{T5}=(\bar{s}\sigma^{\mu\nu} b)(\bar{\ell} \sigma_{\mu\nu}\gamma_5 \ell).\nonumber	\\
	\end{aligned}
\end{equation}
 with eletromagnetic (gluon) filed strength tensor $F_{\mu\nu}$ ($G_{\mu\nu}$).  
 Other notations, including 
  $P_{L,R}=(1\mp\gamma_5)/2$, Fermi coupling constant $G_F$, 
  electron charge magnitude  $e$, and the element of CKM matrix $V_{ij}$
  are  self-evident.
  In SM, only operators $\mathcal{O}_{7,8,9,10}$  appear while their chiral-flipped 
dual operators denoted with a prime as well as scalar, tensor operators give negligible contributions\footnote{
	Other WCs are numerically suppressed: $C_7'\sim m_s/m_b C_7$; 
	$C_S^{(\prime)}$, $C_P^{(\prime)}\sim m_bm_\ell/m_W^2$.
}.  
The WCs, containing high energy information, is obtained by 
integrating out heavy particles at electroweak (EW) scale and
running to $B$ meson scale.
These WCs have been calculated precisely in SM and can be found in \cite{Altmannshofer:2008dz,Du:2015tda,Hou:2014dza,Blake:2016olu}.  
The presence of new physics (NP), within the framework of effective theory, can be classified into two categories. One possibility is the appearance of new operators, with the tensor current being a typical example. Another possibility is the deviation of Wilson coefficients (WCs) from their Standard Model (SM) predictions. 
Specifically, such deviations of 
WCs\footnote{
Here we restrict our attention to $CP$-conserving WCs, which are assumed to be real. 
For discussions involving complex WCs, see \cite{Biswas:2020uaq,Altmannshofer:2021qrr}.} 
can be written as
\begin{equation}
	\begin{aligned}
		C_{i;\text{NP}}^{(\prime)\ell} \equiv \Delta C_{i}^{(\prime)\ell} 
		= C_{i}^{(\prime)\ell} - C_{i;\text{SM}}^{(\prime)\ell},
	\end{aligned}
\end{equation}
where the lepton flavor $\ell$ is explicitly distinguished.

In the previous work \cite{Wen:2023pfq}, tensor-current operators were not included in the analysis. 
The central task of the present study is therefore to address the remaining question: 
how are the constraints on tensor-current operators affected with sufficient experimental data? 
To investigate the role of tensor-current operators and their correlations with other operators, 
five NP scenarios are considered in the following analysis:

\begin{enumerate}[label=\Roman*., start=1]
	\item \textbf{Tensor-only 2D scenario.}\\
	In this case, $C_{T}$ and $C_{T5}$ are treated as free parameters, 
	while lepton-flavor universality is assumed.
	
	\item \textbf{$C_9$-tensor 3D scenario.}\\
	Building on Scenario~I, one additional parameter, $C_9$, is introduced, 
	in view of its well-known NP effects \cite{Wen:2023pfq}.
	
	\item \textbf{$C_{9,10}$-tensor interplay scenario.}\\
	Extending Scenario~II, another frequently discussed left-handed (LH) 
	pseudovector WC, $C_{10}$, is included.
	
	\item \textbf{$C_{9,10}^{(\prime)}$-tensor scenario.}\\
	In this case, the chirality-flipped counterparts $C_{9,10}^{\prime}$ 
	are also taken into account.
	
	\item \textbf{Full scenario.}\\
	All the parameters $\Delta C_{7,8,9,10,S,P,T,T5}^{(\prime)}$ are left unconstrained.
\end{enumerate}

In the following numerical analysis, unless otherwise specified, all WCs are assumed to be lepton-universal ($\mu = e$), 
consistent with the current experimental status.

\subsection{Theoretical and experimental updates on observables}\label{sub: fit_obs}

Two theoretical improvements have been implemented in this work compared with previous one \cite{Wen:2023pfq}.
First, tensor-current operators are incorporated into all relevant processes.
Second, the validation of form factors has been extended, enabling the theoretical framework 
to be applied over the entire $q^2$ region.
For each type of process, 
the corresponding modifications are outlined below and the details are summarized in 
\cref{app:Theo}.

\begin{enumerate}[label=\Roman*., start=1]

\item  $ B\to V\ell^+\ell^- $\\

To account for the effects of tensor-current operators in the process $B \to V \ell^+\ell^-$, new rank-two transversity amplitudes $A_{ij}$ are introduced, where the indices $i$ and $j$ denote the transversity states $t, \perp, \parallel, 0$. In the high-$q^2$ region, where the $K^\ast$ meson is treated as soft, additional assumptions are required to render the problem tractable. A standard approach is to employ a combination of improved Isgur–Wise relations and the OPE, which ensures that long-distance contributions remain under control up to $\mathcal{O}(\alpha_s \Lambda_{\text{QCD}}/m_b)$~\cite{Bobeth:2010wg,Bobeth:2012vn}. This manifests in a reduction of the number of independent form factors. The corresponding modifications to the amplitudes and form factors, together with the angular coefficients, are summarized in \cref{app: BVll_highq2}.

\item $B\to P \ell^+\ell^-$ \\

The Isgur-Wise relations and low-recoil OPE can be applied not only to the process $B \to V \ell \ell$, but also to $B \to P \ell \ell$~\cite{Bobeth:2011nj}. Similarly, the number of form factors for $B \to K$ decays is reduced, as shown in Eq. \eqref{eq:BP_highq2ffs}. The modified angular coefficients for different dilepton invariant masses $q^2$ are presented in \cref{eq:ang_coef_BVlowq2,eq:ang_coef_BVhighq2}.

\item{$B\to X_s\ell^+\ell^-$}\\	

The tensor-current WCs can also be indirectly constrained through the inclusive process $B \to X_s \ell^+ \ell^-$ (for further details, see \cite{Fukae:1998qy}). The modified differential branching ratio is obtained by adding a term involving the tensor WCs, as shown below,
\begin{equation}
\frac{\dd \mathcal{B}(B \to X_s \ell^+ \ell^-)}{\dd q^2} = \frac{\dd \mathcal{B}(B \to X_s \ell^+ \ell^-)}{\dd q^2} \Bigg|_{\text{w/o } C_{T(5)}} + \frac{\dd \mathcal{B}_{T(5)}}{\dd q^2},
\end{equation}
where the expression for $\dd \mathcal{B}_{T(5)}/\dd q^2$ is provided in \cref{app: BXsll_CT5}.	
	
\end{enumerate}

The experimental data used in this study comprise an updated and comprehensive dataset, covering 
both low-$q^2$ and high-$q^2$ regions. Recent results from CMS are also included \cite{CMS:2024syx,CMS:2024atz}. 
As a result, the total number of observations has increased to around 500, providing a robust foundation 
for performing the numerical fit.

\subsection{Fitting framework}
\label{sub:fit_fw}

Our numerical analysis is based on Bayesian statistics, which have been widely applied in previous works \cite{Wen:2023pfq,Ciuchini:2022wbq,Ciuchini:2020gvn,Ciuchini:2019usw,Kowalska:2019ley}. The main procedure in Bayesian inference is to construct the posterior function:
\begin{equation}
	\mathcal{P}(\vec{\theta}|\mathcal{O}_{\text{expt.}}) \propto \mathcal{L}(\mathcal{O}|\vec{\theta}) \pi(\vec{\theta}),
\end{equation}
where $\mathcal{L}(\mathcal{O}|\vec{\theta})$ and $\pi(\vec{\theta})$ represent the likelihood function and prior, respectively. Similar to our previous model-independent fitting, we adopt a negative log-likelihood (NLL) function:
\begin{equation}\label{eq:nll}
	-2\log\mathcal{L}(\mathcal{O}|\vec{\theta}) = \chi^2(\vec{\theta}) = (\mathcal{O}_{\text{theo}}(\vec{\theta}) - \mathcal{O}_{\text{exp}})^\top (V_{\text{exp}} + V_{\text{theo}})^{-1} (\mathcal{O}_{\text{theo}}(\vec{\theta}) - \mathcal{O}_{\text{exp}}),
\end{equation}
where $\mathcal{O}_{\text{theo}}$ and $\mathcal{O}_{\text{exp}}$ denote the theoretical predictions and the corresponding experimental values of various observables. The covariance matrices $V_{\text{theo}}$ and $V_{\text{exp}}$ are constructed from the correlation matrices and errors of the observables, respectively. For the experimental covariance matrix, correlations among relevant experiments are considered \cite{CMS:2024atz,CMS:2024syx,
LHCb:2022vje,
CMS:2022mgd,LHCb:2021awg,LHCb:2019vks,LHCb:2020dof,ATLAS:2018gqc,LHCb:2020gog}, including the most recent updates. In cases of asymmetrical errors, the larger error is used for each source of uncertainty, such as statistical and systematic errors. The final error is computed as the sum of squares. 

The theoretical covariance matrix is derived under the assumption of a multivariate Gaussian distribution for the input parameters, which typically accounts for the dominant error sources (e.g., form factor errors). The parameter vector $\vec{\theta}$ in \cref{eq:nll} consists of the Wilson coefficients (WCs) of interest, leading to the following expression:
\begin{equation}
	\vec{\theta}^{\top} = \left[\Delta C_{7}^{(\prime)}, \Delta C_{8}^{(\prime)}, \Delta C_{9}^{(\prime)\ell}, \Delta C_{10}^{(\prime)\ell}, \Delta C_{S}^{(\prime)\ell}, \Delta C_{P}^{(\prime)\ell}, \Delta C_{T(5)}^{\ell}\right],
\end{equation}
where $\ell = e \text{ or } \mu$ represents the flavor of the dilepton, and the dimension of the parameter space can be up to 14.

In Bayesian inference, the prior function plays a crucial role, typically reflecting our prior knowledge about the problem at hand. In this analysis, the prior represents the known ranges of the parameters of interest. Specifically, the best-fit points and corresponding confidence intervals obtained from our previous work \cite{Wen:2023pfq} are adopted. For the central values of the tensor $C_{T(5)}^\ell$, we assume them to be zero, as there is no definitive evidence after the latest 2022 $R_{K^{(\ast)}}$ measurements. Using the aforementioned information, a prior with a multidimensional Gaussian distribution is assumed, based on our previous LFU fit (S-II), with a common standard deviation of $\sqrt{2}$.


\section{Numerical Analysis}
\label{sec: num}	
In this section we present the main procedure and results 
in numerical analyses. 

\subsection{Inputs and Outputs}
\label{sub: num_input}
Here we collect all the necessary theoretical and experimental inputs.
The detailed theoretical input parameters are summarized in \cref{app:input}.
Parameters, including particle masses, lifetimes, decay constants, Wolfenstein parameters, etc. are sorted by their dimensions in \cref{tab:Input_para} , while the other parameters used in the simplified series expansion (SSE) of form factors such as sub-threshold resonances mass, expansion coefficients are presented in \cref{tab:FFs} therein. 
The SM WCs, defined in \cref{sub: eff_Ham}, have been precisely calculated at $\mu_b$ scale up to two-loop RGE running as adopted in \cite{Wen:2023pfq}.

The about 500 experimental observables, serving as part of inputs, 
are comprehensively summarized in tables of \cref{app:exp_input}.
Moreover,  as comparisons, in these tables we also incorporate predictions in different  theoretical approaches. These predictions can be made
by framework in previous work \cite{Wen:2023pfq}, \textit{Flavio} \cite{Straub:2018kue}
as well as analyses in this current work and will be discussed in below.
In the analyses, the reduced chi-squared $\tilde{\chi}^2=\chi^2_{\text{min}}/\text{d.o.f.}$ which is utilized by frequentists is also used and provided for the sake of capturing the goodness of fit of different scenarios.
To estimate the fitted parameters, namely those parameters of interest in $\vec{\theta}$, the median is adopted as our estimation of the center value of the parameter for its robustness. The boundary of  one standard deviation confidence interval is estimated by the 16th and 84th percentiles. Then the region has the same coverage probability as the standard normal distribution.

\subsection{Numerical Results}
\label{sub: num_res}
We present our main results in form of  tables and figures.

\begin{sidewaystable}[htbp!]
	\caption{Estimated Wilson coefficients in different fitting schemes. S-I  to S-V are our latest Global fits extracted from the full invariant dilepton masses $q^2$ data, while WCs in the last four columns are control groups from other collaborations. }
	\setlength{\tabcolsep}{3.8pt}
	\renewcommand\arraystretch{1.2}
		\label{tab:fullq2_res}
		\begin{tabular}{lrrrrrrrrrr}
			\hline 
			\hline
			Params.&S-II\cite{Wen:2023pfq}&S-I&S-II&S-III&S-IV&S-V&LHCb23\cite{LHCb:2023gel}&LHCb24\cite{LHCb:2024onj}&ABCDMN\cite{Alguero:2023jeh}&HMN\cite{Hurth:2023jwr}
			\\ 
			\hline 
			$\tilde{\chi}^2$&0.931&$2.10^{}_{}$&$1.48^{}_{}$&$1.46^{}_{}$&$1.46^{}_{}$&$1.43^{}_{}$&-&-&-&-
			\\
			\hline 
			$\Delta C_7$&$-0.00^{+0.02}_{-0.02}$&-&-&-&-&$0.01^{+0.01}_{-0.01}$&-&-&$0.00^{+0.02}_{-0.01}$&$0.07^{+0.03}_{-0.03}$
			\\ 
			\hline 
			$\Delta C_7^{\prime}$&$0.02^{+0.02}_{-0.02}$&-&-&-&-&$0.01^{+0.01}_{-0.01}$&-&-&$0.01^{+0.02}_{-0.01}$&$-0.01^{+0.01}_{-0.01}$
			\\ 
			\hline 
			$\Delta C_8$&$-0.92^{+0.44}_{-0.38}$&-&-&-&-&$-0.63^{+1.30}_{-0.70}$&-&-&-&$-0.70^{+0.50}_{-0.50}$
			\\ 
			\hline 
			$\Delta C_8^{\prime}$&$-0.08^{+0.89}_{-0.83}$&-&-&-&-&$-0.03^{+1.03}_{-0.99}$&-&-&-&$-0.50^{+1.20}_{-1.20}$
			\\ 
			\hline 
			$\Delta C_{9}$&$-0.79^{+0.20}_{-0.21}$&-&$-1.14^{+0.06}_{-0.06}$&$-1.05^{+0.07}_{-0.07}$&$-1.05^{+0.08}_{-0.08}$&$-1.13^{+0.10}_{-0.10}$&$-0.93^{+0.53}_{-0.57}$&$-0.71^{+0.33}_{-0.33}$&$-1.21^{+0.18}_{-0.17}$&$-1.18^{+0.19}_{-0.19}$
			\\ 
			\hline 
			$\Delta C_{9}^{\prime}$&$0.05^{+0.34}_{-0.35}$&-&-&-&$-0.08^{+0.11}_{-0.11}$&$-0.11^{+0.11}_{-0.11}$&$0.48^{+0.49}_{-0.55}$&$0.28^{+0.43}_{-0.43}$&$-0.04^{+0.37}_{-0.36}$&$0.06^{+0.31}_{-0.31}$
			\\ 
			\hline 
			$\Delta C_{10}$&$0.16^{+0.17}_{-0.16}$&-&-&$0.22^{+0.06}_{-0.06}$&$0.23^{+0.06}_{-0.06}$&$0.26^{+0.07}_{-0.07}$&$0.48^{+0.29}_{-0.31}$&$0.15^{+0.24}_{-0.24}$&$0.07^{+0.15}_{-0.16}$&$0.23^{+0.20}_{-0.20}$
			\\ 
			\hline 
			$\Delta C_{10}^{\prime}$&$-0.09^{+0.18}_{-0.18}$&-&-&-&$-0.10^{+0.07}_{-0.07}$&$-0.07^{+0.08}_{-0.07}$&$0.38^{+0.28}_{-0.25}$&$-0.09^{+0.22}_{-0.22}$&$-0.06^{+0.19}_{-0.19}$&$-0.05^{+0.19}_{-0.19}$
			\\ 
			\hline 
			$\Delta C_{S}$&$0.06^{+1.19}_{-1.23}$&-&-&-&-&$0.25^{+1.18}_{-1.55}$&-&-&-&$-0.07^{+0.03}_{-0.03}$
			\\ 
			\hline 
			$\Delta C_{S}^{\prime}$&$0.06^{+1.19}_{-1.23}$&-&-&-&-&$0.27^{+1.19}_{-1.53}$&-&-&-&$-0.04^{+0.03}_{-0.03}$
			\\ 
			\hline 
			$\Delta C_{P}$&$0.48^{+0.81}_{-0.90}$&-&-&-&-&$1.31^{+0.56}_{-0.88}$&-&-&-&$0.00^{+0.00}_{-0.00}$
			\\ 
			\hline 
			$\Delta C_{P}^{\prime}$&$0.37^{+0.80}_{-0.90}$&-&-&-&-&$1.17^{+0.57}_{-0.87}$&-&-&-&$-0.01^{+0.02}_{-0.02}$
			\\ 
			\hline 
			$\Delta C_{T}$&-&$-0.02^{+0.02}_{-0.02}$&$0.03^{+0.04}_{-0.04}$&$0.02^{+0.03}_{-0.03}$&$0.03^{+0.03}_{-0.03}$&$0.02^{+0.04}_{-0.04}$&-&-&-&-
			\\ 
			\hline 
			$\Delta C_{T5}$&-&$-0.02^{+0.02}_{-0.02}$&$0.02^{+0.04}_{-0.04}$&$0.01^{+0.03}_{-0.03}$&$0.01^{+0.03}_{-0.04}$&$0.02^{+0.03}_{-0.04}$&-&-&-&-
			\\ 
			\hline 
			\hline 
		\end{tabular}
\end{sidewaystable}

In \cref{tab:fullq2_res}, we summarize the numerical results in the full-$q^2$ region for the five scenarios discussed above. For comparison, we also include the latest fits from LHCb \cite{LHCb:2023gel,LHCb:2024onj} as well as results from other groups \cite{Alguero:2023jeh,Hurth:2023jwr}. In the full-$q^2$ scenario, the Standard Model (SM) $\chi^2$ value is found to be $1025.95$, obtained by switching off all new physics (NP) effects ($C_{i;\text{NP}}^{(\prime)\ell}=0$).  
The second column reveals a relatively large value of $\tilde{\chi}^2\big|_{\text{S-I}}\approx 2.10$. From the perspective of closing the gap between experiment and theory, tensor-type couplings appear to play only a negligible role. In other words, tensor Wilson coefficients are strongly constrained by the first-hand information provided in S-I. The best-fit values $(C_T, C_{T5})=(-0.02,-0.02)$ lie within one standard deviation of the SM, indicating no significant deviation.  
 In the scenario II (S-II), shown as the third column, a sharp decrease in the level of $\tilde{\chi}^2$ is conspicuous, reaching a value of 1.48. Its central value of $C_{9}$ is in close proximity to the well-known  $-1$, while the range of tensor type WCs moves towards the first quadrant. 
When considering our additional independent  1-D full-$q^2$ fit of $C_{9}$, it is notable that
 $\tilde{\chi}^2$ of this scenario is 1.48 too. Furthermore, the fitting value of $C_{9}$ is $-1.12\pm0.06$. These findings indicate that current experiments on the $b\to s\ell \ell $ process do not provide a NP evidence to tensor couplings. This will become clearer and more solid when all the scenarios are taken into account later. 
A comparison between S-II and S-III reveals that, under the LFU global setting, from the view of $\tilde{\chi}^2$, the role of $C_{9}$ is the most prominent, while the contribution of $C_{10}$ is not significant in relieving the tension between experiment and theory. 
With regard to the identification of NP signals, the discrepancy between the theoretical prediction of $C_9$ and the data-driven results persists, with a margin of error exceeding four standard deviations. In this regard, it is also observed from last four columns that the outcomes of the experimental collaborations will manifest a greater degree of conservatism in comparison to those derived from the theoretical fitting groups. On the $C_{10}$ side, our fitting here implicates that the discrepancy holds approximately 3 $\sigma$, while the work of LHCb23\cite{LHCb:2023gel} and the fit of HMN\cite{Hurth:2023jwr} regard it as a slightly positive deviation. This alteration of $C_{10}$ from our previous S-II to the more recent iterations will be elucidated subsequently.
The effect of chirality-flipped operators is discussed in S-IV.
From numerical evaluation, the decreasing trend of $\tilde{\chi}^2$ gradually becomes gentle with the narrow range of both $C_{9}^{\prime}$ and $C_{10}^\prime$ within 1 $\sigma$. It implies that constructing left-handed gauge interaction in model building is still  the main solution.
In the full scenario S-V, the majority of 14 fittings of the WCs are consistent with the previous S-II in \cite{Wen:2023pfq}, all within two standard deviations. We hence summarize that
except left-handed vector and axial-vector WCs, other types of WCs do not deviate from their SM predictions with a 95\% C.L..

Therefore we can conclude that 
the range of $C_T$ and $C_{T5}$ value, 
of which the order of magnitude is approximately 0.03, has been
consistently constrained in all our LFU scenarios using the full-$q^2$ data. 
This is a feature that has been conspicuously absent from the discourse of numerous global fits articles\cite{Alguero:2023jeh,Hurth:2023jwr} since the 2022 LHCb $R_{K^{(\ast)}}$ release. 


\begin{figure}[t]
	\centering
	\includegraphics[scale=0.45]{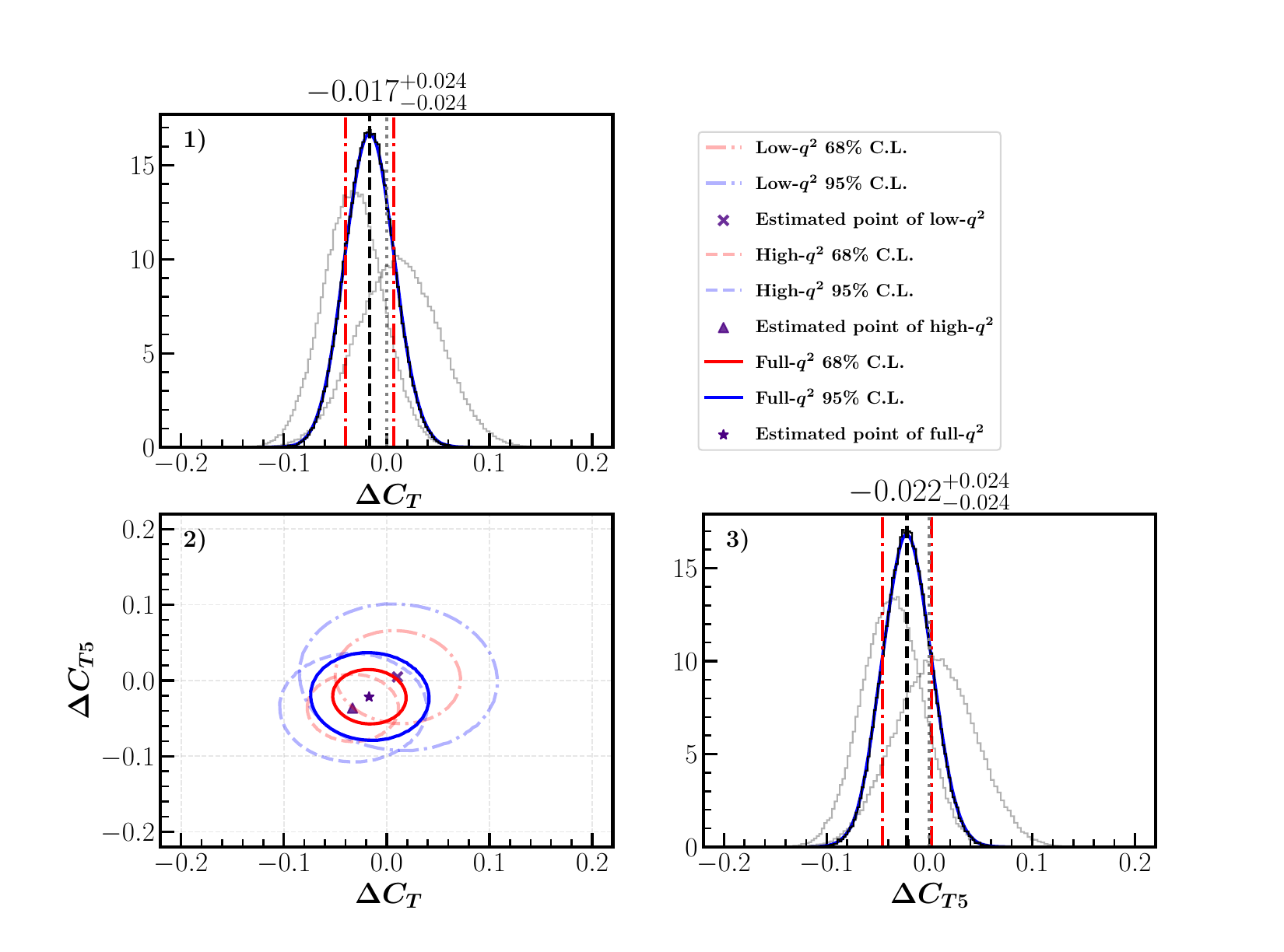}
	\caption{Plots of the confidence region of $C_T$ and $C_{T5}$ under the terms of scenario I. Panels on the diagonal denote their own density (black steps), where the darker steps represent the full-$q^2$ data, and the others are from both low-$q^2$ and high-$q^2$. The panel in the lower left corner of the diagonal represents the correlations between $C_T$ and $C_{T5}$ and preferred regions. The other description is detailed in the \cref{app:sup_materials} for reference. 
	}
	\label{fig:lfus1}
\end{figure}

\begin{table}[b]
	\centering
	\captionof{table}{Estimated WCs in (axial)vector type global fits from S-I to S-IV. 
		The number of observables in the low-, high-, and full-$q^2$  datasets is 272, 218, and 490, respectively.
	}
	\label{tab:diffq2_res}
	\setlength{\tabcolsep}{3.5pt}
	\renewcommand\arraystretch{1.305}
	\begin{adjustbox}{scale=0.60,center}
		\begin{tabular}{l|rrr|rrr|rrr|rrr}
			\toprule
			Scenarios&\multicolumn{3}{c|}{S-I}&\multicolumn{3}{c|}{S-II}&\multicolumn{3}{c|}{S-III}&\multicolumn{3}{c}{S-IV}\\
			\hline
			\resizebox{50pt}{14pt}{
				\diagbox[dir=SE]{{\large{Params}}}{\normalsize{$q^2$-regions}}}&Low&High&Full&Low&High&Full&Low&High&Full&Low&High&Full
			\\ 
			\hline 
			$\tilde{\chi}^2_{\text{SM}}$&$1.59^{}_{}$&$2.74^{}_{}$&$2.10^{}_{}$&$1.18^{}_{}$&$1.85^{}_{}$&$1.48^{}_{}$&$1.09^{}_{}$&$1.85^{}_{}$&$1.46^{}_{}$&$1.09^{}_{}$&$1.86^{}_{}$&$1.46^{}_{}$
			\\ 
			\hline 
			$\Delta C_{9}$&-&-&-&$-1.01^{+0.10}_{-0.10}$&$-1.25^{+0.09}_{-0.09}$&$-1.14^{+0.06}_{-0.06}$&$-0.79^{+0.10}_{-0.10}$&$-1.23^{+0.09}_{-0.09}$&$-1.05^{+0.07}_{-0.07}$&$-0.81^{+0.12}_{-0.12}$&$-1.31^{+0.11}_{-0.11}$&$-1.05^{+0.08}_{-0.08}$
			\\ 
			\hline 
			$\Delta C_{9}^{\prime}$&-&-&-&-&-&-&-&-&-&$-0.08^{+0.20}_{-0.20}$&$0.17^{+0.15}_{-0.15}$&$-0.08^{+0.11}_{-0.11}$
			\\ 
			\hline 
			$\Delta C_{10}$&-&-&-&-&-&-&$0.56^{+0.11}_{-0.11}$&$0.07^{+0.07}_{-0.07}$&$0.22^{+0.06}_{-0.06}$&$0.54^{+0.13}_{-0.12}$&$0.07^{+0.07}_{-0.07}$&$0.23^{+0.06}_{-0.06}$
			\\ 
			\hline 
			$\Delta C_{10}^{\prime}$&-&-&-&-&-&-&-&-&-&$-0.06^{+0.13}_{-0.13}$&$0.02^{+0.08}_{-0.08}$&$-0.10^{+0.07}_{-0.07}$
			\\ 
			\hline 
			$\Delta C_{T}$&$0.01^{+0.04}_{-0.04}$&$-0.03^{+0.03}_{-0.03}$&$-0.02^{+0.02}_{-0.02}$&$0.04^{+0.04}_{-0.04}$&$-0.01^{+0.09}_{-0.09}$&$0.03^{+0.04}_{-0.04}$&$0.04^{+0.04}_{-0.04}$&$-0.01^{+0.08}_{-0.08}$&$0.02^{+0.03}_{-0.03}$&$0.04^{+0.04}_{-0.04}$&$-0.01^{+0.08}_{-0.08}$&$0.03^{+0.03}_{-0.03}$
			\\ 
			\hline 
			$\Delta C_{T5}$&$0.01^{+0.04}_{-0.04}$&$-0.04^{+0.03}_{-0.03}$&$-0.02^{+0.02}_{-0.02}$&$0.04^{+0.04}_{-0.04}$&$-0.06^{+0.10}_{-0.08}$&$0.02^{+0.04}_{-0.04}$&$0.03^{+0.04}_{-0.05}$&$-0.05^{+0.09}_{-0.08}$&$0.01^{+0.03}_{-0.03}$&$0.03^{+0.04}_{-0.04}$&$-0.05^{+0.09}_{-0.08}$&$0.01^{+0.03}_{-0.04}$
			\\ 
			\toprule
		\end{tabular}
	\end{adjustbox}
\end{table}

It is also interesting to expolre whether exist  prominent deviations
 in specific local $q^2$ regions.
This search and its results have been presented in 
 \cref{tab:diffq2_res}, in which the individual effects for each $q^2$ interval has been calculated. 
 The specific $q^2$ regions are defined as follows: 1) low-$q^2$: $m_\ell^2\leq q^2 \leq 8.68 \text{ GeV}$; 2) high-$q^2$: $q^2\geq 8.68 \text{ GeV}$; 3) full-$q^2$. The $\tilde{\chi}^2_{\text{SM}}$ values of the first two schemes are 430.12 and 596.65, respectively, while the value of the last scheme has been discussed in \cref{tab:fullq2_res}.
 As an auxiliary illustration, 
  \cref{fig:lfus1}  explicitly shows  of the impact of data from different regions on WCs in S-I, while the other plots of  are summarized as supporting details in \cref{app:sup_materials}.

In scenario I of \cref{tab:diffq2_res}, it is found that the data from high-$q^2$ prefer a negative contribution in the fourth quadrant on the tensor WCs, while  the data from low-$q^2$ choose a positive contribution in the first quadrant for preference. Such a trend also persists in all subsequent three scenarios. In the lower left corner of \cref{fig:lfus1}, it can be seen that the high-$q^2$ data give a tighter constraint on $(C_{T},C_{T5})$ than the low-$q^2$ data do. This finding is consistent with the conclusion that the majority of other 2-D tensor-fit works have been identified. However, the constraint on $C_T$ and $C_{T5}$ from high-$q^2$ become loosen, resulting in a low-$q^2$ dominance when considering more (axial)vector type couplings concurrently, as shown in \cref{fig:lfus2,fig:lfus3,fig:lfus4}, even in the impure vector-like scenario (depicted in \cref{fig:lfus5}).

When the other WCs are taken into consideration, as outlined above, it is possible to combine the old S-II fits, with a particular focus on the $C_9$ and $C_{10}$ changing. The central value of $C_{9}$ remains almost unchanged, while the value of $C_{10}$ moves towards a positive value, with their errors overall getting narrower. Two primary factors contributing to this can be elucidated as follows. Firstly, an increase in the total number of observables helps to reduce errors. The more measurements that go into the fits, the more accurate the fits tend to be. Secondly, on the theoretical side, the form factors with lower uncertainties have been adopted. This can be seen, for an example, in the predictions of the $B\to K\ell \ell $ processes. This present analysis consolidates the preceding argument, which mentioned the crucial role of the left-handed (axial)vector WCs. 
Therefore, finding an observable that is both clean (with fewer theoretical assumptions) and focused (preferably by only a single parameter of interest, $C_{10}$ in this instance,) is theoretically appealing. 
The identification of such an observable is currently being pursued, and will be introduced  by a separated angular distribution analysis.

To the end, we propose a set of simple closed curves, 
obeying the equation $F(x,y)=0$, 
that can maximally preserve the 95\% C.L. confidence interval. 
One can easily find that among last four scenarios in \cref{tab:diffq2_res}, their constraints on $C_T$ and $C_{T5}$  are closed to each other, while the S-I are independent of this. Therefore, we choose these two scenarios to drive the constraints for simplicity. 
\begin{figure}[t]
	\centering
	\includegraphics[scale=0.45]{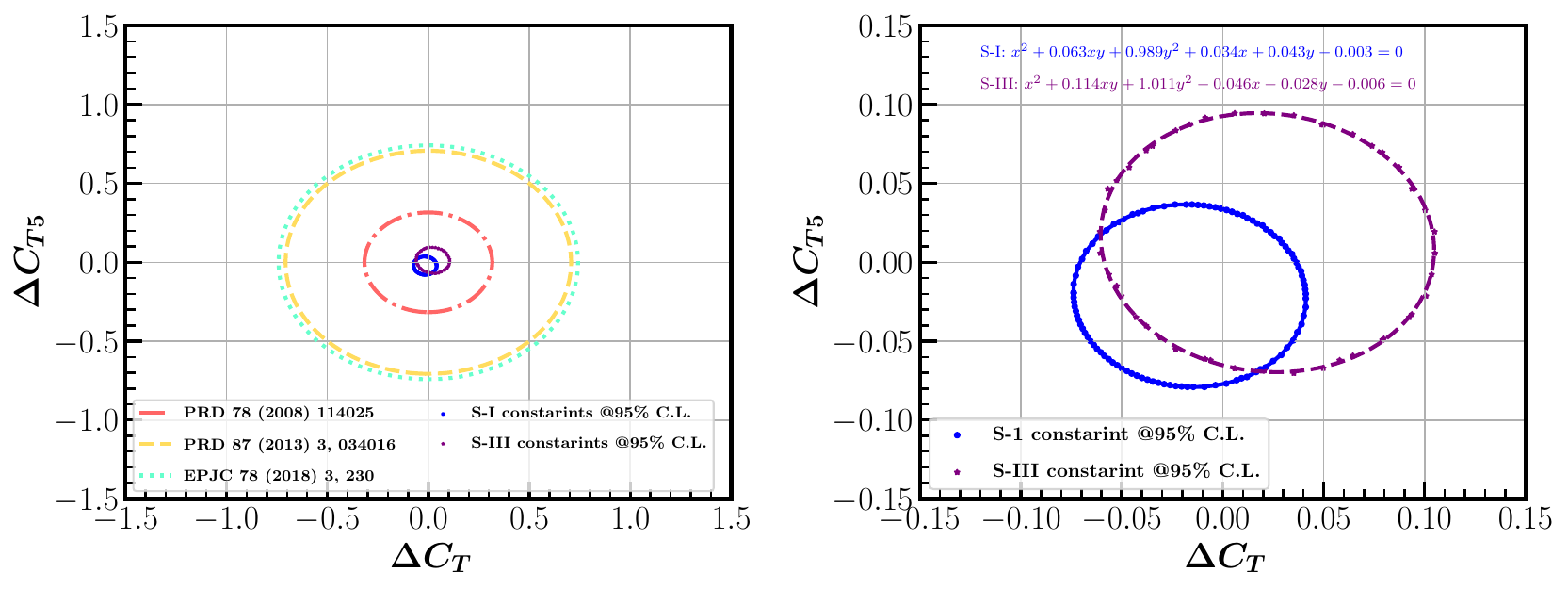}
	\caption{Constraints on the Wilson coefficients $C_{T}$ and $C_{T5}$. 
		The panel on the left provides the comparison of other works prior to the 2022-$R_{K^{(\ast)}}$ release and our this analysis. 
		The right panel presents a zoomed-in image, exclusively depicting the boundary points of 95\% C.L. 
		and elliptic curves in S-I (solid blue) as well as S-III (dashed purple).
	}
	\label{fig:constraints_S13}
\end{figure}
In \cref{fig:constraints_S13}, constraints on WCs are presented intuitively.
From the left side of \cref{fig:constraints_S13}, one can learn that the experimental data accumulated are adequate to restrict the range of current $C_{T}$ and $C_{T5}$ to a smaller scope, which exceeds the constraints of previous works shown in the plot. This is partly due to their using of a minimal set of observables, which serves to simplify their analysis. As demonstrated on the right panel of \cref{fig:constraints_S13}, the order of magnitude of $C_T$ and $C_{T5}$ in both scenarios reach the level of around 0.1. 
In order to derive the closed form of such an elliptic curve, it is sufficient to substitute the boundary points of 95\% C.L. into the general formula of the elliptic curve,
fitting sequentially its six coefficients utilizing the least squares method. 
In this instance, the error of each unknown coefficient is no longer taken into account; rather, their expressions are listed as follows:
\begin{equation}\label{eq:cons_S13}
	\begin{aligned}
		&F(x,y)\Big|_{x=\Delta C_T,y=\Delta C_{T5}}^\text{S-I}= x^2 + 0.063 xy + 0.989y^2+0.034x+0.043y - 0.003,\\
		&F(x,y)\Big|_{x=\Delta C_T,y=\Delta C_{T5}}^\text{S-III}= x^2 + 0.114 xy + 1.011y^2 - 0.046x - 0.028y -0.006.
	\end{aligned}
\end{equation}
Despite of the magnitude of 0.1, the tensor WCs remain consistent with the SM within two standard deviations, so these formulas should be strongly restricted to models in which the tensor-current operators appear.

\section{Concluding Remarks}
\label{sec:con}

The objective of this work is to examine whether NP contributions from tensor-current interactions are compatible with the currently available data.  
To explicitly explore the impact of the tensor-current WCs, $C_T$ and $C_{T5}$,  
we perform global fits that include the most relevant (axial)vector-type couplings, denoted as S-I to S-IV, along with an additional complementary scenario S-V.  
Based on a comprehensive analysis that incorporates both theoretical and experimental inputs from the high-$q^2$ region, we arrive at the following conclusions:

\begin{enumerate}
	\item The NP signal in the left-handed vector-type coupling, $\Delta C_9$, remains largely unchanged.  
	In contrast, the left-handed axialvector-type interaction, $\Delta C_{10}$, exhibits a positive shift.  
	This behavior can be attributed partly to different treatments of form factors and partly to the enlarged dataset of observables.  

	\item It is difficult to reconcile the tension between experimental measurements and theoretical predictions using only $C_T$ and $C_{T5}$ within the framework of S-I.  
	Even in alternative scenarios with broader NP assumptions, the allowed range for $C_{T(5)}$ remains strongly constrained.  

	\item Among the different scenarios considered, low-$q^2$ data provide particularly valuable information for constraining tensor couplings, especially when additional NP possibilities are included.  

	\item Our analysis yields the most stringent bounds to date on the tensor coefficients $C_{T(5)}$.  
\end{enumerate}

In the post-$R_{K^{\ast}}$ era, we anticipate that the information on $C_{T(5)}$ and $\Delta C_{9,10}$ extracted from data will offer useful guidance for future model building.

\acknowledgments
This work is supported by NSFC under Grant  No. 12475095.
\clearpage

\appendix
\section{MODIFIED FORMULAS FOR INVOLVED OBSERVABLES}
\label{app:Theo}
We summarize the necessary formulas with incorporation
of tensor-current contribution in full-$q^2$ region in this part.

\vspace{-0.5cm}
\subsection{\texorpdfstring{$B \to V\ell^+\ell^-$}{BVll} }
\label{app: BVll_highq2}
The full angular decay distribution of $ \bar{B}^0\to\bar{K}^{*0}\ell^+\ell^- $ is shown as \cite{Altmannshofer:2008dz}
\begin{equation}
	\begin{aligned}
		\frac{\dd^4\Gamma^\ell}{\dd q^2~\dd\cos\theta_\ell~\dd\cos\theta_{K^*}~\dd\phi}=\frac{9}{32\pi}J^\ell\left(q^2,\theta_\ell, \theta_{K^*},\phi\right),		
	\end{aligned}
\end{equation}
where the $ J^\ell $ can be decomposed into the following 12 angular coefficients as below:
\begin{equation}
	\begin{aligned}
		J^\ell\left(q^2,\theta_\ell,\theta_{K^*},\phi\right)&=J_{1;s}^\ell\sin^2\theta_{K^*}+J_{1;c}^\ell\cos^\theta_{K^*}+\left(J_{2;s}^\ell\sin^2\theta_{K^*}+J_{2;c}^\ell\cos^2\theta_{K^*}\right)\cos2\theta_\ell\\
		&+J_{3}^\ell\sin^2\theta_{K^*}\sin^2\theta_\ell\cos2\phi+J_{4}^\ell\sin2\theta_{K^*}\sin2\theta_\ell\cos\phi\\
		&+J_{5}^\ell\sin2\theta_{K^*}\sin\theta_\ell\cos\phi\\
		&+\left(J_{6;s}^\ell\sin^2\theta_{K^*}+J_{6;c}^\ell\cos^2\theta_{K^*}\right)\cos\theta_\ell+J_{7}^\ell\sin2\theta_{K^*}\sin\theta_\ell\sin\phi\\
		&+J_{8}^\ell\sin2\theta_{K^*}\sin2\theta_\ell\sin\phi+J_{9}^\ell\sin^2\theta_{K^*}\sin^2\theta_\ell\sin2\phi
	\end{aligned}\label{eq:angular_BVll}
\end{equation}
The angular coefficients $ J_i $ in \eqref{eq:angular_BVll}are  modified as outlined below:
\begin{equation}
	\begin{aligned}
		J_{1;s}^\ell&=\frac{(2+\beta_\ell^2)}{4}\left[|A_{\perp;\ell}^\text{L}|^2+|A_{\parallel;\ell}^{\text{L}}|^2+\left(\text{L}\to\text{R}\right)\right]+\frac{4m_\ell^2}{q^2}\Re\left(A_{\perp;\ell}^{\text{L}}A_{\perp;\ell}^{\text{R$^\ast$}}+A_{\parallel;\ell}^{\text{L}}A_{\parallel;\ell}^{\text{R$^\ast$}}\right)
		\\
		&+4\beta_\ell^2\left(|A_{0\perp;\ell}|^2 + |A_{0\parallel;\ell}|^2\right) + 4(4-3\beta_\ell^2)\left(|A_{t\perp;\ell}|^2 + |A_{t\parallel;\ell}|^2\right)
		\\
		&+8\sqrt{2}\frac{m_\ell}{\sqrt{q^2}}\Re\left[A_{t\parallel}^\ast\left(A_{\parallel;\ell}^{\text{L}} + A_{\parallel;\ell}^{\text{R}}\right) 
		+ A_{t\perp}^\ast\left(A_{\perp;\ell}^{\text{L}} + A_{\perp;\ell}^{\text{R}}\right) \right],
		\\
		J_{1;c}^\ell&=|A_{0;\ell}^\text{L}|^2+|A_{0;\ell}^\text{R}|^2+\frac{4m_\ell^2}{q^2}\left[|A_{t;\ell}|^2+2\Re\left(A_{0;\ell}^\text{L}A_{0;\ell}^\text{R$^\ast$}\right)\right]+\beta^2_{\ell}|A_{\text{S};\ell}|^2
		\\
		&+8(2-\beta_\ell^2)|A_{t0;\ell}|^2 + 8\beta_\ell^2|A_{\parallel \perp;\ell}|^2 
		+16 \frac{m_\ell}{\sqrt{q^2}}\Re\left[A_{t0;\ell}^\ast\left(A_{0;\ell}^{\text{L}} + A_{0;\ell}^{\text{R}}\right)\right],
		\\
		J_{2;s}^\ell&=\frac{\beta_\ell^2}{4}\left[|A_{\perp;\ell}^\text{L}|^2+|A_{\parallel;\ell}^\text{L}|^2+\left(\text{L}\to\text{R}\right) 
		- 16 \left(|A_{t\perp;\ell}|^2 + |A_{t\parallel;\ell}|^2 + |A_{0\perp;\ell}|^2 + |A_{0\parallel;\ell}|^2 \right)\right],
		\\
		 J_{2;c}^\ell&=-\beta_\ell^2\left[|A_{0;\ell}^\text{L}|^2+\left(\text{L}\to\text{R}\right)
		 - 8 \left(|A_{t0;\ell}|^2 + |A_{\parallel\perp;\ell}|^2  \right)\right],
		 \\
		 J_{3}^\ell&=\frac{1}{2}\beta_\ell^2\left[|A_{\perp;\ell}^\text{L}|^2-|A_{\parallel;\ell}^{\text{L}}|^2+\left(\text{L}\to\text{R}\right)
		 + 16 \left(|A_{t\parallel;\ell}|^2 - |A_{t\perp;\ell}|^2 + |A_{0\parallel;\ell}|^2 - |A_{0\perp;\ell}|^2		 
		 \right)\right],
		 \\
		 J_{4}^\ell&=\frac{1}{\sqrt{2}}\beta_\ell^2\left[\Re\left(A_{0;\ell}^\text{L}A_{\parallel;\ell}^\text{L$^\ast$}\right)+\left(\text{L}\to\text{R}\right)
		 -8\sqrt{2}\left(A_{t\parallel;\ell}^\ast A_{t0;\ell} + A_{0\parallel;\ell}^\ast A_{\parallel\perp;\ell}\right)
		 \right],
	\end{aligned}
\end{equation}
\begin{equation}
	\begin{aligned}
		J_{5}^\ell&=\sqrt{2}\beta_\ell\Bigg[\Re\left(A_{0;\ell}^\text{L}A_{\perp;\ell}^\text{L$^\ast$}\right)-\left(\text{L}\to\text{R}\right)
		-\frac{m_\ell}{\sqrt{q^2}}\Re\Big(
		A_{\parallel;\ell}^\text{L} A_{\text{S};\ell}^\ast + A_{\parallel;\ell}^\text{R}A_{\text{S};\ell}^\ast
		+ 4\sqrt{2}A_{t;\ell}^\ast A_{0\parallel;\ell} 
		\\
		& \hspace{1.5cm}
		- 4\sqrt{2}\left(A_{0;\ell}^\text{L} - A_{0;\ell}^\text{R}\right)A_{t\perp;\ell}^\ast
		-4\left(A_{\perp;\ell}^\text{L} - A_{\perp;\ell}^\text{R}\right)A_{t0;\ell}^\ast
		\Big)\Bigg],
		\\
		J_{6;s}^\ell&=2\beta_\ell\left[\Re\left(A_{\parallel;\ell}^{\text{L}}A_{\perp;\ell}^{\text{L$^\ast$}}\right)-\left(\text{L}\to\text{R}\right)
		+ 4\sqrt{2} \frac{m_\ell}{\sqrt{q^2}}\Re\left(A_{\perp;\ell}^{\text{L} }  A_{t\parallel;\ell}^\ast - A_{\perp;\ell}^{\text{R} }  A_{t\parallel;\ell}^\ast \right)\right. 
		\\
		&\left.\hspace{1.5cm} + 4\sqrt{2} \frac{m_\ell}{\sqrt{q^2}}  \Re\left(A_{\parallel;\ell}^{\text{L} }  A_{t\perp;\ell}^\ast - A_{\parallel;\ell}^{\text{R} }  A_{t\perp;\ell}^\ast \right)
		\right],
		\\ J_{6;c}^\ell&=4\beta_\ell\frac{m_\ell}{\sqrt{q^2}}\left[\Re\left(A_{0;\ell}^{\text{L}}A_{\text{S};\ell}^{\text{$^\ast$}}\right)+\left(\text{L}\to\text{R}\right)\right]
		+ 8\beta_\ell \Re\left[A_{t0;\ell} A_{\text{S};\ell}^\ast + 2 A_{\parallel\perp;\ell} A_{t;\ell}^\ast\right],
		\\
		J_{7}^\ell&=\sqrt{2}\beta_\ell\left[\Im\left(A_{0;\ell}^\text{L}A_{\parallel;\ell}^{\text{L$^\ast$}}\right)-\left(\text{L}\to\text{R}\right)
		+ 2\sqrt{2} \Im(A_{t\perp;\ell}A_{\text{S};\ell}^\ast )
		+ \frac{m_\ell}{\sqrt{q^2}}\Im\left(A_{\perp;\ell}^\text{L}A_{\text{S};\ell}^\ast+A_{\perp;\ell}^\text{R}A_{\text{S};\ell}^\ast\right)\right.
		\\
		&\left. \hspace{1.5cm} +4\sqrt{2}\frac{m_\ell}{\sqrt{q^2}}\Im
		\left(A_{0\perp;\ell}A_{t;\ell}^\ast + A_{0;\ell}^{\text{L}} A_{t\parallel;\ell}^\ast - A_{0;\ell}^{\text{R}} A_{t\parallel;\ell}^\ast
		-\frac{1}{\sqrt{2}}(A_{\parallel;\ell}^{\text{L}} - A_{\parallel;\ell}^{\text{R}}) A_{t0;\ell}^\ast
		\right)
		\right],
		\\
		J_{8}^\ell&=\frac{1}{\sqrt{2}}\beta_\ell^2\left[\Im\left(A_{0;\ell}^\text{L}A_{\perp;\ell}^{\text{L$^\ast$}}\right)+\left(\text{L}\to\text{R}\right)\right],
		\quad J_{9}^\ell=\beta_\ell^2\left[\Im\left(A_{\parallel;\ell}^\text{L$^\ast$}A_{\perp;\ell}^\text{L}\right)+\left(\text{L}\to\text{R}\right)\right],
	\end{aligned}
\end{equation}
where six more transversity amplitudes are also updated in the consecutive context. In the large recoil region, the six additional transversity amplitudes are demonstrated in the work of Bobeth\cite{Bobeth:2012vn} giving that
\begin{equation}
	\begin{aligned}
A_{\parallel\perp;\ell(t0;\ell)}&=\pm N_0^\ell\frac{C_{T(5)}}{M_{K^\ast}}\left[(M_B^2+3M_{K^\ast}-q^2)T_2-\frac{\lambda^2M_B^4}{M_B^2-M_{K^\ast}^2}T_3\right],\\
A_{t\perp;\ell(0\perp;\ell)}&=\pm2N_0^\ell\frac{\lambda M_B^2}{\sqrt{q^2}}C_{T(5)}T_1,\\
A_{0\parallel;\ell(t\parallel;\ell)}&=\pm2N_0^\ell\frac{M_B^2-M_{K^\ast}^2}{\sqrt{q^2}}C_{T(5)}T_2.
	\end{aligned}
	\label{eq:amps_BVtensor}
\end{equation}
As for the formulas in the high-$q^2$ region, 12 transversity amplitudes need to be modified and written as 
\begin{equation}
	\begin{aligned}
				A_{\perp;\ell}^{\text{L,R}}&=+N_0^\ell\left[\left(C_9^{\text{eff};\ell}+C_9^{\prime\ell}\right)\mp\left(C_{10}^{\ell}+C_{10}^{\prime\ell}\right)
				+\frac{2\kappa m_bM_B}{q^2}\left(C_{7}^{\text{eff}}+C_7^{\prime}\right)\right]F_{\perp},
				\\
				A_{\parallel;\ell}^\text{L,R}&=-N_0^\ell\left[\left(C_9^{\text{eff};\ell}-C_9^{\prime\ell}\right)\mp\left(C_{10}^{\ell}-C_{10}^{\prime\ell}\right)
				+\frac{2\kappa m_bM_B}{q^2}\left(C_7^{\text{eff}}-C_7^{\prime}\right)\right]F_{\parallel},
				\\
				A_{0;\ell}^{\text{L,R}}&=-N_0^\ell\left[\left(C_9^{\text{eff};\ell}-C_9^{\prime\ell}\right)\mp\left(C_{10}^{\ell}-C_{10}^{\prime\ell}\right)
				+\frac{2\kappa m_bM_B}{q^2}\left(C^{\text{eff}}_7-C_7^{\prime}\right)\right]F_{0},
	\end{aligned}\label{eq:highq2_BVamps1}
\end{equation}
\begin{equation}
	\begin{aligned}
		A_{\parallel\perp;\ell(t0;\ell)}&=\pm 2N_0^\ell C_{T(5)} \frac{\kappa M_B}{\sqrt{q^2}}\left[1+\mathcal{O}\left(\frac{\Lambda_{\text{QCD}}}{M_B}\right)\right]F_{0},
		\\
		A_{t\perp;\ell(0\perp;\ell)}&=\pm\sqrt{2} N_0^\ell C_{T(5)} \frac{\kappa M_B}{\sqrt{q^2}}\left[1+\mathcal{O}\left(\frac{\Lambda_{\text{QCD}}}{M_B}\right)\right]F_{\perp},
		\\
		A_{0\parallel;\ell(t\parallel;\ell)}&=\pm \sqrt{2}N_0^\ell C_{T(5)} \frac{\kappa M_B}{\sqrt{q^2}}\left[1+\mathcal{O}\left(\frac{\Lambda_{\text{QCD}}}{M_B}\right)\right]F_{\parallel}.
	\end{aligned}\label{eq:highq2_BVamps2}
\end{equation}
The plus and minus signs in above eqns. \eqref{eq:amps_BVtensor} and \eqref{eq:highq2_BVamps2} represent the WCs $C_{T} $ and $C_{T5}$, respectively. In eq. \eqref{eq:highq2_BVamps1}, the plus and minus signs denote $L$ as well as $R$, respectively. $N_0^\ell$ is referred to as the normalization factor. This factor which, in conjunction with the definitions of  parameters $ \lambda $ and $ \beta_\ell $  has been delineated in previous work and will not be repeated here, while the other factor will be explained later.

In the previous work, the number of form factors was reduced to two by applying the large energy limit (LEL) assumption in the large recoil. However, in this study, seven form factors are retained in the low-$q^2$ region in order to preserve the 
continuity of form factors in whole $q^2$ region. In the high-$q^2$, the effective number of form factors is four because three of them are reduced by the improved Isgur-Wise relations\cite{Bobeth:2010wg} that
\begin{equation}
	\begin{aligned}
		T_1(q^2)=\kappa V(q^2),\qquad T_2(q^2)=\kappa A_1(q^2),\qquad T_3(q^2)=\kappa A_{2}(q^2)\frac{M_B^2}{q^2}.
	\end{aligned}
\end{equation}
To reduce the length of equations further, three form factors $F_{\perp, \parallel, 0}$ are introduced  following their application to the improved Isgur-Wise relations\cite{Bobeth:2010wg,Bobeth:2012vn}, as follows:
\begin{equation}
	\begin{aligned}
		F_\perp&\equiv \frac{\sqrt{2}\lambda M_B^2}{M_B + M_V}V,
		\\
		F_\parallel&\equiv \sqrt{2}\left(M_B + M_V\right)A_1,
		\\
		F_0&\equiv \frac{\left(M_B^2 - M_V^2 - q^2\right) \left(M_B + M_V\right)^2 A_1 - \lambda^2M_B^4A_2}{2 M_V \left(M_B + M_V\right)\sqrt{q^2}}.
	\end{aligned}
\end{equation}

To ensure the continuity and consistency of work, form factors from LCSR and LQCD combined fit \cite{Bharucha:2015bzk}, which are in the form of a  simplified series expansion (SSE), are also employed  in our practical analysis. Corresponding inputs are summarized in tables (\cref{tab:Input_para,tab:FFs}).

\subsection{\texorpdfstring{$ B\to P\ell^+\ell^- $}{BPll} }
\label{app: BPll_highq2}
In previous work dealing with formulas of large recoil, the form factor $f_+(q^2)$ was considered to be an overall factor. Other two form factors, $f_T$ and $f_0$, were reduced to $\xi_P \equiv f_+$ by symmetry relations\cite{Bobeth:2007dw}. 
To take tensor WCs into consideration, the angular coefficient $I_c^\ell$ is revised as follows:
\begin{equation}\label{eq:ang_coef_BVlowq2}
	\begin{aligned}
		I_c^\ell\left(q^2;C_{7,8,9,10,T(5)}^{\prime}\right)&=\frac{q^2}{M_B^4}\left(\beta_\ell|F_{T5}^\ell|^2 + \beta_\ell^3|F_T^\ell|^2\right)
		+ \frac{2\lambda m_\ell\beta_\ell^2}{M_B^2}\Re(F_{T5}F_A^\ast)
		-\frac{\lambda^2}{4}\beta_\ell^3\left(|F_A^\ell|^2 + |F_V^\ell|^2 \right),
	\end{aligned}
\end{equation}
where form factors related to tensor WCs are defined as 
\begin{equation}
	\begin{aligned}
		F^\ell_T=\frac{2\lambda M_B^2\beta_\ell}{M_B+M_P}\frac{f_T(q^2)}{f_+(q^2)}C_{T}^\ell,
		 \qquad F^\ell_{T5}=F_{T}^\ell \left(C_{T}^\ell\rightarrow C_{T5}^\ell\right),
	\end{aligned}
\end{equation}
while the remaining form factors follow the previous conventions.

With regard to the application of the formulas in low recoil region, it is important to note that they are also applied to Isgur-Wise relations which are examined in this study\cite{Bobeth:2011nj,Bobeth:2012vn}. The effective number of form factors is reduced to two through a relation that 
\begin{equation}\label{eq:BP_highq2ffs}
	\begin{aligned}
		f_T(q^2)=\frac{M_B(M_B + M_P)}{q^2}\kappa f_+(q^2).
	\end{aligned}
\end{equation}
Those scattered formulas are collected below, starting with the angular coefficients:
\begin{equation}\label{eq:ang_coef_BVhighq2}
	\begin{aligned}
		I_a^\ell\left(q^2;C_{7,8,9,10,S,P,T}^{\prime}\right)&=\frac{\lambda^2}{4}\left[
		\rho_1^+ 
		+ \left(\frac{f_0}{f_+}\right)^2\left(\rho^{S+P} + \frac{m_\ell}{\sqrt{q^2}} \rho^{P\times10}\right)
		+ \frac{m_\ell}{M_B}\rho^{T\times79}
		\right],
		\\
		I_c^\ell\left(q^2;C_{7,8,9,10,T(5)}^{\prime}\right)&=\frac{\lambda^2}{4}\left[ \rho_1^T - \rho_1^+ \right],
	\end{aligned}
\end{equation}
as well as 
\begin{equation}
	\begin{aligned}
		\rho_1^+&\equiv\frac{1}{2}\left(
		\left|\kappa\frac{2m_bM_B}{q^2}(C_{7}^{\text{eff}}+ C_{7}^\prime) + (C_9^{\text{eff};\ell} + C_9^{\prime \ell}) +(C_{10}^\ell + C_{10}^{\prime\ell})
		\right|^2 \right.\\
		&\hspace{0.75cm}\left. + \left|
		\kappa\frac{2m_bM_B}{q^2}(C_{7}^{\text{eff}}+ C_{7}^\prime) + (C_9^{\text{eff};\ell} + C_9^{\prime \ell}) - (C_{10}^\ell + C_{10}^{\prime\ell})
		\right|^2\right),
		\\
		\rho^{S+P}&\equiv \frac{q^2\left(M_B^2 - M_P^2\right)^2}{m_b^2\left(m_b-m_s\right)^2\lambda^2M_B^4} 
		\left(\left|C_{S}^\ell+C_{S}^{\prime\ell}\right|^2 + \left|C_{P}^\ell+C_{P}^{\prime\ell}\right|^2\right),
		\\
		\rho^{P\times10}&\equiv \frac{4\sqrt{q^2}\left(M_B^2 - M_P^2\right)^2}{m_b(m_b-m_s)\lambda^2M_B^4} \Re
		\left[\left(C_P^\ell +C_{P}^{\prime\ell}\right)\left(C_{10}^\ell + C_{10}^{\prime\ell}\right)^\ast\right],
		\\
		\rho^{T\times79}&\equiv 16\kappa \frac{M_B^2}{q^2}\Re\left[C_{T}^\ell \left((C_{9}^{\text{eff};\ell} + C_{9}^{\prime\ell}) + \kappa \frac{2m_bM_B}{q^2} (C_{7}^{\text{eff}} + C_7^\prime)\right)^\ast\right],
		\\
		\rho_1^T&\equiv 16\kappa^2\frac{M_B^2}{q^2}\left(|C_{T}^\ell|^2 + |C_{T5}^\ell|^2\right),
		\\
		\kappa&\equiv 1 - \frac{2\alpha_s(\mu)}{3\pi}\log(\frac{\mu}{m_b})+\mathcal{O}\left(\alpha_s^2\right),
	\end{aligned}
\end{equation}
where $\kappa$ is obtained form the lowest order improved Isgur-Wise relations. Effective coefficients $C_{7, 9}^{\text{eff}}$ here have been calculated to include the effects of the matrix element of the four-quark operators\cite{Altmannshofer:2008dz}.

In the entire $q^2$ region, the  form factors of the $B\to P$  transition are adopted from \cite{Gubernari:2018wyi} 
while those of the $B\to V$ transition are extracted from \cite{Bharucha:2015bzk}, and are always consistent. It is no need to reiterate  that the definitions of the element of matrices which can be found in the relevant literature\cite{Wirbel:1985ji, Isgur:1990kf, Bobeth:2007dw,Bobeth:2011nj,Gubernari:2018wyi}. The parameterization of extrapolated form factors $f\in\{ f_+,f_0,f_T, A_{0,1,12}, T_{1,2,23}, V\}$ based on SSE is specifically defined as follows:
\begin{equation}
	\begin{aligned}
		f(q^2)
		&\equiv\frac{1}{1-q^2/m^2_{\text{res},f}}\sum_{k=0}^{2}
		\alpha_k^f \left[z(q^2)-z(0)\right]^k
		,\\
		z(q^2)=\frac{\sqrt{\tau_+-q^2}-\sqrt{\tau_+-\tau_0}}{\sqrt{\tau_+-q^2}+\sqrt{\tau_+-\tau_0}}&,
		~\tau_0=\sqrt{\tau_+}(\sqrt{\tau_+}-\sqrt{\tau_+-\tau_-}),
		~\tau_\pm=\left(M_B\pm M_{P;V}\right)^2,
	\end{aligned}
\end{equation} 
in which $m_{\text{res},f}$ represents the mass of sub-threshold resonances. These are listed, in conjunction with SSE expansion coefficients $\alpha_k^f$, in Table \ref{tab:FFs}.

\subsection{\texorpdfstring{$B\to X_s\ell^+\ell^-$}{BXsll}}
\label{app: BXsll_CT5}
The branching fraction of the inclusive process $B\to X_s\ell\ell$ can be modified by the addition of tensor WCs, as illustrated below:
\begin{equation}\label{eq:BXsll_tensor}
	\begin{aligned}
		\frac{\dd \mathcal{B}_{T(5)}}{\dd q^2}& = \frac{8\mathcal{B}_0}{m_b^8}M_9(q^2)\left(|C_T|^2 + |C_{T5}|^2\right),
	\end{aligned}
\end{equation}
\begin{equation}
	\begin{aligned}
		\mathcal{B}_0\equiv \frac{\mathcal{B}(B\to X_c \ell \nu)}{f_{X_s}(\hat{m}_c)\kappa_{X_s}(\hat{m}_c)}\left(\frac{3\alpha^2}{16\pi^2}\right)\left|\frac{V_{tb}V_{ts}^\ast}{V_{cb}}\right|^2,
	\end{aligned}
\end{equation}
where the space factor $f_{X_s}(z)$ and the QCD correction factor $\kappa_{X_s}$ are given by
\begin{equation}
	\begin{aligned}
		f_{X_s}(z)&=1-8z^2 + 8z^6 - z^8 - 24z^4\ln z,
		\\
		\kappa_{X_s}(z)&= 1-\frac{2\alpha_s(m_b)}{3\pi}\left[\left(\pi^2 - \frac{31}{4}\right) \left(1-z\right)^2 + \frac{3}{2}\right].
	\end{aligned}
\end{equation}
As demonstrated in the work of \cite{Fukae:1998qy}, the kinematic function $M_9$ is defined as follows: 
\begin{equation}
	\begin{aligned}
		M_9(s)&= 2u(s)\left[-\frac{2}{3}u(s)^2 - 2\left(m_b^2 + m_s^2\right)s + 2(m_b^2 - m_s^2)^2\right],\\
		u(s)&=\sqrt{
			\left[s-\left(m_b + m_s\right)^2\right] 
			\left[s-\left(m_b - m_s\right)^2\right]
			\left(1-4m_\ell^2/s\right)}.
	\end{aligned}
\end{equation}

\newpage

\section{SUMMARY OF INPUT PARAMETERS}
\label{app:input}

In this part, we summarize all the theoretical inputs
adopted in the analysis.

\begin{table}[ht]
	\centering
	\setlength{\tabcolsep}{6pt}
	\caption{\label{tab:Input_para}
		Input parameters I: some basic parameters in the numerical analysis. 	}
	\renewcommand\arraystretch{1.1}
	\begin{tabular}{lccc}
		\hline\hline
		Parameters& Values& Parameters & Values \\
		\hline
		$ m_{b} $& 4.18( $\!^{+4}_{-3} $)~GeV \cite{ParticleDataGroup:2024cfk}& $ m_{t} $ & 172.57(29)~GeV\cite{ParticleDataGroup:2024cfk} \\
		$ m_{c} $&1.2730$ (46) $~GeV\cite{ParticleDataGroup:2024cfk}&$ m_{s} $ &93.5$\left(8\right)$~MeV\cite{ParticleDataGroup:2024cfk}\\
		$ m_{d} $&4.70$ (7) $~MeV\cite{ParticleDataGroup:2024cfk}&$ m_{u} $&2.16$( 7 )$~MeV\cite{ParticleDataGroup:2024cfk}\\
		$ m_{e} $&0.51099895000$ (15) $~MeV\cite{ParticleDataGroup:2024cfk}&$ m_{\mu} $&105.6583755$ (23) $~MeV\cite{ParticleDataGroup:2024cfk}\\
		$ m_{B_u} $&5279.41(7)~MeV\cite{ParticleDataGroup:2024cfk}&$ m_{B_s} $&5366.93(10)~MeV\cite{ParticleDataGroup:2024cfk}\\
		$ m_{B_d} $&5279.72(8)~MeV\cite{ParticleDataGroup:2024cfk}&$ m_\phi $&1019.461(16)~MeV\cite{ParticleDataGroup:2024cfk}\\
		$ m_{K^\pm} $&493.677(13)~MeV\cite{ParticleDataGroup:2024cfk}&$ m_{K^0} $&497.611(13)~MeV\cite{ParticleDataGroup:2024cfk}\\

		$ m_{K^{\ast\pm}} $&891.67(26)~MeV\cite{ParticleDataGroup:2024cfk}&$ m_{K^{\ast0}} $&895.55(20)~MeV\cite{ParticleDataGroup:2024cfk}\\
		$ m_{\Lambda_b} $&5619.60(17)~MeV\cite{ParticleDataGroup:2024cfk}&$ m_{\Lambda} $&1115.683(6)~MeV\cite{ParticleDataGroup:2024cfk}\\
		$ \tau_{B_u} $&1.638(4)~ps\cite{ParticleDataGroup:2024cfk}&$ \tau_{B_s} $&1.520(5)~ps\cite{ParticleDataGroup:2024cfk}\\
		
		$ \tau_{B_{d}} $&1.517$\left(4\right)$~ps\cite{ParticleDataGroup:2024cfk}
		&$ \tau_{\Lambda_b} $&1.471(9)~ps\cite{ParticleDataGroup:2024cfk}\\
		$ f_{B_s} $&230.7(1.3)~MeV\cite{ParticleDataGroup:2024cfk}&$ f_{B_{d}} $&190.0(1.3)~MeV\cite{ParticleDataGroup:2024cfk}\\
		$ f_{\Lambda} $&6.0(4)$\times10^{-3}$~GeV$^2$\cite{Aslam:2008hp}&$ f_{\Lambda_b} $&$ 3.9(^{+4}_{-2})\times10^{-3} $~GeV$^2$\cite{Aslam:2008hp}\\
		$ m_t(m_t) $&162.690(6)~GeV\cite{ParticleDataGroup:2024cfk}&$ G_F $&1.1663788(6)~GeV$ ^{-2} $\cite{ParticleDataGroup:2024cfk}\\
		\hline
		$ \alpha_\Lambda $&0.642(13)\cite{Detmold:2016pkz}&$\alpha_{\text{e}}(m_{Z})$&1/127.951(9)\cite{ParticleDataGroup:2024cfk}\\
		$\alpha_s(m_Z)$&0.1180(9)\cite{ParticleDataGroup:2024cfk}&$ \sin^2\theta_W $&0.23122(4)\cite{ParticleDataGroup:2024cfk}\\
		$y_s$&0.064(4)\cite{ParticleDataGroup:2024cfk}&$y_d$&0.0005(50)\cite{ParticleDataGroup:2024cfk}\\
		$ \mu^2_{G} $&0.336$\pm0.064 $\cite{Gambino:2013rza}&&\\
		$ \rho_D^3 $&$ 0.153\pm0.45 $\cite{Gambino:2013rza}&$ \rho_{LS}^3 $&$ -0.145\pm0.098 $\cite{Gambino:2013rza}\\
		$ \mathcal{B}(B\to X_ce\bar{\nu})_{\text{exp}} $&$ 0.997(41)$\cite{Belle-II:2021jlu}&$ \mathcal{B}(B\to X_c\ell\bar{\nu})_{\text{exp}} $&$ 0.975(50) $\cite{Belle-II:2021jlu}\\
		\hline
		$ \lambda $&0.22501(68)\cite{ParticleDataGroup:2024cfk}&$ A $&0.826$ (^{+16}_{-15}) $\cite{ParticleDataGroup:2024cfk}\\
		$\bar{\rho}$&$ 0.1591(94) $\cite{ParticleDataGroup:2024cfk}&$\bar{\eta}$&$ 0.3523(^{+73}_{-71}) $\cite{ParticleDataGroup:2024cfk}\\
		\hline\hline
	\end{tabular}
\end{table}

\begin{table}[!ht]
	\centering
	\setlength{\tabcolsep}{6pt}
	\caption{
		Input parameters II: form factors as well as resonance pole mass used in the numerical analysis of $ B\to P\ell^+\ell^- $\cite{Gubernari:2018wyi} and $ B\to V\ell^+\ell^-$\cite{Gubernari:2018wyi,Bharucha:2015bzk}, respectively. }
	\begin{tabular}{lccccc}
		\hline\hline
		Parameters & Values&Parameters& Values& Parameters & Values \\
		\hline
		$ a_0^{f_0}(K) $&$+0.33\pm0.03$&$ a_0^{A_0}(K^\ast) $&$ +0.34\pm0.03 $&$ a_0^{A_0}(\phi) $&$ +0.42\pm0.02 $\\
		$ a_1^{f_0}(K) $&$+0.20\pm0.17$&$ a_1^{A_0}(K^\ast) $&$ -1.13\pm0.20 $&$ a_1^{A_0}(\phi) $&$ -0.98\pm0.24 $\\
		$ a_2^{f_0}(K) $&$-0.45\pm0.41$&$ a_2^{A_0}(K^\ast) $&$ +2.33\pm1.79 $&$ a_2^{A_0}(\phi) $&$ +3.27\pm1.36 $
		\\
		$ a_0^{f_+}(K) $&$+0.33\pm0.03$&$ a_0^{A_1}(K^\ast) $&$ +0.30\pm0.03 $&$ a_0^{A_1}(\phi) $&$ +0.29\pm0.01 $\\
		$ a_2^{f_+}(K) $&$+0.01\pm0.75$&$ a_1^{A_1}(K^\ast) $&$ +0.50\pm0.15 $&$ a_1^{A_1}(\phi) $&$ +0.35\pm0.10 $\\
		$ a_2^{f_T}(K) $&$+0.01\pm0.87$&$ a_2^{A_1}(K^\ast) $&$ +1.14\pm0.75 $&$ a_2^{A_1}(\phi) $&$ +1.70\pm0.79 $
		\\
		$ a_0^{f_T}(K) $&$+0.30\pm0.03$&$ a_0^{A_{12}}(K^\ast) $&$ +0.30\pm0.02$&$ a_0^{A_{12}}(\phi) $&$ +0.27\pm0.02 $\\
		$ a_1^{f_T}(K) $&$-0.77\pm0.15$&$ a_1^{A_{12}}(K^\ast) $&$ +0.56\pm0.35 $&$ a_1^{A_{12}}(\phi) $&$ +0.95\pm0.13 $\\
		$ a_1^{f_+}(K) $&$-0.87\pm0.14$&$ a_2^{A_{12}}(K^\ast) $&$ +0.66\pm 2.08 $&$ a_2^{A_{12}}(\phi) $&$ +2.15\pm0.48 $
		\\
		&&$ a_0^{V}(K^\ast) $&$ +0.38\pm0.04 $&$ a_0^{V}(\phi) $&$ +0.36\pm0.01 $\\
		&&$ a_1^{V}(K^\ast) $&$ -1.10\pm0.17 $&$ a_1^{V}(\phi) $&$ -1.22\pm0.16 $\\
		&&$ a_2^{V}(K^\ast) $&$ +2.11\pm2.15 $&$ a_2^{V}(\phi) $&$ +3.74\pm1.73 $
		\\
		&&$ a_0^{T_1}(K^\ast) $&$ +0.34\pm0.03 $&$ a_0^{T_1}(\phi) $&$ +0.30\pm0.01 $\\
		&&$ a_1^{T_1}(K^\ast) $&$ -0.87\pm0.39 $&$ a_1^{T_1}(\phi) $&$ -1.10\pm0.08 $\\
		&&$ a_2^{T_1}(K^\ast) $&$ -0.12\pm5.83 $&$ a_2^{T_1}(\phi) $&$ +0.58\pm1.00 $
		\\
		&&$ a_0^{T_2}(K^\ast) $&$ +0.34\pm0.03 $&$ a_0^{T_2}(\phi) $&$ +0.30\pm0.01 $\\
		&&$ a_1^{T_2}(K^\ast) $&$ +0.70\pm0.36 $&$ a_1^{T_2}(\phi) $&$ +0.40\pm0.08 $\\
		&&$ a_2^{T_2}(K^\ast) $&$ -1.11\pm6.28 $&$ a_2^{T_2}(\phi) $&$ +1.04\pm0.61 $
		\\
		&&$ a_0^{T_{23}}(K^\ast) $&$ +0.65\pm0.05 $&$ a_0^{T_{23}}(\phi) $&$ +0.65\pm0.04 $\\
		&&$ a_1^{T_{23}}(K^\ast) $&$ +0.73\pm0.74 $&$ a_1^{T_{23}}(\phi) $&$ +2.10\pm0.33 $\\
		&&$ a_2^{T_{23}}(K^\ast) $&$ -0.31\pm13.28 $&$ a_2^{T_{23}}(\phi) $&$ +6.74\pm1.80 $
		\\
		\hline
		$ m_{\text{res},f_0}(K^\ast) $&5.630 GeV&$ m_{\text{res},A_0}(K^\ast) $&5.366 GeV&$ m_{\text{res},A_0}(\phi) $&5.366 GeV
		\\
		$ m_{\text{res},f_+}(K^\ast) $&5.412 GeV&$ m_{\text{res},T_1;V}(K^\ast) $&5.412 GeV&$ m_{\text{res},T_1}(\phi) $&5.415 GeV
		\\
		$ m_{\text{res},f_T}(K^\ast) $&5.412 GeV&$ m_{\text{res},A_{1(2)};T_{2(3)}}(K^\ast) $&5.829 GeV&$ m_{\text{res},T_1}(\phi) $&5.829 GeV
		\\
		\hline\hline
	\end{tabular}
	\label{tab:FFs}
\end{table}

\section{EXPERIMENTAL DATA FOR RELATED OBSERVABLES}
\label{app:exp_input}
Here  we summarize all the  experimental results  related to our analysis. The number of observables is 490 at total. The detailed values have been presented in the following four tables ( Tables \crefrange{tab:old_data1}{tab:new_data2})  while the SM predictions in last column are calculated by our code supporting this analysis. 
For a comparison,
theoretical predictions in different approaches, including the \textit{Flavio} package  \cite{Straub:2018kue} as well as this analysis, are also provided 
in the tables.

\begin{center}\scriptsize
	\setlength{\tabcolsep}{10pt}
	\renewcommand\arraystretch{1.5}
	\captionof{table}{The differential branching fractions part of in the unit of  $\text{GeV}^{-2} $.}
	\label{tab:old_data1}
	\tablefirsthead{
	\toprule
		\multicolumn{1}{l}{Observable} 
		& \multicolumn{1}{c}{$q^2$ (GeV$^{2}$)} 
		&\multicolumn{1}{r}{Experimental value}  
		&\multicolumn{1}{r}{Previous work \cite{Wen:2023pfq}}  
		&\multicolumn{1}{r}{\textit{Flavio} \cite{Straub:2018kue} }
		&\multicolumn{1}{r}{This analysis}\\
	\hline
		}
	\tablehead{
	\multicolumn{6}{l}{\small \sl Table continued from previous page}\\
	\toprule
	\multicolumn{1}{l}{Observable} 
	& \multicolumn{1}{c}{$q^2$ (GeV$^{2}$)} 
	&\multicolumn{1}{r}{Experimental value}  
	&\multicolumn{1}{r}{Previous work \cite{Wen:2023pfq}}  
	&\multicolumn{1}{r}{\textit{Flavio} \cite{Straub:2018kue} }
	&\multicolumn{1}{r}{This analysis}\\
	\hline
	}
	\tabletail{
		\hline
		\multicolumn{6}{r}{\small \sl Table continued on next page}\\
	}
	\tablelasttail{
	\toprule
	}
	\begin{supertabular}
	{L{0.12\textwidth} L{0.08\textwidth} R{0.18\textwidth} R{0.13\textwidth}  R{0.12\textwidth} R{0.12\textwidth}}
	\multicolumn{6}{c}{LHCb $(B\to K^{(\ast)}\ell^+\ell^-)$\cite{LHCb:2022vje}}\\
	\hline
	$R_{K^{+}}$&$\left[1.1,6.0\right]$
	&$0.949^{+0.042+0.022}_{-0.041-0.022}$   &$1.000\pm0.000$ &$1.001\pm0.000$ &$1.001\pm0.000$\\
	
	$R_{K^{+}}$&$\left[0.1,1.1\right]$
	&$0.994^{+0.090+0.029}_{-0.082-0.027}$  &$0.994\pm0.000$ &$0.993\pm0.000$ &$0.993\pm0.000$\\

	$R_{K^{\ast 0}}$&$\left[1.1,6.0\right]$
	&$1.027^{+0.072+0.027}_{-0.068-0.026}$ &$0.996\pm0.000$ &$0.996\pm0.001$ &$0.996\pm0.001$\\

	$R_{K^{\ast 0}}$&$\left[0.1,1.1\right]$
	&$0.927^{+0.093+0.036}_{-0.087-0.035}$ &$0.983\pm0.000$ &$0.983\pm0.001$ &$0.982\pm0.001$\\
	
	$\frac{10^9\dd \mathcal{B}(K^+e^+e^-)}{\dd q^2}$
	&$\left[1.1,6.0\right]$&$25.5^{+1.3+1.1}_{-1.2-1.1}$ &$36.474\pm11.5$ &$34.841\pm5.392$&$36.577\pm5.876$\\
	
	$\frac{10^9\dd\mathcal{B}(K^{\ast0}e^+e^-)}{\dd q^2}$
	&$\left[1.1,6.0\right]$&$33.3^{+2.7+2.2}_{-2.6-2.2}$  &$41.541\pm8.73$ &$47.580\pm5.727$&$40.106\pm5.898$\\
	
	\hline
	\multicolumn{6}{c}{LHCb $(B\to K^{(\ast)}\ell^+\ell^-)$\cite{LHCb:2021lvy}}\\
	\hline
	$R_{K^{\ast+}}$&$\left[0.045,6.0\right]$
	&$0.70^{+0.18+0.03}_{-0.13-0.04}$ &$0.974\pm0.000$ &$0.972\pm0.003$ &$0.974\pm0.000$\\
	
	$R_{K_S}$&$\left[1.1,6.0\right]$
	&$0.66^{+0.20+0.02}_{-0.14-0.04}$ &$1.000\pm0.000$ &$1.001\pm0.000$ &$1.001\pm0.000$\\
	
	$\frac{10^8\dd\mathcal{B}(K^0e^+e^-)}{\dd q^2}$
	&$\left[1.1,6.0\right]$&$2.6^{+0.6+0.1}_{-0.6-0.1}$ &$3.383\pm1.045$ &$3.230\pm0.513$&$3.382\pm0.543$\\
	
	$\frac{10^8d\mathcal{B}(K^{\ast+}e^+e^-)}{\dd q^2}$
	&$\left[0.045,6.0\right]$&$9.2^{+1.9+0.8}_{-1.8-0.6}$ &$5.639\pm1.036$ &$6.539\pm0.810$&$5.979\pm0.895$\\

	\hline
	\multicolumn{6}{c}{Belle $(B \to K^{\ast}\ell^+\ell^-)$\cite{Belle:2019oag}}\\
	
	\hline
	\multirow{1}*{$R_{K^{\ast+}}$}
	&$[0.045,1.1]$&$0.62^{+0.60}_{-0.36}\pm0.09$ &$0.932\pm0.000$ &$0.926\pm0.006$&$0.920\pm0.002$\\
	
	&$[1.1,6.0]$&$0.72^{+0.99}_{-0.44}\pm0.15$ &$0.996\pm0.000$ &$0.996\pm0.001$&$0.996\pm0.001$\\
	&$[0.1,8.0]$&$0.96^{+0.56}_{-0.35}\pm0.13$ &&$0.994\pm0.001$&$0.992\pm0.001$\\
	&$[15.0,19.0]$&$1.40^{+1.99}_{-0.68}\pm0.11$ & &$0.998\pm0.000$&$0.998\pm0.000$\\
	&$[0.045,19.0]$&$0.70^{+0.24}_{-0.19}\pm0.06$ & &$0.990\pm0.001$&$0.990\pm0.001$\\
	
	$10^7\mathcal{B}(K^{\ast+}e^+e^-)$
	&$[1.1,6.0]$&$1.7^{+1.0}_{-1.0}\pm0.2$ &$2.227\pm0.464$ &$2.546\pm0.470$&$2.119\pm0.312$\\
	
	&$[0.1,8.0]$&$4.6^{+1.6}_{-1.5}\pm0.7$ & &$4.256\pm0.611$&$4.294\pm0.620$\\
	
	&$[15.0,19.0]$&$2.1^{+1.2}_{-1.0}\pm0.2$ & &$2.555\pm0.263$&$2.395\pm0.265$\\
	
	&$[0.045,19.0]$&$14.1^{+3.1}_{-2.8}\pm1.8$ & &$13.353\pm1.449$&$12.577\pm1.533$\\

	$10^7\mathcal{B}(K^{\ast+}\mu^+\mu^-)$
	&$[1.1,6.0]$&$1.2^{+0.9}_{-0.7}\pm0.2$ &$2.219\pm0.465$ &$2.537\pm0.405$&$2.110\pm0.311$\\
	
	&$[0.1,8.0]$&$4.4^{+1.6}_{-1.4}\pm0.5$ & &$4.842\pm0.689$&$4.262\pm0.615$\\
	
	&$[15.0,19.0]$&$2.9^{+1.0}_{-0.8}\pm0.3$ & &$2.550\pm0.291$&$2.390\pm0.264$\\
	
	&$[0.045,19.0]$&$9.9^{+2.4}_{-2.3}\pm1.1$ & &$13.224\pm1.442$&$12.448\pm1.514$\\
	
	\hline
	
	\multirow{1}*{$R_{K^{\ast0}}$}
	&$[0.045,1.1]$&$0.46^{+0.55}_{-0.27}\pm0.13$ &$0.931\pm0.000$ &$0.925\pm0.004$&$0.920\pm0.002$\\
	&$[1.1,6.0]$&$1.06^{+0.63}_{-0.38}\pm0.13$ &$0.996\pm0.000$ &$0.996\pm0.001$&$0.996\pm0.001$\\
	&$[0.1,8.0]$&$0.86^{+0.33}_{-0.24}\pm0.09$ & &$0.994\pm0.001$&$0.992\pm0.001$\\
	&$[15,19]$&$1.12^{+0.61}_{-0.36}\pm0.10$ & &$0.998\pm0.000$&$0.998\pm0.000$\\
	&$[0.045,19]$&$1.12^{+0.27}_{-0.21}\pm0.09$ & &$0.990\pm0.001$&$0.990\pm0.001$\\
	
	$10^7\mathcal{B}(K^{\ast0}e^+e^-)$
	&$[1.1,6.0]$&$1.8^{+0.6}_{-0.6}\pm0.2$ &$2.035\pm0.430$ &$2.331\pm0.286$&$1.965\pm0.289$\\
	
	$10^7\mathcal{B}(K^{\ast0}e^+e^-)$
	&$[0.1,8.0]$&$3.7^{+0.9}_{-0.9}\pm0.4$ & &$4.516\pm0.563$&$3.982\pm0.574$\\
	
	$10^7\mathcal{B}(K^{\ast0}e^+e^-)$
	&$[15.0,19.0]$&$2.0^{+0.6}_{-0.5}\pm0.2$ & &$2.371\pm0.254$&$2.197\pm0.243$\\
	
	$10^7\mathcal{B}(K^{\ast0}e^+e^-)$
	&$[0.045,19.0]$&$9.2^{+1.6}_{-1.6}\pm0.8$ & &$12.367\pm1.410$&$11.588\pm1.415$\\
	
	$10^7\mathcal{B}(K^{\ast0}\mu^+\mu^-)$
	&$[1.1,6.0]$&$1.9^{+0.6}_{-0.5}\pm0.3$ &$2.028\pm0.426$ &$2.323\pm0.315$&$1.957\pm0.288$\\
	
	$10^7\mathcal{B}(K^{\ast0}\mu^+\mu^-)$
	&$[0.1,8.0]$&$3.2^{+0.8}_{-0.8}\pm0.4$ & &$4.487\pm0.577$&$3.952\pm0.570$\\
	
	$10^7\mathcal{B}(K^{\ast0}\mu^+\mu^-)$
	&$[15.0,19.0]$&$2.2^{+0.5}_{-0.4}\pm0.2$ & &$2.367\pm0.260$&$2.193\pm0.243$\\	
	
	$10^7\mathcal{B}(K^{\ast0}\mu^+\mu^-)$
	&$[0.045,19.0]$&$10.3^{+1.3}_{-1.3}\pm1.1$ & &$12.242\pm1.330$&$11.468\pm1.398$\\
	
	\hline
	\multicolumn{6}{c}{Belle $(B\to K^\ast\gamma)$\cite{BelleII:2021tzi}}\\			
	\hline
	$10^5\mathcal{B}\left(K^{\ast0}\gamma\right)$
	&&$4.5\pm0.3\pm0.2$ &$4.146\pm0.420$ &$4.202\pm0.845$&$4.091\pm0.414$\\				
	
	$10^5\mathcal{B}\left(K^{\ast+}\gamma\right)$
	&&$5.2\pm0.4\pm0.3$ &$4.474\pm0.454$ &$4.271\pm0.967$&$4.420\pm0.447$\\			
	\hline
	\multicolumn{6}{c}{Belle $(B^+\to K^{+} \ell^+\ell^-)$\cite{BELLE:2019xld}}\\
	\hline
	$10^7\mathcal{B}(K^{+}\mu^+\mu^-)$&$[0.1,4.0]$
	&$1.76^{+0.41}_{-0.37}\pm0.04$ &$1.444\pm0.434$ &$1.370\pm0.244$&$1.436\pm0.236$\\
	
	$10^7\mathcal{B}(K^{0}_S\mu^+\mu^-)$&$[0.1,4.0]$
	&$0.62^{+0.30}_{-0.23}\pm0.02$ &$0.670\pm0.202$ &$0.635\pm0.109$&$0.663\pm0.109$\\
	
	$10^7\mathcal{B}(K^{+}e^+e^-)$&$[0.1,4.0]$
	&$1.80^{+0.33}_{-0.30}\pm0.05$ &$1.446\pm0.442$ &$1.371\pm0.244$&$1.438\pm0.237$\\
	
	$10^7\mathcal{B}(K^{0}_Se^+e^-)$&$[0.1,4.0]$
	&$0.38^{+0.25}_{-0.19}\pm0.01$ &$0.671\pm0.201$ &$0.636\pm0.111$&$0.664\pm0.109$\\
	
	$R_{K^+}$&$[0.1,4.0]$
	&$0.98^{+0.29}_{-0.26}\pm0.02$ &$0.999\pm0.000$ &$0.999\pm0.000$&$0.999\pm0.000$\\
	
	$R_{K^0_S}$&$[0.1,4.0]$
	&$1.62^{+1.31}_{-1.01}\pm0.02$ &$0.999\pm0.000$ &$0.999\pm0.000$&$0.999\pm0.000$\\

	\hline
	$10^7\mathcal{B}(K^{+}\mu^+\mu^-)$&$[1.0,6.0]$
	&$2.30^{+0.41}_{-0.38}\pm0.05$ &$1.825\pm0.570$ &$1.744\pm0.339$&$1.831\pm0.294$\\
	
	$10^7\mathcal{B}(K^{0}_S\mu^+\mu^-)$&$[1.0,6.0]$
	&$0.31^{+0.22}_{-0.16}\pm0.01$ &$0.846\pm0.264$&$0.808\pm0.140$&$0.846\pm0.136$\\
	
	$10^7\mathcal{B}(K^{+}e^+e^-)$&$[1.0,6.0]$
	&$1.66^{+0.32}_{-0.29}\pm0.04$ &$1.825\pm0.581$ &$1.743\pm0.284$&$1.829\pm0.294$\\
	
	$10^7\mathcal{B}(K^{0}_Se^+e^-)$&$[1.0,6.0]$
	&$0.56^{+0.25}_{-0.20}\pm0.02$ &$0.846\pm0.265$ &$0.808\pm0.130$&$0.846\pm0.136$\\
	
	$R_{K^+}$&$[1.0,6.0]$
	&$1.39^{+0.36}_{-0.33}\pm0.02$ &$1.000\pm0.000$ &$1.001\pm0.000$&$1.001\pm0.000$\\
	
	$R_{K^0_S}$&$[1.0,6.0]$
	&$0.55^{+0.46}_{-0.34}\pm0.01$ &$1.000\pm0.000$ &$1.001\pm0.000$&$1.001\pm0.000$\\
	
	\hline
	$10^7\mathcal{B}(K^{+}\mu^+\mu^-)$&$[4.00,8.12]$
	&$1.24^{+0.28}_{-0.25}\pm0.03$ & &$1.410\pm0.231$&$1.483\pm0.227$\\
	
	$10^7\mathcal{B}(K^{0}_S\mu^+\mu^-)$&$[4.00,8.12]$
	&$0.27^{+0.18}_{-0.13}\pm0.01$ & &$0.654\pm0.114$&$0.686\pm0.105$\\
	
	$10^7\mathcal{B}(K^{+}e^+e^-)$&$[4.00,8.12]$
	&$0.96^{+0.24}_{-0.22}\pm0.03$ & &$1.409\pm0.209$&$1.482\pm0.227$\\
	
	$10^7\mathcal{B}(K^{0}_Se^+e^-)$&$[4.00,8.12]$
	&$0.52^{+0.21}_{-0.17}\pm0.02$ & &$0.653\pm0.108$&$0.685\pm0.105$\\
	
	$R_{K^+}$&$[4.00,8.12]$
	&$1.29^{+0.44}_{-0.39}\pm0.02$ & &$1.001\pm0.000$&$1.001\pm0.000$\\
	
	$R_{K^0_S}$&$[4.00,8.12]$
	&$0.51^{+0.41}_{-0.31}\pm0.01$ & &$1.001\pm0.000$&$1.001\pm0.000$\\

	\hline
	$10^7\mathcal{B}(K^{+}\mu^+\mu^-)$&$[10.2,12.8]$
	&$0.86^{+0.22}_{-0.20}\pm0.02$ & &$0.819\pm0.104$&$0.862\pm0.113$\\
	
	$10^7\mathcal{B}(K^{0}_S\mu^+\mu^-)$&$[10.2,12.8]$
	&$0.29^{+0.17}_{-0.13}\pm0.01$ & &$0.379\pm0.046$&$0.398\pm0.052$\\
	
	$10^7\mathcal{B}(K^{+}e^+e^-)$&$[10.2,12.8]$
	&$0.44^{+0.20}_{-0.17}\pm0.01$ & &$0.818\pm0.097$&$0.861\pm0.113$\\
	
	$10^7\mathcal{B}(K^{0}_Se^+e^-)$&$[10.2,12.8]$
	&$0.06^{+0.19}_{-0.15}\pm0.01$ & &$0.379\pm0.046$&$0.398\pm0.052$\\
	
	$R_{K^+}$&$[10.2,12.8]$
	&$1.96^{+1.03}_{-0.89}\pm0.02$ & &$1.001\pm0.000$&$1.001\pm0.000$\\
	
	$R_{K^0_S}$&$[10.2,12.8]$
	&$5.18^{+17.69}_{-14.32}\pm0.06$ & &$1.001\pm0.000$&$1.001\pm0.000$\\
	
	\hline
	$10^7\mathcal{B}(K^{+}\mu^+\mu^-)$&$[14.18,22.90]$
	&$1.34^{+0.24}_{-0.22}\pm0.03$ & &$1.276\pm0.139$&$1.230\pm0.123$\\
	
	$10^7\mathcal{B}(K^{0}_S\mu^+\mu^-)$&$[14.18,22.90]$
	&$0.49^{+0.22}_{-0.18}\pm0.01$ & &$0.587\pm0.065$&$0.565\pm0.055$\\
	
	$10^7\mathcal{B}(K^{+}e^+e^-)$&$[14.18,22.90]$
	&$1.18^{+0.25}_{-0.22}\pm0.03$ & &$1.272\pm0.158$&$1.230\pm0.123$\\
	
	$10^7\mathcal{B}(K^{0}_Se^+e^-)$&$[14.18,22.90]$
	&$0.32^{+0.21}_{-0.17}\pm0.01$ & &$0.585\pm0.073$&$0.565\pm0.055$\\
	
	$R_{K^+}$&$[14.18,22.90]$
	&$1.13^{+0.31}_{-0.28}\pm0.01$ & &$1.003\pm0.001$&$1.000\pm0.000$\\
	
	$R_{K^0_S}$&$[14.18,22.90]$
	&$1.57^{+1.28}_{-1.00}\pm0.02$ & &$1.003\pm0.001$&$1.000\pm0.000$\\
	\hline
	
	\multicolumn{6}{c}{LHCb $(B^+\to K^{+} \mu^+\mu^-)$\cite{LHCb:2014cxe}}\\
	\hline
	\multirow{1}*{$10^9d\mathcal{B}/dq^2$}
	&$[0.1,0.98]$&$33.2\pm1.8\pm1.7$ & &$35.202\pm6.150$&$36.884\pm6.202$\\
	&$[1.1,2.0]$&$23.3\pm1.5\pm1.2$ &$37.243\pm11.219$ &$35.256\pm5.751$&$36.973\pm6.132$\\
	&$[2.0,3.0]$&$28.2\pm1.6\pm1.4$ &$36.911\pm11.308$ &$35.095\pm5.312$&$36.824\pm6.023$\\
	&$[3.0,4.0]$&$25.4\pm1.5\pm1.3$ &$36.540\pm11.480$ &$34.908\pm5.031$&$36.646\pm5.901$\\
	&$[4.0,5.0]$&$22.1\pm1.4\pm1.1$ &$36.128\pm11.715$ &$34.689\pm4.670$&$36.434\pm5.768$\\
	&$[5.0,6.0]$&$23.1\pm1.4\pm1.2$ &$35.664\pm11.996$ &$34.429\pm5.459$&$36.183\pm5.622$\\
	&$[1.1,6.0]$&$24.2\pm0.7\pm1.2$ &$36.482\pm11.472$ &$34.868\pm6.136$&$36.605\pm5.878$\\
	
	&$[6.0,7.0]$&$24.5\pm1.4\pm1.2$ & &$36.125\pm5.304$&$35.888\pm5.460$\\
	&$[7.0,8.0]$&$23.1\pm1.4\pm1.2$ & &$33.777\pm4.783$&$35.551\pm5.283$\\
	&$[11.0,11.8]$&$17.7\pm1.3\pm0.9$ & &$32.124\pm4.794$&$33.920\pm4.455$\\
	&$[11.8,12.5]$&$19.3\pm1.2\pm1.0$ & &$30.890\pm4.278$&$32.453\pm4.132$\\
	&$[15.0,16.0]$&$16.1\pm1.0\pm0.8$ & &$23.838\pm2.771$&$22.624\pm2.397$\\
	&$[16.0,17.0]$&$16.4\pm1.0\pm0.8$ & &$21.420\pm2.574$&$20.570\pm2.063$\\
	&$[17.0,18.0]$&$20.6\pm1.1\pm1.0$ & &$18.734\pm1.991$&$18.174\pm1.741$\\
	&$[18.0,19.0]$&$13.7\pm1.0\pm0.7$ & &$15.727\pm1.744$&$15.393\pm1.447$\\
	&$[19.0,20.0]$&$7.4\pm0.8\pm0.4$ & &$12.359\pm1.281$&$12.188\pm1.194$\\
	&$[20.0,21.0]$&$5.9\pm0.7\pm0.3$ & &$8.625\pm1.026$&$8.558\pm0.983$\\
	&$[21.0,22.0]$&$4.3\pm0.7\pm0.2$ & &$4.640\pm0.462$&$4.618\pm0.786$\\
	&$[15.0,22.0]$&$12.1\pm0.4\pm0.6$ & &$15.049\pm1.732$&$14.589\pm1.425$\\
	\hline
	\multicolumn{6}{c}{LHCb $(B^0\to K^{0} \mu^+\mu^-)$\cite{LHCb:2014cxe}}  \\
	\hline
	\multirow{1}*{$10^9d\mathcal{B}/dq^2$}
	&$[0.1,2.0]$&$12.2^{+5.9}_{-5.2}\pm0.6$ &$34.658\pm10.247$ &$34.668\pm5.640$&$34.063\pm5.687$\\
	&$[2.0,4.0]$&$18.7^{+5.5}_{-4.9}\pm0.9$ &$34.073\pm10.450$ &$32.448\pm6.038$&$33.969\pm5.512$\\
	&$[4.0,6.0]$&$17.3^{+5.3}_{-4.8}\pm0.9$ &$33.283\pm10.899$ &$32.034\pm5.647$&$33.594\pm5.267$\\
	&$[1.1,6.0]$&$18.7^{+3.5}_{-3.2}\pm0.9$ &$33.842\pm10.537$ &$33.323\pm5.878$&$33.847\pm5.434$\\
	&$[6.0,8.0]$&$27.0^{+5.8}_{-5.3}\pm1.4$ & &$31.465\pm4.859$&$33.047\pm4.968$\\
	&$[11.0,12.5]$&$12.7^{+4.5}_{-4.0}\pm0.6$ & &$29.213\pm4.186$&$30.715\pm3.972$\\
	&$[15.0,17.0]$&$14.3^{+3.5}_{-3.2}\pm0.7$ & &$20.913\pm2.429$&$19.911\pm2.035$\\
	&$[17.0,22.0]$&$7.8^{+1.7}_{-1.5}\pm0.4$ & &$11.031\pm1.258$&$10.789\pm1.012$\\
	&$[15.0,22.0]$&$9.5^{+1.6}_{-1.5}\pm0.5$ & &$13.855\pm1.327$&$13.395\pm1.266$\\
	\hline
	\multicolumn{6}{c}{LHCb $(B^+\to K^{\ast+} \mu^+\mu^-)$\cite{LHCb:2014cxe}}  \\
	\hline
	\multirow{1}*{$10^9d\mathcal{B}/dq^2$}
	&$[0.1,2.0]$&$59.2^{+14.4}_{-13.0}\pm4.0$ &$68.174\pm11.994$ &$79.748\pm9.700$&$78.697\pm12.736$\\
	
	&$[2.0,4.0]$&$55.9^{+15.9}_{-14.4}\pm3.8$ &$42.981\pm9.597$ &$48.903\pm7.539$&$40.124\pm6.046$\\
	
	&$[4.0,6.0]$&$24.9^{+11.0}_{-9.6}\pm1.7$ &$47.412\pm9.431$ &$54.486\pm7.413$&$45.076\pm6.505$\\
	
	&$[1.1,6.0]$&$36.6^{+8.3}_{-7.6}\pm2.6$ &$45.294\pm9.577$ &$51.772\pm7.654$&$43.053\pm6.338$\\
	
	&$[6.0,8.0]$&$33.0^{+11.3}_{-10.0}\pm2.3$ & &$62.964\pm10.402$&$53.119\pm7.393$\\
	
	&$[11.0,12.5]$&$82.8^{+15.8}_{-14.1}\pm5.6$ & &$84.039\pm9.012$&$83.186\pm10.457$\\
	
	&$[15.0,17.0]$&$64.4^{+12.9}_{-11.5}\pm4.4$ & &$75.801\pm8.515$&$71.106\pm7.844$\\
	
	&$[17.0,19.0]$&$11.6^{+9.1}_{-7.6}\pm0.8$ & &$51.707\pm6.117$&$48.410\pm5.554$\\
	
	&$[15.0,19.0]$&$39.5^{+8.0}_{-7.3}\pm2.8$ & &$63.754\pm7.576$&$59.758\pm6.607$\\
	\hline
	\multicolumn{6}{c}{LHCb $(B^0\to K^{\ast0}\mu^+\mu^-)$\cite{LHCb:2016ykl}}\\
	\hline
	\multirow{1}*{$10^7d\mathcal{B}/dq^2$}
	&$\left[0.10,0.98\right]$&$1.016^{+0.067+0.029+0.069}_{-0.073-0.029-0.069}$ &$0.881\pm0.144$ &$1.063\pm0.149$&$1.081\pm0.183$\\
	
	&$\left[1.1,2.5\right]$&$ 0.326^{+0.032+0.010+0.022}_{-0.031-0.010-0.022}$ &$0.405\pm0.088$ &$0.465\pm0.069$&$0.402\pm0.062$\\
	
	&$\left[2.5,4.0\right]$&$ 0.334^{+0.031+0.009+0.023}_{-0.033-0.009-0.023} $ &$0.393\pm0.085$ &$0.448\pm0.067$&$0.372\pm0.056$\\
	
	&$\left[4.0,6.0\right]$&$ 0.354^{+0.027+0.009+0.024}_{-0.026-0.009-0.024}$ &$0.435\pm0.085$ &$0.500\pm0.068$&$0.417\pm0.060$\\
	
	&$\left[1.0,6.0\right]$&$ 0.342^{+0.017+0.009+0.023}_{-0.017-0.009-0.023}$ &$0.414\pm0.086$ &$0.474\pm0.072$&$0.399\pm0.059$\\
	
	&$\left[6.0,8.0\right]$&$ 0.429^{+0.028+0.010+0.029}_{-0.027-0.010-0.029}$ & &$0.581\pm0.083$&$0.492\pm0.068$\\
	
	&$\left[11.0,12.5\right]$&$ 0.487^{+0.031+0.012+0.033}_{-0.032-0.012-0.033}$ & &$0.780\pm0.074$&$0.769\pm0.097$\\
	
	&$\left[15.0,17.0\right]$&$ 0.534^{+0.027+0.020+0.036}_{-0.037-0.020-0.036}$ & &$0.703\pm0.068$&$0.655\pm0.072$\\
	
	&$\left[17.0,19.0\right]$&$ 0.355^{+0.027+0.017+0.024}_{-0.022-0.017-0.024}$ & &$0.480\pm0.072$&$0.441\pm0.051$\\
	
	&$\left[15.0,19.0\right]$&$ 0.436^{+0.018+0.007+0.030}_{-0.019-0.007-0.030}$ & &$0.592\pm0.059$&$0.548\pm0.061$\\
	\hline
	\multicolumn{6}{c}{CMS $(B^0\to K^{\ast0}\mu^+\mu^-)$ \cite{CMS:2015bcy}}  \\
	\hline
	\multirow{1}*{$10^8d\mathcal{B}/dq^2$}
	&$\left[1.0,2.0\right]$&$ 4.6^{+0.7}_{-0.7}\pm0.30 $ &$4.216\pm0.090$ &$4.855\pm0.724$&$4.277\pm0.661$\\
	
	&$\left[2.0,4.30\right]$&$ 3.3^{+0.5}_{-0.5}\pm0.2 $ &$3.939\pm0.087$ &$4.492\pm0.690$&$3.746\pm0.562$\\
	
	&$\left[4.30,6.00\right]$&$ 3.4^{+0.5}_{-0.5}\pm0.3 $ &$4.398\pm0.086$ &$5.056\pm0.834$&$4.222\pm0.608$\\
	
	&$\left[1.0,6.0\right]$&$ 3.6^{+0.3}_{-0.3}\pm0.2 $ &$4.151\pm0.087$ &$4.756\pm0.650$&$4.014\pm0.590$\\
	
	&$\left[6.00,8.68\right]$&$ 4.7^{+0.4}_{-0.4}\pm0.3 $ & &$5.960\pm0.965$&$5.277\pm0.727$\\
	
	&$\left[10.09,12.86\right]$&$ 6.2^{+0.4}_{-0.4}\pm0.5 $ & &$7.637\pm0.731$&$7.521\pm0.950$\\
	
	&$\left[14.18,16.00\right]$&$ 6.7^{+0.6}_{-0.6}\pm0.5 $ & &$7.461\pm0.735$&$6.968\pm0.780$\\
	
	&$\left[16.00,19.00\right]$&$ 4.2^{+0.3}_{-0.3}\pm0.3 $ & &$5.448\pm0.635$&$5.030\pm0.563$\\
	\hline
	\multicolumn{6}{c}{LHCb $(B^0_s\to \phi\mu^+\mu^-)$ \cite{LHCb:2021zwz}}\rule{0pt}{15pt}\\
	\hline
	\multirow{1}*{$10^{8}d\mathcal{B}/dq^2$}
	&$\left[0.1,0.98\right]$&$7.74\pm0.53\pm 0.12\pm 0.37$ &$10.448\pm1.652$ &$11.424\pm1.166$&$11.084\pm0.777$\\
	
	&$\left[1.1,2.5\right]$&$3.15\pm0.29\pm0.07\pm0.15$ &$4.625\pm0.985$ &$5.473\pm0.637$&$5.408\pm0.515$\\
	
	&$\left[2.5,4.0\right]$&$2.34\pm0.26\pm 0.05\pm0.11$ &$4.405\pm0.942$ &$5.166\pm0.641$&$5.111\pm0.494$\\
	
	&$\left[4.0,6.0\right]$&$3.11\pm0.24\pm0.06\pm0.15$ &$4.820\pm0.922$ &$5.529\pm0.670$&$5.463\pm0.491$\\
	
	&$\left[1.1,6.0\right]$&$2.88\pm0.15\pm0.05\pm0.14$ &$4.637\pm0.944$ &$5.402\pm0.705$&$5.339\pm0.497$\\
	
	&$\left[6.0,8.0\right]$&$3.15\pm0.24\pm0.06\pm0.15$ & &$6.176\pm0.953$&$6.092\pm0.502$\\
	
	&$\left[11.0,12.5\right]$&$4.78\pm0.30\pm0.08\pm0.23$ & &$7.918\pm0.622$&$8.916\pm0.622$\\
	
	&$\left[15.0,17.0\right]$&$5.25\pm0.29\pm0.10\pm0.25$ & &$6.998\pm0.606$&$7.523\pm0.510$\\
	
	&$\left[17.0,19.0\right]$&$4.19\pm0.29\pm0.12\pm0.20$ & &$4.371\pm0.483$&$4.756\pm0.356$\\
	
	&$\left[15.0,19.0\right]$&$4.63\pm0.20\pm0.11\pm0.22$ & &$5.718\pm0.552$&$6.175\pm0.432$\\
	\hline
	\multicolumn{6}{c}{LHCb $(\Lambda_b^0\to \Lambda\mu^+\mu^-)$\cite{LHCb:2015tgy}}  \\
	\hline
	\multirow{1}*{$10^7d\mathcal{B}/dq^2$}
	&$[0.1,2.0]$&$ 0.36^{+0.12+0.02}_{-0.11-0.02}\pm0.07 $ &$0.201\pm0.064$ &$0.136\pm0.071$&$0.200\pm0.048$\\
	&$[2.0,4.0]$&$ 0.11^{+0.12+0.01}_{-0.09-0.01}\pm0.02 $ &$0.192\pm0.062$ &$0.088\pm0.051$&$0.190\pm0.043$\\
	&$[4.0,6.0]$&$ 0.02^{+0.09+0.01}_{-0.00-0.01}\pm0.01 $ &$0.256\pm0.066$ &$0.128\pm0.060$&$0.255\pm0.047$\\
	&$[1.1,6.0]$&$0.09^{+0.06+0.01}_{-0.05-0.01}\pm0.02$ &$0.213\pm0.063$ &$0.103\pm0.057$&$0.212\pm0.044$\\
	&$[6.0,8.0]$&$0.25^{+0.12+0.01}_{-0.11-0.01}\pm0.05$ & &$0.195\pm0.063$&$0.347\pm0.051$\\
	&$[11.0,12.5]$&$0.75^{+0.15+0.03}_{-0.14-0.05}\pm0.15$ & &$0.473\pm0.089$&$0.720\pm0.058$\\
	&$[15.0,16.0]$&$1.12^{+0.19+0.05}_{-0.18-0.05}\pm0.23$ & &$0.714\pm0.080$&$0.955\pm0.051$\\
	&$[16.0,18.0]$&$1.22^{+0.14+0.03}_{-0.14-0.06}\pm0.25$ & &$0.770\pm0.078$&$0.981\pm0.053$\\
	&$[18.0,20.0]$&$1.24^{+0.14+0.06}_{-0.14-0.05}\pm0.26$ & &$0.647\pm0.082$&$0.777\pm0.051$\\
	&$[15.0,20.0]$&$1.20^{+0.09+0.02}_{-0.09-0.04}\pm0.25$ & &$0.709\pm0.066$&$0.894\pm0.050$\\
	\hline
	\multicolumn{6}{c}{BaBar $(B\to X_S\ell^+\ell^-)$\cite{BaBar:2013qry}}  \\
	\hline
	\multirow{1}*{$\frac{10^6\dd\mathcal{B}(X_s e^+e^-)}{\dd q^2}$}
	&$[0.1,2.0]$&$3.05^{+0.52+0.29}_{-0.49-0.21}\pm0.35$ &$0.779\pm0.038$ &$0.656\pm0.065$&$0.779\pm0.080$\\
	&$[2.0,4.3]$&$0.69^{+0.31+0.11}_{-0.28-0.07}\pm0.07$ &$0.355\pm0.017$ &$0.345\pm0.033$&$0.355\pm0.036$\\
	&$[4.3,6.8]$&$0.69^{+0.31+0.13}_{-0.29-0.10}\pm0.05$ &$0.298\pm0.014$ &$0.294\pm0.033$&$0.298\pm0.030$\\
	&$[1.0,6.0]$&$1.93^{+0.47+0.21}_{-0.45-0.16}\pm0.18$ &$0.361\pm0.017$ &$0.347\pm0.036$&$0.361\pm0.037$\\
	&$[10.1,12.9]$&$1.14^{+0.42+0.22}_{-0.40-0.10}\pm0.04$ & &$0.175\pm0.015$&$0.218\pm0.022$\\
	\hline
	\multirow{1}*{$\frac{10^6\dd\mathcal{B}(X_s \mu^+\mu^-)}{\dd q^2}$}
	&$[0.1,2.0]$&$1.83^{+0.90+0.30}_{-0.80-0.24}\pm0.20$ &$0.782\pm0.038$ &$0.622\pm0.062$&$0.782\pm0.080$\\
	&$[2.0,4.3]$&$-0.15^{+0.50+0.26}_{-0.43-0.14}\pm0.01$ &$0.355\pm0.017$ &$0.331\pm0.032$&$0.355\pm0.036$\\
	&$[4.3,6.8]$&$0.34^{+0.54+0.19}_{-0.50-0.15}\pm0.03$ &$0.298\pm0.014$ &$0.289\pm0.030$&$0.298\pm0.030$\\
	&$[1.0,6.0]$&$0.66^{+0.82+0.30}_{-0.76-0.24}\pm0.07$ &$0.361\pm0.017$ &$0.334\pm0.033$&$0.361\pm0.037$\\
	&$[10.1,12.9]$&$0.87^{+0.51+0.11}_{-0.47-0.08}\pm0.03$ & &$0.185\pm0.016$&$0.218\pm0.022$\\
	\hline
	\multicolumn{6}{c}{LHCb $(B^0\to \ell^+\ell^-)$ \cite{LHCb:2021awg}}\\
	\hline
	\multicolumn{2}{l}{$10^9\mathcal{B}(B^0_s\to \mu^+\mu^-)$}
	&$3.09^{+0.46+0.15}_{-0.43-0.11}$ &$3.681\pm0.020$ &$3.672\pm0.152$&$3.792\pm0.024$\\
	\multicolumn{2}{l}{$10^{10}\mathcal{B}(B^0_d\to \mu^+\mu^-)$}
	&$1.20^{+0.83+0.14}_{-0.74-0.14}$ &$0.997\pm0.007$ &$1.024\pm0.074$&$0.991\pm0.007$\\
	\hline
	\multicolumn{6}{c}{CMS $(B^0\to\ell^+\ell^-)$ \cite{CMS:2022mgd}}  \\
	\hline
	\multicolumn{2}{l}{$10^9\mathcal{B}(B^0_s\to\mu^+\mu^-)$}
	&$3.83^{+0.38+0.19+0.14}_{-0.36-0.16-0.13}$ &$3.681\pm0.020$ &$3.672\pm0.152$&$3.792\pm0.024$\\
	\multicolumn{2}{l}{$10^{10}\mathcal{B}(B^0_d\to\mu^+\mu^-)$}
	&$0.37^{+0.75+0.08}_{-0.67-0.09}$ &$0.997\pm0.007$ &$1.024\pm0.074$&$0.991\pm0.007$\\
	\hline
	\multicolumn{6}{c}{Belle $(B\to X_s\gamma)$\cite{Belle:2014nmp}}\\
	\hline
	Observable& $E_{\gamma}$ (GeV) & Experimental value &Previous work &\textit{Flavio}\cite{Straub:2018kue} &this work\\
	\hline
	$10^4\mathcal{B}$
	&$>1.9$&$3.51\pm0.17\pm0.33$\\
	$10^6\text{(EXPLT. result)}$
	&$>1.6$&$375\pm18\pm35$  &$296.1\pm38.0$ &$330.8\pm22.1$ &$296.2\pm38.0$\\
	\hline
	\multicolumn{6}{c}{Belle $(B\to \phi\gamma)$\cite{Belle:2014sac}}\\
	\hline
	Observable&  & Experimental value &Previous work  &\textit{Flavio}\cite{Straub:2018kue} &this work\\
	\hline
	$10^5\mathcal{B}$
	&&$3.6\pm0.5\pm0.3\pm0.6$ &$3.348\pm0.526$ &$4.072\pm0.497$&$3.318\pm0.518$\\
	\hline
\end{supertabular}
\end{center}

\vspace{3cm}
\begin{center}\scriptsize
	\setlength{\tabcolsep}{10pt}
	\renewcommand\arraystretch{1.5}
	\captionof{table}{The set of measured angular distribution observables.}
	\label{tab:old_data2}
	\tablefirsthead{
		\toprule
			\multicolumn{1}{l}{Observable} 
		&\multicolumn{1}{c}{$q^2$ (GeV$^{2}$)} 
		&\multicolumn{1}{r}{Experimental value}  
		&\multicolumn{1}{r}{Previous work \cite{Wen:2023pfq}}  
		&\multicolumn{1}{r}{\textit{Flavio} \cite{Straub:2018kue} }
		&\multicolumn{1}{r}{This analysis}\\
		\hline
	}
	\tablehead{
		\multicolumn{6}{l}{\small \sl Table continued from previous page}\\
		\toprule
	    	\multicolumn{1}{l}{Observable} 
		& \multicolumn{1}{c}{$q^2$ (GeV$^{2}$)} 
		&\multicolumn{1}{r}{Experimental value}  
		&\multicolumn{1}{r}{Previous work \cite{Wen:2023pfq}}  
		&\multicolumn{1}{r}{\textit{Flavio} \cite{Straub:2018kue} }
		&\multicolumn{1}{r}{This analysis}\\
		\hline
	}
	\tabletail{
		\hline
		\multicolumn{6}{r}{\small \sl Table continued on next page}\\
	}
	\tablelasttail{
	\toprule
	}
\begin{supertabular}
	{L{0.12\textwidth} L{0.08\textwidth} R{0.18\textwidth} R{0.13\textwidth}  R{0.12\textwidth} R{0.12\textwidth}}
	\multicolumn{6}{c}{LHCb $(B^0_s\to \phi\mu^+\mu^-)$ \cite{LHCb:2021xxq}}\\
	\hline
	\multirow{1}*{$F_{\text{L}}$}
	&$\left[0.1,0.98\right]$&$0.254\pm0.045\pm 0.017$ &$0.301\pm0.060$ &$0.345\pm0.035$&$0.354\pm0.028$\\
	
	&$\left[1.1,4.0\right]$&$0.723\pm0.053\pm0.015$ &$0.793\pm0.044$ &$0.811\pm0.017$&$0.808\pm0.015$\\
	
	&$\left[4.0,6.0\right]$&$0.701\pm0.050\pm0.016$ &$0.749\pm0.050$ &$0.750\pm0.031$&$0.741\pm0.018$\\
	
	&$\left[1.1,6.0\right]$&$0.715\pm0.036\pm0.013$ &$0.774\pm0.047$ &$0.785\pm0.018$&$0.780\pm0.016$\\
	
	&$\left[6.0,8.0\right]$&$0.624\pm0.051\pm0.012$ & &$0.644\pm0.037$&$0.635\pm0.021$\\
	
	&$\left[11.0,12.5\right]$&$0.353\pm0.044\pm0.012$ & &$0.450\pm0.021$&$0.451\pm0.019$\\
	
	&$\left[15.0,18.9\right]$&$0.359\pm0.031\pm0.019$ & &$0.341\pm0.027$&$0.338\pm0.014$\\
	\hline
	\multicolumn{6}{c}{LHCb $(B^0\to K^{\ast0}\mu^+\mu^-)$\cite{LHCb:2020lmf}}\\
	\hline
	\multirow{1}*{$F_L$}
	&$[0.1,0.98]$&$ 0.255\pm0.032\pm0.007 $ & &$0.294\pm0.045$&$0.233\pm0.063$\\
	&$[1.1,2.5]$&$ 0.655\pm0.046\pm0.017 $ &$0.776\pm0.051$ &$0.760\pm0.044$&$0.694\pm0.069$\\
	&$[2.5,4.0]$&$ 0.756\pm0.047\pm0.023 $ &$0.826\pm0.043$ &$0.797\pm0.038$&$0.748\pm0.053$\\
	&$[4.0,6.0]$&$ 0.684\pm0.035\pm0.015 $ &$0.762\pm0.054$ &$0.712\pm0.051$&$0.658\pm0.064$\\
	&$[1.1,6.0]$&$ 0.700\pm0.025\pm0.013$ &$0.785\pm0.050$ &$0.750\pm0.038$&$0.694\pm0.060$\\
	&$[6.0,8.0]$&$ 0.645\pm0.030\pm0.011$ & &$0.608\pm0.061$&$0.547\pm0.071$\\
	&$[11.0,12.5]$&$ 0.461\pm0.031\pm0.010$ & &$0.435\pm0.035$&$0.388\pm0.064$\\
	&$[15.0,17.0]$&$ 0.352\pm0.026\pm0.009$ & &$0.349\pm0.031$&$0.313\pm0.041$\\
	&$[17.0,19.0]$&$ 0.344\pm0.032\pm0.025$ & &$0.328\pm0.029$&$0.306\pm0.058$\\
	&$[15.0,19.0]$&$ 0.345\pm0.020\pm0.007$ & &$0.340\pm0.030$&$0.310\pm0.045$\\
	\hline
	\multirow{1}*{$P_1$}
	&$[0.1,0.98]$&$ 0.090\pm0.119\pm0.009 $ & &$0.044\pm0.023$&$0.034\pm0.012$\\
	&$[1.1,2.5]$&$ -0.617\pm0.296\pm0.023 $ &$-0.001\pm0.001$ &$0.024\pm0.044$&$0.019\pm0.033$\\
	&$[2.5,4.0]$&$ 0.168\pm0.371\pm0.043 $ &$-0.064\pm0.021$&$-0.116\pm0.040$&$-0.103\pm0.078$\\
	&$[4.0,6.0]$&$ 0.088\pm0.235\pm0.029 $ &$-0.103\pm0.032$ &$-0.178\pm0.049$&$-0.166\pm0.111$\\
	&$[1.1,6.0]$&$ -0.079\pm0.159\pm0.021$ &$-0.066\pm0.021$ &$-0.113\pm0.032$&$-0.101\pm0.071$\\
	&$[6.0,8.0]$&$ -0.071\pm0.211\pm0.020$ & &$-0.206\pm0.060$&$-0.195\pm0.117$\\
	&$[11.0,12.5]$&$ -0.460\pm0.132\pm0.015$ & &$-0.307\pm0.041$&$-0.288\pm0.113$\\
	&$[15.0,17.0]$&$ -0.511\pm0.096\pm0.020$ & &$-0.534\pm0.051$&$-0.501\pm0.074$\\
	&$[17.0,19.0]$&$ -0.763\pm0.152\pm0.094$ & &$-0.750\pm0.041$&$-0.727\pm0.055$\\
	&$[15.0,19.0]$&$ -0.577\pm0.090\pm0.031$ & &$-0.624\pm0.050$&$-0.593\pm0.066$\\
	\hline
	\multirow{1}*{$P_2$}
	&$[0.1,0.98]$&$ -0.003\pm0.038\pm0.003 $ & &$-0.132\pm0.009$&$-0.128\pm0.011$\\
	&$[1.1,2.5]$&$ -0.443\pm0.100\pm0.027 $ &$-0.452\pm0.145$ &$-0.451\pm0.013$&$-0.455\pm0.005$\\
	&$[2.5,4.0]$&$ -0.191\pm0.116\pm0.043 $ &$-0.114\pm0.033$ &$-0.064\pm0.096$&$-0.129\pm0.093$\\
	&$[4.0,6.0]$&$ 0.105\pm0.068\pm0.009 $ &$0.268\pm0.086$ &$0.293\pm0.063$&$0.250\pm0.066$\\
	&$[1.1,6.0]$&$ -0.162\pm0.050\pm0.012$ &$-0.014\pm0.005$ &$0.025\pm0.087$&$-0.030\pm0.077$\\
	&$[6.0,8.0]$&$ 0.207\pm0.048\pm0.013$ & &$0.415\pm0.040$&$0.393\pm0.042$\\
	&$[11.0,12.5]$&$ 0.411\pm0.033\pm0.008$ & &$0.464\pm0.008$&$0.473\pm0.018$\\
	&$[15.0,17.0]$&$ 0.396\pm0.022\pm0.004$ & &$0.413\pm0.019$&$0.424\pm0.021$\\
	&$[17.0,19.0]$&$ 0.328\pm0.032\pm0.017$ & &$0.317\pm0.028$&$0.330\pm0.028$\\
	&$[15.0,19.0]$&$ 0.359\pm0.018\pm0.009$ & &$0.373\pm0.020$&$0.386\pm0.024$\\
	\hline
	\multirow{1}*{$P_3$}
	&$[0.1,0.98]$&$ 0.073\pm0.057\pm0.003 $ & &$0.001\pm0.014$&$0.001\pm0.000$\\
	&$[1.1,2.5]$&$ 0.324\pm0.147\pm0.014 $ &$0.001\pm0.001$ &$0.004\pm0.026$&$0.002\pm0.001$\\
	&$[2.5,4.0]$&$ 0.049\pm0.195\pm0.014 $ &$0.002\pm0.001$ &$0.004\pm0.011$&$0.001\pm0.001$\\
	&$[4.0,6.0]$&$ -0.090\pm0.139\pm0.006 $ &$0.001\pm0.0004$ &$0.003\pm0.015$&$0.001\pm0.001$\\
	&$[1.1,6.0]$&$ 0.085\pm0.090\pm0.005$ &$0.001\pm0.0004$ &$0.003\pm0.008$&$0.001\pm0.001$\\
	&$[6.0,8.0]$&$ -0.068\pm0.104\pm0.007$ & &$0.002\pm0.020$&$0.001\pm0.001$\\
	&$[11.0,12.5]$&$ -0.078\pm0.077\pm0.007$ & &$-0.001\pm0.001$&$-0.001\pm0.005$\\
	&$[15.0,17.0]$&$ -0.000\pm0.056\pm0.003$ & &$-0.000\pm0.013$&$-0.000\pm0.000$\\
	&$[17.0,19.0]$&$ 0.085\pm0.068\pm0.004$ & &$0.000\pm0.018$&$0.000\pm0.000$\\
	&$[15.0,19.0]$&$ 0.048\pm0.045\pm0.002$ & &$0.000\pm0.018$&$0.000\pm0.000$\\
	\hline
	\multirow{1}*{$P_4'$}
	&$[0.1,0.98]$&$ 0.135\pm0.118\pm0.010 $ & &$0.240\pm0.012$&$0.222\pm0.008$\\
	&$[1.1,2.5]$&$ -0.080\pm0.142\pm0.019 $ &$-0.056\pm0.016$ &$-0.061\pm0.044$&$-0.067\pm0.041$\\
	&$[2.5,4.0]$&$ -0.435\pm0.169\pm0.035 $ &$-0.374\pm0.101$ &$-0.392\pm0.042$&$-0.390\pm0.042$\\
	&$[4.0,6.0]$&$ -0.312\pm0.115\pm0.013 $ &$-0.489\pm0.122$ &$-0.503\pm0.028$&$-0.505\pm0.032$\\
	&$[1.1,6.0]$&$ -0.298\pm0.087\pm0.016 $ &$-0.338\pm0.090$ &$-0.353\pm0.034$&$-0.349\pm0.039$\\
	&$[6.0,8.0]$&$ -0.574\pm0.091\pm0.018$ & &$-0.536\pm0.022$&$-0.538\pm0.029$\\
	&$[11.0,12.5]$&$ -0.491\pm0.095\pm0.013$ & &$-0.570\pm0.011$&$-0.567\pm0.026$\\
	&$[15.0,17.0]$&$ -0.626\pm0.069\pm0.018$ & &$-0.619\pm0.012$&$-0.612\pm0.015$\\
	&$[17.0,19.0]$&$ -0.647\pm0.086\pm0.057$ & &$-0.661\pm0.009$&$-0.657\pm0.010$\\
	&$[15.0,19.0]$&$ -0.638\pm0.055\pm0.020$ & &$-0.636\pm0.009$&$-0.630\pm0.013$\\
	\hline
	\multirow{1}*{$P_5'$}
	&$[0.1,0.98]$&$ 0.521\pm0.095\pm0.024 $ & &$0.660\pm0.029$&$0.708\pm0.017$\\
	&$[1.1,2.5]$&$ 0.365\pm0.122\pm0.013 $ &$0.208\pm0.055$ &$0.139\pm0.092$&$0.242\pm0.093$\\
	&$[2.5,4.0]$&$ -0.150\pm0.144\pm0.032 $ &$-0.451\pm0.126$ &$-0.501\pm0.101$&$-0.393\pm0.124$\\
	&$[4.0,6.0]$&$ -0.439\pm0.111\pm0.036 $ &$-0.752\pm0.191$ &$-0.759\pm0.082$&$-0.694\pm0.098$\\
	&$[1.1,6.0]$&$-0.114\pm0.068\pm0.026$ &$-0.406\pm0.110$ &$-0.447\pm0.097$&$-0.348\pm0.113$\\
	&$[6.0,8.0]$&$ -0.583\pm0.090\pm0.030$ & &$-0.835\pm0.054$&$-0.796\pm0.084$\\
	&$[11.0,12.5]$&$ -0.622\pm0.088\pm0.017$ & &$-0.822\pm0.029$&$-0.837\pm0.069$\\
	&$[15.0,17.0]$&$ -0.714\pm0.074\pm0.021$ & &$-0.670\pm0.038$&$-0.694\pm0.051$\\
	&$[17.0,19.0]$&$ -0.590\pm0.084\pm0.059$ & &$-0.483\pm0.043$&$-0.505\pm0.052$\\
	&$[15.0,19.0]$&$ -0.667\pm0.053\pm0.029$ & &$-0.595\pm0.040$&$-0.618\pm0.051$\\
	\hline
	\multirow{1}*{$P_6'$}
	&$[0.1,0.98]$&$ 0.015\pm0.094\pm0.007 $ & &$-0.054\pm0.036$&$-0.036\pm0.004$\\
	&$[1.1,2.5]$&$ -0.226\pm0.128\pm0.005 $ &$-0.068\pm0.018$ &$-0.069\pm0.075$&$-0.049\pm0.005$\\
	&$[2.5,4.0]$&$ -0.155\pm0.148\pm0.024 $ &$-0.051\pm0.013$ &$-0.052\pm0.097$&$-0.039\pm0.003$\\
	&$[4.0,6.0]$&$ -0.293\pm0.117\pm0.004 $ &$-0.028\pm0.007$ &$-0.030\pm0.110$&$-0.022\pm0.002$\\
	&$[1.1,6.0]$&$-0.197\pm0.075\pm0.009$ &$-0.045\pm0.011$ &$-0.045\pm0.092$&$-0.034\pm0.003$\\
	&$[6.0,8.0]$&$ -0.155\pm0.098\pm0.009$ & &$-0.018\pm0.127$&$-0.011\pm0.002$\\
	&$[11.0,12.5]$&$ -0.193\pm0.100\pm0.003$ & &$-0.005\pm0.005$&$0.000\pm0.000$\\
	&$[15.0,17.0]$&$ 0.061\pm0.085\pm0.003$ & &$-0.003\pm0.062$&$-0.000\pm0.000$\\
	&$[17.0,19.0]$&$ 0.103\pm0.105\pm0.016$ & &$-0.001\pm0.068$&$0.000\pm0.000$\\
	&$[15.0,19.0]$&$ 0.073\pm0.067\pm0.006$ & &$-0.002\pm0.057$&$0.000\pm0.000$\\
	\hline
	\multirow{1}*{$P_8'$}
	&$[0.1,0.98]$&$ -0.007\pm0.122\pm0.002 $ & &$-0.006\pm0.023$&$0.011\pm0.001$\\
	&$[1.1,2.5]$&$ -0.366\pm0.158\pm0.005 $ &$-0.015\pm0.004$ &$-0.018\pm0.039$&$0.001\pm0.002$\\
	&$[2.5,4.0]$&$ 0.037\pm0.169\pm0.007 $ &$-0.016\pm0.004$ &$-0.017\pm0.036$&$-0.005\pm0.001$\\
	&$[4.0,6.0]$&$ 0.166\pm0.127\pm0.004 $ &$-0.010\pm0.002$ &$-0.012\pm0.037$&$-0.005\pm0.001$\\
	&$[1.1,6.0]$&$-0.020\pm0.089\pm0.009$ &$-0.013\pm0.003$ &$-0.015\pm0.034$&$-0.003\pm0.001$\\
	&$[6.0,8.0]$&$ -0.129\pm0.098\pm0.005$ & &$-0.008\pm0.039$&$-0.003\pm0.001$\\
	&$[11.0,12.5]$&$ 0.018\pm0.099\pm0.009$ & &$0.002\pm0.002$&$0.003\pm0.003$\\
	&$[15.0,17.0]$&$ 0.007\pm0.086\pm0.002$ & &$0.001\pm0.022$&$0.000\pm0.000$\\
	&$[17.0,19.0]$&$ -0.055\pm0.099\pm0.006$ & &$0.000\pm0.020$&$0.000\pm0.000$\\
	&$[15.0,19.0]$&$ 0.011\pm0.069\pm0.003$ & &$0.001\pm0.020$&$0.000\pm0.000$\\
	\hline
	\multicolumn{6}{c}{CMS $(B^0\to K^{\ast0}\mu^+\mu^-)$\cite{CMS:2017rzx}}\\
	\hline
	\multirow{1}*{$P_1$}
	&$\left[1.0,2.0\right]$&$ 0.12^{+0.46}_{-0.47}\pm0.10 $ &$0.007\pm0.002$ &$0.045\pm0.049$&$0.036\pm0.033$\\
	&$\left[2.0,4.30\right]$&$ -0.69^{+0.58}_{-0.27}\pm0.023 $ &$-0.059\pm0.020$ &$-0.105\pm0.037$&$-0.093\pm0.071$\\
	&$\left[4.30,6.00\right]$&$ 0.53^{+0.24}_{-0.33}\pm0.19 $ &$-0.104\pm0.033$ &$-0.180\pm0.046$&$-0.169\pm0.113$\\	
	&$\left[6.00,8.68\right]$&$ -0.47^{+0.27}_{-0.23}\pm0.15 $ & &$-0.212\pm0.053$&$-0.200\pm0.116$\\	
	&$\left[10.09,12.86\right]$&$ -0.53^{+0.20}_{-0.14}\pm0.15 $ & &$-0.301\pm0.043$&$-0.282\pm0.113$\\	
	&$\left[14.18,16.00\right]$&$ -0.33^{+0.24}_{-0.23}\pm0.20 $ & &$-0.465\pm0.048$&$-0.430\pm0.076$\\	
	&$\left[16.00,19.00\right]$&$ -0.53^{+0.19}_{-0.19}\pm0.16 $ & &$-0.680\pm0.049$&$-0.652\pm0.063$\\	
	\hline
	\multirow{1}*{$P_5'$}
	&$\left[1.0,2.0\right]$&$ 0.10^{+0.32}_{-0.31}\pm0.07 $ &$0.352\pm0.101$ &$0.289\pm0.064$&$0.379\pm0.074$\\
	&$\left[2.0,4.30\right]$&$ -0.57^{+0.34}_{-0.31}\pm0.18 $ &$-0.398\pm0.108$ &$-0.450\pm0.096$&$-0.340\pm0.124$\\
	&$\left[4.30,6.00\right]$&$ -0.96^{+0.22}_{-0.21}\pm0.25 $ &$-0.766\pm0.191$ &$-0.771\pm0.068$&$-0.709\pm0.096$\\
	&$\left[6.00,8.68\right]$&$ -0.64^{+0.15}_{-0.19}\pm0.13 $ & &$-0.837\pm0.057$&$-0.820\pm0.079$\\
	&$\left[10.09,12.86\right]$&$ -0.69^{+0.11}_{-0.14}\pm0.13 $ & &$-0.823\pm0.027$&$-0.839\pm0.068$\\
	&$\left[14.18,16.00\right]$&$ -0.66^{+0.13}_{-0.20}\pm0.18 $ & &$-0.719\pm0.035$&$-0.741\pm0.049$\\
	&$\left[16.00,19.00\right]$&$ -0.56^{+0.12}_{-0.12}\pm0.07 $ & &$-0.547\pm0.041$&$-0.570\pm0.052$\\
	\hline
	\multicolumn{6}{c}{CMS $(B^0\to K^{\ast0}\mu^+\mu^-)$ \cite{CMS:2015bcy}}  \\
	\hline
	\multirow{1}*{$F_L$}
	&$\left[1.0,2.0\right]$&$ 0.64^{+0.10}_{-0.09}\pm0.07 $ &$0.739\pm0.057$ &$0.724\pm0.046$&$0.652\pm0.076$\\
	&$\left[2.0,4.30\right]$&$ 0.80^{+0.08}_{-0.08}\pm0.06 $ &$0.822\pm0.043$ &$0.794\pm0.033$&$0.744\pm0.054$\\
	&$\left[4.30,6.00\right]$&$ 0.62^{+0.10}_{-0.09}\pm0.07 $ &$0.756\pm0.054$ &$0.704\pm0.051$&$0.649\pm0.065$\\
	&$\left[1.0,6.0\right]$&$ 0.73^{+0.05}_{-0.05}\pm0.04 $ &$0.781\pm0.050$ &$0.747\pm0.041$&$0.690\pm0.061$\\
	&$\left[6.0,8.68\right]$&$ 0.50^{+0.06}_{-0.06}\pm0.06 $ & &$0.591\pm0.054$&$0.532\pm0.071$\\
	&$\left[10.09,12.86\right]$&$ 0.39^{+0.05}_{-0.05}\pm0.04 $ & &$0.443\pm0.045$&$0.395\pm0.064$\\
	&$\left[14.18,16.00\right]$&$ 0.48^{+0.05}_{-0.06}\pm0.04 $ & &$0.362\pm0.036$&$0.320\pm0.041$\\
	&$\left[16.00,19.00\right]$&$ 0.38^{+0.05}_{-0.06}\pm0.04 $ & &$0.334\pm0.029$&$0.307\pm0.051$\\
	\hline
	\multirow{1}*{$A_{\text{FB}}$}
	&$\left[1.0,2.0\right]$&$ -0.27^{+0.17}_{-0.40}\pm0.07 $ &$-0.143\pm0.038$ &$-0.156\pm0.033$&$-0.201\pm0.046$\\
	&$\left[2.0,4.30\right]$&$ -0.12^{+0.15}_{-0.17}\pm0.05 $ &$-0.034\pm0.009$ &$-0.026\pm0.029$&$-0.056\pm0.037$\\
	&$\left[4.30,6.00\right]$&$ 0.01^{+0.15}_{-0.15}\pm0.03 $ &$0.100\pm0.024$ &$0.132\pm0.041$&$0.137\pm0.043$\\
	&$\left[1.0,6.0\right]$&$ -0.16^{+0.10}_{-0.09}\pm0.05 $ &$-0.008\pm0.002$ &$0.005\pm0.031$&$-0.018\pm0.034$\\
	&$\left[6.0,8.68\right]$&$ 0.03^{+0.10}_{-0.10}\pm0.02 $ & &$0.256\pm0.047$&$0.290\pm0.053$\\
	&$\left[10.09,12.86\right]$&$ 0.16^{+0.06}_{-0.06}\pm0.01 $ & &$0.384\pm0.024$&$0.426\pm0.050$\\
	&$\left[14.18,16.00\right]$&$ 0.39^{+0.04}_{-0.06}\pm0.01 $ & &$0.412\pm0.029$&$0.450\pm0.034$\\
	&$\left[16.00,19.00\right]$&$ 0.35^{+0.07}_{-0.07}\pm0.01 $ & &$0.350\pm0.026$&$0.377\pm0.038$\\
	\hline
	\multicolumn{6}{c}{ATLAS $(B^0\to K^{\ast0}\mu^+\mu^-)$\cite{ATLAS:2018gqc}}\\
	\hline
	\multirow{1}*{$F_{\text{L}}$}
	&$\left[0.04,2.0\right]$&$0.44^{+0.08}_{-0.08}\pm0.07$ & &$0.394\pm0.059$&$0.320\pm0.075$\\
	&$\left[0.04,4.0\right]$&$0.52^{+0.07}_{-0.07}\pm0.06$ & &$0.538\pm0.059$&$0.456\pm0.082$\\
	&$\left[0.04,6.0\right]$&$0.50^{+0.06}_{-0.06}\pm0.04$ & &$0.588\pm0.052$&$0.509\pm0.078$\\
	&$\left[2.0,4.0\right]$&$ 0.64^{+0.11}_{-0.11}\pm0.05 $ &$0.825\pm0.042$ &$0.799\pm0.032$&$0.749\pm0.054$\\
	&$\left[4.0,6.0\right]$&$ 0.42^{+0.13}_{-0.13}\pm0.12 $ &$0.762\pm0.053$ &$0.712\pm0.043$&$0.658\pm0.064$\\
	&$\left[1.1,6.0\right]$&$ 0.56^{+0.07}_{-0.07}\pm0.06 $ &$0.785\pm0.050$ &$0.750\pm0.038$&$0.694\pm0.060$\\
	\hline
	\multirow{1}*{$P_1$}
	&$\left[0.04,2.0\right]$&$-0.05^{+0.30}_{-0.30}\pm0.08$ & &$0.044\pm0.027$&$0.034\pm0.015$\\
	&$\left[0.04,4.0\right]$&$-0.22^{+0.26}_{-0.26}\pm0.16$ & &$0.015\pm0.026$&$0.011\pm0.014$\\
	&$\left[0.04,6.0\right]$&$-0.15^{+0.23}_{-0.23}\pm0.10$ & &$-0.035\pm0.020$&$-0.031\pm0.032$\\
	&$\left[2.0,4.0\right]$&$ -0.78^{+0.51}_{-0.51}\pm0.34 $ &$-0.053\pm0.018$ &$-0.095\pm0.039$&$-0.083\pm0.067$\\
	&$\left[4.0,6.0\right]$&$ 0.14^{+0.43}_{-0.43}\pm0.26 $ &$-0.103\pm0.033$ &$-0.178\pm0.043$&$-0.166\pm0.111$\\
	&$\left[1.1,6.0\right]$&$ -0.17^{+0.31}_{-0.31}\pm0.13 $ &$-0.066\pm0.022$ &$-0.113\pm0.034$&$-0.101\pm0.071$\\
	\hline
	\multirow{1}*{$P_{4}'$}
	&$\left[0.04,2.0\right]$&$0.31^{+0.40}_{-0.40}\pm0.20$ & &$0.151\pm0.020$&$0.137\pm0.015$\\
	&$\left[0.04,4.0\right]$&$-0.30^{+0.24}_{-0.24}\pm0.17$ & &$-0.022\pm0.028$&$-0.025\pm0.029$\\
	&$\left[0.04,6.0\right]$&$0.05^{+0.20}_{-0.20}\pm0.14$ & &$-0.166\pm0.040$&$-0.162\pm0.036$\\
	&$\left[2.0,4.0\right]$&$ -0.76^{+0.31}_{-0.31}\pm0.21 $ &$-0.330\pm0.088$ &$-0.347\pm0.046$&$-0.344\pm0.045$\\
	&$\left[4.0,6.0\right]$&$ 0.64^{+0.33}_{-0.33}\pm0.18 $ &$-0.489\pm0.125$ &$-0.503\pm0.026$&$-0.505\pm0.032$\\
	&$\left[1.1,6.0\right]$&$ 0.05^{+0.22}_{-0.22}\pm0.14 $ &$-0.338\pm0.088$ &$-0.353\pm0.038$&$-0.349\pm0.039$\\
	\hline
	\multirow{1}*{$P_{5}'$}
	&$\left[0.04,2.0\right]$&$0.67^{+0.26}_{-0.26}\pm0.16$ & &$0.494\pm0.033$&$0.553\pm0.034$\\
	&$\left[0.04,4.0\right]$&$0.32^{+0.21}_{-0.21}\pm0.11$ & &$0.168\pm0.062$&$0.248\pm0.069$\\
	&$\left[0.04,6.0\right]$&$0.27^{+0.19}_{-0.19}\pm0.06$ & &$-0.111\pm0.082$&$-0.022\pm0.090$\\
	&$\left[2.0,4.0\right]$&$ -0.33^{+0.31}_{-0.31}\pm0.13 $ &$-0.353\pm0.096$ &$-0.410\pm0.096$&$-0.297\pm0.126$\\
	&$\left[4.0,6.0\right]$&$ 0.26^{+0.35}_{-0.35}\pm0.18 $ &$-0.752\pm0.196$ &$-0.759\pm0.063$&$-0.694\pm0.098$\\
	&$\left[1.1,6.0\right]$&$ 0.01^{+0.21}_{-0.21}\pm0.08 $ &$-0.406\pm0.108$ &$-0.447\pm0.092$&$-0.348\pm0.113$\\
	\hline
	\multirow{1}*{$P_6'$}
	&$\left[0.04,2.0\right]$&$-0.18^{+0.21}_{-0.21}\pm0.04$ & &$-0.055\pm0.045$&$-0.037\pm0.004$\\
	&$\left[0.04,4.0\right]$&$0.01^{+0.17}_{-0.17}\pm0.10$ & &$-0.052\pm0.055$&$-0.036\pm0.003$\\
	&$\left[0.04,6.0\right]$&$0.03^{+0.15}_{-0.15}\pm0.10$ & &$-0.045\pm0.071$&$-0.032\pm0.003$\\
	&$\left[2.0,4.0\right]$&$ 0.31^{+0.28}_{-0.28}\pm0.19 $ &$-0.055\pm0.014$ &$-0.056\pm0.092$&$-0.041\pm0.003$\\
	&$\left[4.0,6.0\right]$&$ 0.06^{+0.27}_{-0.27}\pm0.13 $ &$-0.028\pm0.006$ &$-0.030\pm0.111$&$-0.022\pm0.002$\\
	&$\left[1.1,6.0\right]$&$ 0.03^{+0.17}_{-0.17}\pm0.12 $ &$-0.045\pm0.011$ &$-0.046\pm0.089$&$-0.034\pm0.003$\\
	\hline
	\multirow{1}*{$P_8'$}
	&$\left[0.04,2.0\right]$&$-0.29^{+0.48}_{-0.48}\pm0.18$ & &$-0.008\pm0.025$&$0.008\pm0.001$\\
	&$\left[0.04,4.0\right]$&$0.38^{+0.33}_{-0.33}\pm0.24$ & &$-0.011\pm0.025$&$0.004\pm0.001$\\
	&$\left[0.04,6.0\right]$&$0.14^{+0.27}_{-0.27}\pm0.17$ & &$-0.011\pm0.026$&$0.001\pm0.001$\\
	&$\left[2.0,4.0\right]$&$ 1.07^{+0.41}_{-0.41}\pm0.39 $ &$-0.016\pm0.004$ &$-0.018\pm0.039$&$-0.004\pm0.001$\\
	&$\left[4.0,6.0\right]$&$ -0.24^{+0.42}_{-0.42}\pm0.09 $ &$-0.010\pm0.002$ &$-0.012\pm0.037$&$-0.005\pm0.001$\\
	&$\left[1.1,6.0\right]$&$ 0.23^{+0.28}_{-0.28}\pm0.20 $ &$-0.013\pm0.003$ &$-0.015\pm0.034$&$-0.003\pm0.001$\\
	\hline
	\multicolumn{6}{c}{Belle $(B^0\to K^{\ast0}e^+e^-)$\cite{Belle:2016fev}}\\
	\hline
	\multirow{1}*{$P_4^{\mu'}$}
	&$\left[0.10,4.00\right]$&$ -0.38^{+0.50}_{-0.48}\pm0.12 $ &$-0.026\pm0.008$ &$-0.028\pm0.032$&$-0.031\pm0.030$\\
	&$\left[4.00,8.00\right]$&$ -0.07^{+0.32}_{-0.31}\pm0.07 $ &$-0.503\pm0.124$ &$-0.518\pm0.021$&$-0.520\pm0.029$\\
	&$\left[1.0,6.0\right]$&$ -0.22^{+0.35}_{-0.34}\pm0.15 $ &$-0.326\pm0.086$ &$-0.341\pm0.038$&$-0.337\pm0.039$\\
	&$\left[10.09,12.90\right]$&$ -0.40^{+0.33}_{-0.29}\pm0.09 $ & &$-0.568\pm0.009$&$-0.565\pm0.026$\\
	&$\left[14.18,19.00\right]$&$ -0.10^{+0.39}_{-0.39}\pm0.07 $ & &$-0.627\pm0.011$&$-0.622\pm0.014$\\
	\hline
	\multirow{1}*{$P_4^{e'}$}
	&$\left[0.10,4.00\right]$&$ 0.34^{+0.41}_{-0.45}\pm0.11 $ &$-0.004\pm0.002$ &$-0.004\pm0.027$&$-0.008\pm0.028$\\
	&$\left[4.00,8.00\right]$&$ -0.52^{+0.24}_{-0.22}\pm0.03 $ &$-0.503\pm0.124$ &$-0.518\pm0.023$&$-0.520\pm0.029$\\
	&$\left[1.0,6.0\right]$&$ -0.72^{+0.40}_{-0.39}\pm0.06 $ &$-0.323\pm0.093$ &$-0.338\pm0.039$&$-0.334\pm0.040$\\
	&$\left[14.18,19.00\right]$&$ -0.15^{+0.41}_{-0.40}\pm0.04 $ & &$-0.627\pm0.010$&$-0.622\pm0.014$\\
	\hline
	\multirow{1}*{$P_5^{\mu'}$}
	&$\left[0.10,4.00\right]$&$ 0.42^{+0.39}_{-0.39}\pm0.14 $ &$0.205\pm0.061$ &$0.156\pm0.064$&$0.238\pm0.072$\\
	&$\left[4.00,8.00\right]$&$ -0.03^{+0.31}_{-0.30}\pm0.09 $ &$-0.802\pm0.198$ &$-0.795\pm0.064$&$-0.746\pm0.089$\\
	&$\left[1.0,6.0\right]$&$ 0.43^{+0.26}_{-0.28}\pm0.10 $ &$-0.382\pm0.104$ &$-0.423\pm0.082$&$-0.324\pm0.113$\\
	&$\left[10.09,12.90\right]$&$ 0.09^{+0.29}_{-0.29}\pm0.02 $ & &$-0.822\pm0.023$&$-0.839\pm0.069$\\
	&$\left[14.18,19.00\right]$&$ -0.13^{+0.39}_{-0.35}\pm0.06 $ & &$-0.626\pm0.041$&$-0.649\pm0.050$\\
	\hline
	\multirow{1}*{$P_5^{e'}$}
	&$\left[0.10,4.00\right]$&$ 0.51^{+0.39}_{-0.46}\pm0.09 $ &$0.219\pm0.063$ &$0.174\pm0.055$&$0.249\pm0.064$\\
	&$\left[4.00,8.00\right]$&$ -0.52^{+0.28}_{-0.26}\pm0.03 $ &$-0.799\pm0.197$ &$-0.792\pm0.061$&$-0.743\pm0.089$\\
	&$\left[1.0,6.0\right]$&$ -0.22^{+0.39}_{-0.41}\pm0.03 $ &$-0.375\pm0.107$ &$-0.416\pm0.086$&$-0.317\pm0.112$\\
	&$\left[14.18,19.00\right]$&$ -0.91^{+0.36}_{-0.30}\pm0.03 $ & &$-0.625\pm0.038$&$-0.648\pm0.050$\\
	\hline
	\multirow{1}*{$Q_4$}
	&$\left[0.10,4.00\right]$&$ -0.723^{+0.676}_{-0.676}\pm0.163 $ &$-0.022\pm0.008$ &$-0.024\pm0.003$&$-0.023\pm0.003$\\
	&$\left[4.00,8.00\right]$&$ 0.448^{+0.392}_{-0.392}\pm0.076 $ &$-0.000\pm0.175$ &$-0.000\pm0.000$&$-0.000\pm0.000$\\
	&$\left[1.0,6.0\right]$&$ 0.498^{+0.527}_{-0.527}\pm0.166 $ &$-0.003\pm0.127$ &$-0.003\pm0.000$&$-0.003\pm0.000$\\
	&$\left[14.18,19.00\right]$&$ 0.041^{+0.565}_{-0.565}\pm0.082 $ & &$-0.000\pm0.000$&$0.000\pm0.000$\\
	\hline
	\multirow{1}*{$Q_5$}
	&$\left[0.10,4.00\right]$&$ -0.097^{+0.601}_{-0.601}\pm0.164 $ &$-0.014\pm0.088$ &$-0.018\pm0.005$&$-0.011\pm0.008$\\
	&$\left[4.00,8.00\right]$&$ 0.498^{+0.410}_{-0.410}\pm0.095 $ &$-0.003\pm0.279$ &$-0.003\pm0.000$&$-0.003\pm0.000$\\
	&$\left[1.0,6.0\right]$&$ 0.656^{+0.485}_{-0.485}\pm0.103 $ &$-0.007\pm0.149$ &$-0.007\pm0.001$&$-0.007\pm0.001$\\
	&$\left[14.18,19.00\right]$&$ 0.778^{+0.502}_{-0.502}\pm0.065 $ & &$-0.001\pm0.000$&$-0.001\pm0.000$\\
	\hline
	
	\multicolumn{6}{c}{LHCb $(B^+\to K^{\ast+}\mu^+\mu^-)$\cite{LHCb:2020gog}}\\  
	\hline
	\multirow{1}*{$F_L$}
	&$[0.10,0.98]$&$0.34^{+0.10}_{-0.10}\pm0.06$ & &$0.305\pm0.051$&$0.229\pm0.062$\\
	&$[1.1,2.5]$&$0.54^{+0.18}_{-0.19}\pm0.03$ &$0.784\pm0.049$ &$0.768\pm0.042$&$0.692\pm0.070$\\
	&$[2.5,4.0]$&$0.17^{+0.24}_{-0.14}\pm0.04$ &$0.829\pm0.042$ &$0.800\pm0.033$&$0.748\pm0.054$\\
	&$[4.0,6.0]$&$0.67^{+0.11}_{-0.14}\pm0.03$ &$0.764\pm0.053$ &$0.714\pm0.057$&$0.658\pm0.064$\\
	&$[1.1,6.0]$&$0.59^{+0.10}_{-0.10}\pm0.03$ &$0.788\pm0.049$ &$0.754\pm0.041$&$0.693\pm0.060$\\
	&$[6.0,8.0]$&$0.39^{+0.20}_{-0.21}\pm0.02$ & &$0.610\pm0.061$&$0.548\pm0.071$\\
	&$[11.0,12.5]$&$0.39^{+0.23}_{-0.16}\pm0.03$ & &$0.435\pm0.037$&$0.388\pm0.063$\\
	&$[15.0,17.0]$&$0.41^{+0.18}_{-0.14}\pm0.02$ & &$0.349\pm0.032$&$0.313\pm0.041$\\
	&$[17.0,19.0]$&$0.34^{+0.11}_{-0.12}\pm0.04$ & &$0.328\pm0.028$&$0.306\pm0.058$\\
	&$[15.0,19.0]$&$0.40^{+0.13}_{-0.11}\pm0.02$ & &$0.340\pm0.031$&$0.310\pm0.045$\\
	\hline
	\multirow{1}*{$P_1$}
	&$[0.10,0.98]$&$0.44^{+0.38}_{-0.40}\pm0.11$ & &$0.045\pm0.028$&$0.034\pm0.012$\\
	&$[1.1,2.5]$&$1.60^{+4.92}_{-1.75}\pm0.32$ &$-0.001\pm0.0004$ &$0.022\pm0.052$&$0.020\pm0.033$\\
	&$[2.5,4.0]$&$-0.29^{+1.43}_{-1.04}\pm0.22$ &$-0.064\pm0.021$ &$-0.118\pm0.042$&$-0.101\pm0.078$\\
	&$[4.0,6.0]$&$-1.24^{+0.99}_{-1.17}\pm0.29$ &$-0.102\pm0.032$ &$-0.178\pm0.048$&$-0.164\pm0.111$\\
	&$[1.1,6.0]$&$-0.51^{+0.56}_{-0.54}\pm0.08$ &$-0.066\pm0.022$ &$-0.115\pm0.034$&$-0.099\pm0.071$\\
	&$[6.0,8.0]$&$-0.78^{+0.61}_{-0.69}\pm0.10$ & &$-0.206\pm0.056$&$-0.192\pm0.117$\\
	&$[11.0,12.5]$&$-0.32^{+0.44}_{-0.52}\pm0.09$ & &$-0.307\pm0.044$&$-0.285\pm0.113$\\
	&$[15.0,17.0]$&$-0.88^{+0.41}_{-0.67}\pm0.07$ & &$-0.535\pm0.051$&$-0.496\pm0.074$\\
	&$[17.0,19.0]$&$-0.40^{+0.58}_{-0.57}\pm0.09$ & &$-0.751\pm0.037$&$-0.722\pm0.056$\\
	&$[15.0,19.0]$&$-0.70^{+0.35}_{-0.43}\pm0.07$ & &$-0.624\pm0.052$&$-0.588\pm0.067$\\
	\hline
	\multirow{1}*{$P_2$}
	&$[0.10,0.98]$&$-0.05^{+0.12}_{-0.12}\pm0.03$ & &$-0.135\pm0.008$&$-0.127\pm0.011$\\
	&$[1.1,2.5]$&$-0.28^{+0.24}_{-0.42}\pm0.15$ &$-0.453\pm0.154$ &$-0.451\pm0.016$&$-0.455\pm0.006$\\
	&$[2.5,4.0]$&$0.03^{+0.26}_{-0.25}\pm0.11$ &$-0.109\pm0.033$ &$-0.055\pm0.103$&$-0.131\pm0.093$\\
	&$[4.0,6.0]$&$-0.15^{+0.19}_{-0.20}\pm0.06$ &$0.268\pm0.087$ &$0.295\pm0.072$&$0.250\pm0.066$\\
	&$[1.1,6.0]$&$-0.13^{+0.13}_{-0.13}\pm0.05$ &$-0.011\pm0.005$ &$0.032\pm0.096$&$-0.031\pm0.077$\\
	&$[6.0,8.0]$&$-0.06^{+0.12}_{-0.13}\pm0.05$ & &$0.415\pm0.039$&$0.393\pm0.042$\\
	&$[11.0,12.5]$&$0.62^{+0.55}_{-0.14}\pm0.04$ & &$0.464\pm0.006$&$0.473\pm0.017$\\
	&$[15.0,17.0]$&$0.45^{+0.03}_{-0.07}\pm0.03$ & &$0.413\pm0.022$&$0.425\pm0.021$\\
	&$[17.0,19.0]$&$0.14^{+0.10}_{-0.10}\pm0.04$ & &$0.317\pm0.023$&$0.333\pm0.029$\\
	&$[15.0,19.0]$&$0.34^{+0.09}_{-0.07}\pm0.04$ & &$0.373\pm0.023$&$0.388\pm0.024$\\
	\hline
	\multirow{1}*{$P_3$}
	&$[0.10,0.98]$&$-0.42^{+0.20}_{-0.21}\pm0.05$ & &$0.002\pm0.013$&$0.001\pm0.000$\\
	&$[1.1,2.5]$&$-0.09^{+0.70}_{-0.99}\pm0.18$ &$0.001\pm0.0004$ &$0.004\pm0.025$&$0.001\pm0.001$\\
	&$[2.5,4.0]$&$-0.45^{+0.50}_{-0.62}\pm0.20$ &$0.002\pm0.0005$ &$0.004\pm0.011$&$0.001\pm0.001$\\
	&$[4.0,6.0]$&$0.52^{+0.82}_{-0.62}\pm0.15$ &$0.001\pm0.0004$ &$0.003\pm0.018$&$0.001\pm0.001$\\
	&$[1.1,6.0]$&$0.12^{+0.27}_{-0.28}\pm0.04$ &$0.001\pm0.0004$ &$0.003\pm0.008$&$0.001\pm0.001$\\
	&$[6.0,8.0]$&$0.17^{+0.33}_{-0.31}\pm0.06$ & &$0.002\pm0.021$&$0.001\pm0.001$\\
	&$[11.0,12.5]$&$-0.32^{+0.29}_{-0.65}\pm0.05$& &$-0.001\pm0.001$&$-0.001\pm0.005$\\
	&$[15.0,17.0]$&$-0.23^{+0.16}_{-0.20}\pm0.02$ & &$0.000\pm0.013$&$0.000\pm0.000$\\
	&$[17.0,19.0]$&$0.12^{+0.21}_{-0.21}\pm0.02$ & &$0.000\pm0.017$&$0.000\pm0.000$\\
	&$[15.0,19.0]$&$-0.07^{+0.12}_{-0.13}\pm0.03$ & &$0.000\pm0.015$&$0.000\pm0.000$\\
	\hline
	\multirow{1}*{$P_4'$}
	&$[0.10,0.98]$&$-0.09^{+0.36}_{-0.35}\pm0.12$ & &$0.235\pm0.011$&$0.217\pm0.008$\\
	&$[1.1,2.5]$&$0.58^{+0.62}_{-0.56}\pm0.11$ &$-0.052\pm0.015$ &$-0.063\pm0.046$&$-0.069\pm0.041$\\
	&$[2.5,4.0]$&$-0.81^{+1.09}_{-0.84}\pm0.14$ &$-0.371\pm0.098$ &$-0.391\pm0.046$&$-0.389\pm0.042$\\
	&$[4.0,6.0]$&$-0.79^{+0.47}_{-0.28}\pm0.09$ &$-0.487\pm0.120$ &$-0.502\pm0.028$&$-0.505\pm0.032$\\
	&$[1.1,6.0]$&$-0.41^{+0.28}_{-0.28}\pm0.07$ &$-0.335\pm0.096$ &$-0.353\pm0.041$&$-0.350\pm0.039$\\
	&$[6.0,8.0]$&$-0.43^{+0.41}_{-0.45}\pm0.06$ & &$-0.536\pm0.021$&$-0.537\pm0.029$\\
	&$[11.0,12.5]$&$-0.63^{+0.30}_{-0.34}\pm0.07$ & &$-0.570\pm0.010$&$-0.566\pm0.026$\\
	&$[15.0,17.0]$&$-0.32^{+0.23}_{-0.22}\pm0.08$ & &$-0.619\pm0.012$&$-0.611\pm0.015$\\
	&$[17.0,19.0]$&$-0.57^{+0.29}_{-0.36}\pm0.13$ & &$-0.661\pm0.009$&$-0.656\pm0.011$\\
	&$[15.0,19.0]$&$-0.39^{+0.18}_{-0.21}\pm0.10$ & &$-0.636\pm0.010$&$-0.629\pm0.013$\\
	\hline
	\multirow{1}*{$P_5'$}
	&$[0.10,0.98]$&$0.51^{+0.30}_{-0.28}\pm0.12$ & &$0.656\pm0.656$&$0.717\pm0.018$\\
	&$[1.1,2.5]$&$0.88^{+0.70}_{-0.71}\pm0.10$ &$0.180\pm0.050$ &$0.113\pm0.113$&$0.248\pm0.093$\\
	&$[2.5,4.0]$&$-0.87^{+1.00}_{-1.68}\pm0.09$ &$-0.467\pm0.125$ &$-0.517\pm0.517$&$-0.391\pm0.125$\\
	&$[4.0,6.0]$&$-0.25^{+0.32}_{-0.40}\pm0.09$ &$-0.756\pm0.187$ &$-0.764\pm0.764$&$-0.694\pm0.098$\\
	&$[1.1,6.0]$&$-0.07^{+0.25}_{-0.25}\pm0.04$ &$-0.421\pm0.123$ &$-0.461\pm0.461$&$-0.347\pm0.113$\\
	&$[6.0,8.0]$&$-0.15^{+0.40}_{-0.41}\pm0.06$ & &$-0.836\pm0.836$&$-0.797\pm0.085$\\
	&$[11.0,12.5]$&$-0.88^{+0.28}_{-0.34}\pm0.05$ & &$-0.822\pm0.822$&$-0.839\pm0.069$\\
	&$[15.0,17.0]$&$-0.14^{+0.21}_{-0.20}\pm0.06$ & &$-0.670\pm0.670$&$-0.697\pm0.051$\\
	&$[17.0,19.0]$&$-0.66^{+0.36}_{-0.80}\pm0.13$ & &$-0.483\pm0.483$&$-0.510\pm0.052$\\
	&$[15.0,19.0]$&$-0.24^{+0.16}_{-0.16}\pm0.05$ & &$-0.595\pm0.595$&$-0.621\pm0.051$\\
	\hline
	\multirow{1}*{$P_6'$}
	&$[0.10,0.98]$&$-0.02^{+0.40}_{-0.34}\pm0.06$ & &$-0.047\pm0.040$&$-0.033\pm0.003$\\
	&$[1.1,2.5]$&$0.25^{+1.22}_{-1.32}\pm0.08$ &$-0.059\pm0.017$ &$-0.054\pm0.072$&$-0.044\pm0.004$\\
	&$[2.5,4.0]$&$-0.37^{+1.59}_{-3.91}\pm0.05$ &$-0.049\pm0.012$ &$-0.044\pm0.115$&$-0.035\pm0.003$\\
	&$[4.0,6.0]$&$-0.09^{+0.40}_{-0.41}\pm0.05$ &$-0.029\pm0.007$ &$-0.028\pm0.113$&$-0.019\pm0.002$\\
	&$[1.1,6.0]$&$-0.21^{+0.23}_{-0.23}\pm0.04$ &$-0.043\pm0.012$ &$-0.039\pm0.082$&$-0.030\pm0.002$\\
	&$[6.0,8.0]$&$-0.74^{+0.29}_{-0.40}\pm0.03$ & &$-0.018\pm0.126$&$-0.009\pm0.002$\\
	&$[11.0,12.5]$&$-0.11^{+0.28}_{-0.29}\pm0.03$ & &$-0.005\pm0.005$&$0.000\pm0.000$\\
	&$[15.0,17.0]$&$-0.48^{+0.21}_{-0.21}\pm0.02$ & &$-0.003\pm0.061$&$0.000\pm0.000$\\
	&$[17.0,19.0]$&$0.12^{+0.33}_{-0.33}\pm0.04$ & &$-0.001\pm0.066$&$0.000\pm0.000$\\
	&$[15.0,19.0]$&$-0.28^{+0.19}_{-0.14}\pm0.03$ & &$-0.002\pm0.054$&$0.000\pm0.000$\\
	\hline
	\multirow{1}*{$P_8'$}
	&$[0.10,0.98]$&$0.45^{+0.50}_{-0.39}\pm0.09$ & &$-0.031\pm0.022$&$0.016\pm0.002$\\
	&$[1.1,2.5]$&$0.12^{+0.75}_{-0.76}\pm0.05$ &$-0.021\pm0.005$ &$-0.027\pm0.035$&$0.003\pm0.001$\\
	&$[2.5,4.0]$&$0.12^{+7.89}_{-4.95}\pm0.07$ &$-0.016\pm0.004$ &$-0.018\pm0.031$&$-0.003\pm0.001$\\
	&$[4.0,6.0]$&$-0.15^{+0.44}_{-0.48}\pm0.05$ &$-0.011\pm0.002$ &$-0.011\pm0.033$&$-0.004\pm0.001$\\
	&$[1.1,6.0]$&$0.03^{+0.26}_{-0.28}\pm0.06$ &$-0.015\pm0.004$ &$-0.017\pm0.033$&$-0.002\pm0.001$\\
	&$[6.0,8.0]$&$-0.39^{+0.30}_{-0.39}\pm0.02$ & &$-0.008\pm0.036$&$-0.003\pm0.001$\\
	&$[11.0,12.5]$&$0.13^{+0.29}_{-0.30}\pm0.04$ & &$0.002\pm0.002$&$0.003\pm0.003$\\
	&$[15.0,17.0]$&$-0.34^{+0.23}_{-0.22}\pm0.04$ & &$0.001\pm0.022$&$0.000\pm0.000$\\
	&$[17.0,19.0]$&$0.36^{+0.37}_{-0.33}\pm0.07$ & &$0.000\pm0.018$&$0.000\pm0.000$\\
	&$[15.0,19.0]$&$-0.11^{+0.19}_{-0.18}\pm0.03$ & &$0.001\pm0.020$&$0.000\pm0.000$\\
	\hline
	\multicolumn{6}{c}{LHCb $(B^0\to K^{\ast0}e^+e^-)$\cite{LHCb:2020dof}}\\  
	\hline
	$F_L$&$[0.0008,0.257]$&$0.044\pm0.026\pm0.014$ &$0.077\pm0.026$ &$0.050\pm0.012$&$0.036\pm0.013$\\
	$A_T^{\Re}$&$[0.0008,0.257]$&$-0.06\pm0.08\pm0.02$ &$-0.032\pm0.012$ &$-0.024\pm0.001$&$-0.023\pm0.002$\\
	$A_T^2$&$[0.0008,0.257]$&$0.11\pm0.10\pm0.02$ &$-0.000\pm0.000$ &$-0.002\pm0.020$&$-0.002\pm0.000$\\
	$A_T^{\Im}$&$[0.0008,0.257]$&$0.02\pm0.10\pm0.01$ &$0.001\pm0.000$ &$0.032\pm0.023$&$0.024\pm0.012$\\
	\hline
	\multicolumn{6}{c}{LHCb $(B^0_s\to \phi\gamma)$\cite{LHCb:2019vks} }\\  
	\hline
	$S_{\phi\gamma}$&&$0.43\pm0.30\pm0.11$ &$0.001\pm0.000$ &$0.000\pm0.000$&$0.001\pm0.0002$\\
	$A_{\text{CP}}$&&$0.11\pm0.29\pm0.11$ &$0.000\pm0.000$ &$0.004\pm0.002$&$0.000\pm0.000$\\
	$A_{\Delta\Gamma}$&&$-0.67^{+0.37}_{-0.41}\pm0.17$ &$0.029\pm0.000$ &$0.031\pm0.020$&$0.029\pm0.001$\\
	\hline
\end{supertabular}
\end{center}

\begin{center}\scriptsize
	\setlength{\tabcolsep}{10pt}
	\renewcommand\arraystretch{1.5}
	\captionof{table}{The latest results of angular distribution observables measured by the CMS collaboration.}
	\label{tab:new_data1}
	\tablefirsthead{
		\toprule
		\multicolumn{1}{l}{Observable} 
		& \multicolumn{1}{c}{$q^2$ (GeV$^{2}$)} 
		&\multicolumn{1}{r}{Experimental value}  
		&\multicolumn{1}{r}{Previous work \cite{Wen:2023pfq}}  
		&\multicolumn{1}{r}{\textit{Flavio} \cite{Straub:2018kue} }
		&\multicolumn{1}{r}{This analysis}\\
		\hline
	}
	\tablehead{
		\multicolumn{6}{l}{\small \sl Table continued from previous page}\\
		\hline
	}
	\tabletail{
		\hline
		\multicolumn{6}{r}{\small \sl Table continued on next page}\\
	}
	\tablelasttail{
	\toprule
	}
	\begin{supertabular}%
		{L{0.12\textwidth} L{0.08\textwidth} R{0.18\textwidth} R{0.13\textwidth}  R{0.12\textwidth} R{0.12\textwidth}}
		\multicolumn{6}{c}{CMS $(B^0\to K^{\ast 0}\ell^+\ell^-)$\cite{CMS:2024atz}}\\ 
		\hline
		\multirow{1}*{$F_L$}
		&$[1.1,2.0]$&$0.709^{+0.073+0.021}_{-0.054-0.021}$ & &$0.736\pm0.050$&$0.666\pm0.075$\\
		&$[2.0,4.3]$&$0.810^{+0.036+0.016}_{-0.030-0.016}$ & &$0.794\pm0.037$&$0.744\pm0.054$\\
		&$[4.3,6.0]$&$0.714^{+0.032+0.012}_{-0.030-0.012}$ & &$0.704\pm0.052$&$0.649\pm0.065$\\
		&$[6.00,8.68]$&$0.627^{+0.016+0.011}_{-0.016-0.011}$ & &$0.591\pm0.053$&$0.532\pm0.071$\\
		&$[10.09,12.86]$&$0.474^{+0.011+0.009}_{-0.011-0.009}$ & &$0.443\pm0.034$&$0.395\pm0.064$\\
		&$[14.18,16.00]$&$0.394^{+0.012+0.009}_{-0.012-0.009}$ & &$0.362\pm0.037$&$0.320\pm0.041$\\
		\hline
		\multirow{1}*{$P_1$}
		&$[1.1,2.0]$&$0.09^{+0.23+0.04}_{-0.20-0.04}$ & &$0.042\pm0.054$&$0.034\pm0.034$\\
		&$[2.0,4.3]$&$-0.29^{+0.19+0.05}_{-0.21-0.05}$ & &$-0.105\pm0.036$&$-0.093\pm0.071$\\
		&$[4.3,6.0]$&$-0.30^{+0.15+0.04}_{-0.17-0.04}$ & &$-0.180\pm0.052$&$-0.169\pm0.113$\\
		&$[6.00,8.68]$&$-0.06^{+0.10+0.05}_{-0.10-0.05}$ & &$-0.212\pm0.053$&$-0.200\pm0.116$\\
		&$[10.09,12.86]$&$-0.439^{+0.051+0.030}_{-0.047-0.030}$ & &$-0.301\pm0.043$&$-0.282\pm0.113$\\
		&$[14.18,16.00]$&$-0.465^{+0.037+0.025}_{-0.037-0.025}$ & &$-0.465\pm0.047$&$-0.430\pm0.076$\\
		\hline
		\multirow{1}*{$P_2$}
		&$[1.1,2.0]$&$-0.37^{+0.17+0.10}_{-0.13-0.10}$ & &$-0.461\pm0.013$&$-0.454\pm0.019$\\
		&$[2.0,4.3]$&$-0.244^{+0.094+0.039}_{-0.077-0.039}$ & &$-0.093\pm0.105$&$-0.156\pm0.089$\\
		&$[4.3,6.0]$&$0.121^{+0.080+0.030}_{-0.076-0.030}$ & &$0.309\pm0.078$&$0.268\pm0.063$\\
		&$[6.00,8.68]$&$0.188^{+0.039+0.014}_{-0.040-0.014}$ & &$0.424\pm0.039$&$0.418\pm0.033$\\
		&$[10.09,12.86]$&$0.386^{+0.021+0.018}_{-0.019-0.018}$ & &$0.462\pm0.007$&$0.472\pm0.017$\\
		&$[14.18,16.00]$&$0.440^{+0.008+0.008}_{-0.010-0.008}$ & &$0.432\pm0.018$&$0.443\pm0.018$\\
		\hline
		\multirow{1}*{$P_3$}
		&$[1.1,2.0]$&$-0.05^{+0.21+0.04}_{-0.22-0.04}$ & &$0.004\pm0.025$&$0.002\pm0.001$\\
		&$[2.0,4.3]$&$-0.19^{+0.20+0.09}_{-0.22-0.09}$ & &$0.004\pm0.012$&$0.001\pm0.001$\\
		&$[4.3,6.0]$&$-0.03^{+0.14+0.08}_{-0.14-0.08}$ & &$0.002\pm0.016$&$0.001\pm0.001$\\
		&$[6.00,8.68]$&$0.099^{+0.092+0.014}_{-0.090-0.014}$ & &$0.002\pm0.022$&$0.000\pm0.000$\\
		&$[10.09,12.86]$&$0.013^{+0.041+0.007}_{-0.043-0.007}$ & &$-0.001\pm0.001$&$-0.001\pm0.005$\\
		&$[14.18,16.00]$&$-0.034^{+0.037+0.010}_{-0.038-0.010}$ & &$-0.001\pm0.012$&$0.000\pm0.000$\\
		\hline
		\multirow{1}*{$P_4^\prime$}
		&$[1.1,2.0]$&$-0.44^{+0.29+0.11}_{-0.32-0.11}$ & &$0.006\pm0.033$&$-0.004\pm0.036$\\
		&$[2.0,4.3]$&$-0.43^{+0.16+0.08}_{-0.19-0.08}$ & &$-0.366\pm0.049$&$-0.363\pm0.043$\\
		&$[4.3,6.0]$&$-0.72^{+0.15+0.07}_{-0.16-0.07}$ & &$-0.509\pm0.026$&$-0.510\pm0.031$\\
		&$[6.00,8.68]$&$-0.95^{+0.10+0.06}_{-0.10-0.06}$ & &$-0.539\pm0.019$&$-0.540\pm0.028$\\
		&$[10.09,12.86]$&$-1.025^{+0.064+0.059}_{-0.066-0.059}$ & &$-0.568\pm0.010$&$-0.565\pm0.026$\\
		&$[14.18,16.00]$&$-1.159^{+0.042+0.041}_{-0.038-0.041}$ & &$-0.605\pm0.012$&$-0.598\pm0.016$\\
		\hline
		\multirow{1}*{$P_5^\prime$}
		&$[1.1,2.0]$&$0.36^{+0.17+0.03}_{-0.13-0.03}$ & &$0.261\pm0.261$&$0.355\pm0.078$\\
		&$[2.0,4.3]$&$-0.14^{+0.10+0.04}_{-0.09-0.04}$ & &$-0.450\pm0.450$&$-0.340\pm0.124$\\
		&$[4.3,6.0]$&$-0.44^{+0.10+0.03}_{-0.10-0.03}$ & &$0.771\pm0.771$&$-0.709\pm0.096$\\
		&$[6.00,8.68]$&$-0.495^{+0.067+0.023}_{-0.067-0.023}$ & &$-0.837\pm0.838$&$-0.820\pm0.079$\\
		&$[10.09,12.86]$&$-0.746^{+0.033+0.014}_{-0.032-0.014}$ & &$-0.823\pm0.823$&$-0.839\pm0.068$\\
		&$[14.18,16.00]$&$-0.688^{+0.038+0.021}_{-0.036-0.021}$ & &$-0.719\pm0.719$&$-0.741\pm0.049$\\
		\hline
		\multirow{1}*{$P_6^\prime$}
		&$[1.1,2.0]$&$0.000^{+0.094+0.021}_{-0.097-0.021}$ & &$-0.070\pm0.066$&$-0.049\pm0.005$\\
		&$[2.0,4.3]$&$0.108^{+0.075+0.018}_{-0.071-0.018}$ & &$-0.054\pm0.098$&$-0.039\pm0.003$\\
		&$[4.3,6.0]$&$0.129^{+0.074+0.011}_{-0.071-0.011}$ & &$-0.029\pm0.121$&$-0.021\pm0.002$\\
		&$[6.00,8.68]$&$0.010^{+0.052+0.016}_{-0.052-0.016}$ & &$-0.017\pm0.148$&$-0.007\pm0.002$\\
		&$[10.09,12.86]$&$0.080^{+0.037+0.011}_{-0.041-0.011}$ & &$-0.004\pm0.006$&$0.000\pm0.000$\\
		&$[14.18,16.00]$&$0.121^{+0.040+0.011}_{-0.039-0.011}$ & &$-0.004\pm0.057$&$0.000\pm0.000$\\
		\hline
		\multirow{1}*{$P_8^\prime$}
		&$[1.1,2.0]$&$0.16^{+0.37+0.11}_{-0.37-0.11}$ & &$-0.017\pm0.034$&$0.002\pm0.001$\\
		&$[2.0,4.3]$&$0.73^{+0.18+0.06}_{-0.19-0.06}$ & &$-0.017\pm0.040$&$-0.004\pm0.001$\\
		&$[4.3,6.0]$&$-0.01^{+0.22+0.04}_{-0.22-0.04}$ & &$-0.011\pm0.037$&$-0.004\pm0.001$\\
		&$[6.00,8.68]$&$0.06^{+0.14+0.04}_{-0.14-0.04}$ & &$-0.007\pm0.040$&$-0.002\pm0.000$\\
		&$[10.09,12.86]$&$0.09^{+0.09+0.03}_{-0.10-0.03}$ & &$0.003\pm0.003$&$0.003\pm0.003$\\
		&$[14.18,16.00]$&$0.011^{+0.089+0.022}_{-0.086-0.022}$ & &$0.001\pm0.021$&$0.000\pm0.000$\\
		\hline
	\end{supertabular}
\end{center}

\begin{table}[h]\scriptsize
	\caption{The newly measured branching ratio and $R_K$ related to CMS new results.} 
	\label{tab:new_data2}
	\setlength{\tabcolsep}{9pt}
	\renewcommand\arraystretch{1.2}
	\begin{tabular}%
		{L{0.14\textwidth} C{0.12\textwidth} R{0.16\textwidth} R{0.12\textwidth}  R{0.12\textwidth} R{0.12\textwidth}}
		\toprule
		Observable & $q^2$ (GeV$^{2}$) 
		&Exp.  value & Previously \cite{Wen:2023pfq}
		&\textit{Flavio} \cite{Straub:2018kue} &This analysis\\
		\hline
		\multicolumn{6}{c}{CMS $(B^+\to K^{+}\ell^+\ell^-)$\cite{CMS:2024syx}}\\ 
		\hline
		$R_{K^+}$&$[1.1,6.0]$&$0.780^{+0.470}_{-0.230}$ & &$1.001\pm0.000$&$1.001\pm0.000$\\
		\multirow{1}*{$10^8\mathcal{B}(K^+ \mu^+\mu^-)$}
		&$[1.1,6.0]$&$12.42^{+0.68}_{-0.68}$ & &$17.085\pm3.376$&$17.936\pm2.880$\\
		\multirow{1}*{$\frac{10^8\dd\mathcal{B}(K^+ \mu^+\mu^-)}{\dd q^2}$}
		&$[0.10,0.98]$&$2.91^{+0.24}_{-0.24}$ & &$3.098\pm0.514$&$3.246\pm0.546$\\
		&$[1.1,2.0]$&$1.93^{+0.20}_{-0.20}$ & &$3.173\pm0.589$&$3.328\pm0.552$\\
		&$[2.0,3.0]$&$3.06^{+0.25}_{-0.25}$ & &$3.510\pm0.551$&$3.682\pm0.602$\\
		&$[3.0,4.0]$&$2.54^{+0.23}_{-0.23}$ & &$3.491\pm0.613$&$3.665\pm0.590$\\
		&$[4.0,5.0]$&$2.47^{+0.24}_{-0.24}$ & &$3.469\pm0.577$&$3.672\pm0.577$\\
		&$[5.0,6.0]$&$2.53^{+0.27}_{-0.27}$ & &$3.443\pm0.601$&$3.672\pm0.562$\\		
		&$[6.0,7.0]$&$2.50^{+0.23}_{-0.23}$ & &$3.412\pm0.542$&$3.672\pm0.546$\\
		&$[7.0,8.0]$&$2.34^{+0.25}_{-0.25}$ & &$3.378\pm0.618$&$3.672\pm0.528$\\
		&$[11.0,11.8]$&$1.62^{+0.18}_{-0.18}$ & &$2.570\pm0.309$&$3.672\pm0.356$\\
		&$[11.8,12.5]$&$1.26^{+0.14}_{-0.14}$ & &$2.162\pm0.307$&$3.672\pm0.289$\\
		&$[14.82,16.00]$&$1.83^{+0.17}_{-0.17}$ & &$2.837\pm0.314$&$3.672\pm0.286$\\
		&$[16.0,17.0]$&$1.57^{+0.15}_{-0.15}$ & &$2.142\pm0.264$&$3.672\pm0.206$\\
		&$[17.0,18.0]$&$2.11^{+0.16}_{-0.16}$ & &$1.873\pm0.224$&$3.672\pm0.174$\\
		&$[18.0,19.24]$&$1.74^{+0.15}_{-0.15}$ & &$1.901\pm0.197$&$3.672\pm0.175$\\
		&$[19.24,22.90]$&$2.02^{+0.30}_{-0.30}$ & &$2.336\pm0.237$&$3.672\pm0.297$\\
		\toprule
	\end{tabular}
\end{table}

\newpage
\section{SUPPORTING MATERIALS}
\label{app:sup_materials}

As a continuity of \cref{fig:lfus1}, more figures are summarized here to
take a further support.

\begin{figure}[htp]
	\centering
	\subfigure[S-II]{
	\label{fig:lfus2}
	\includegraphics[width=0.45\textwidth]{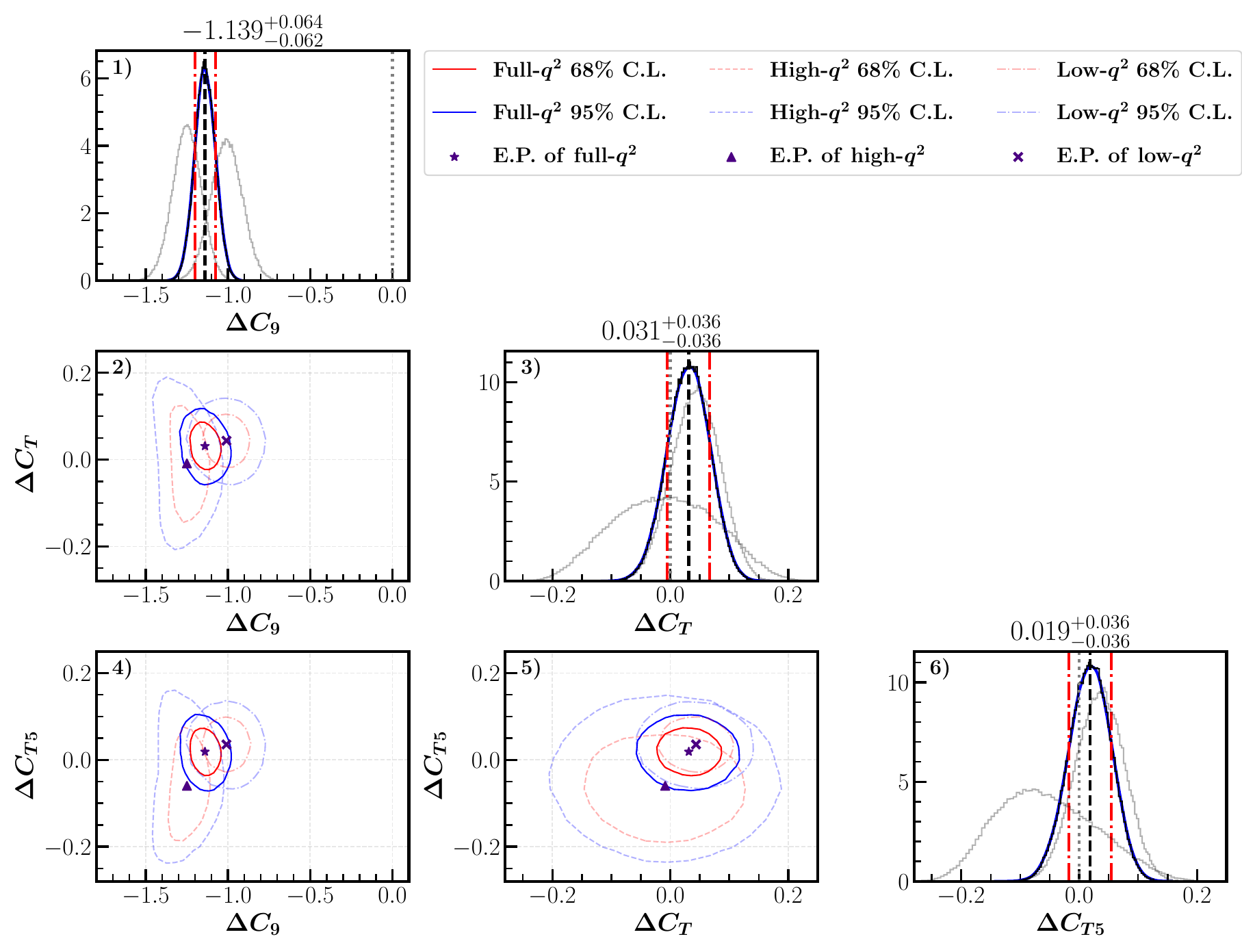}
	}
	\subfigure[S-III]{
	\label{fig:lfus3}
	\includegraphics[width=0.45\textwidth]{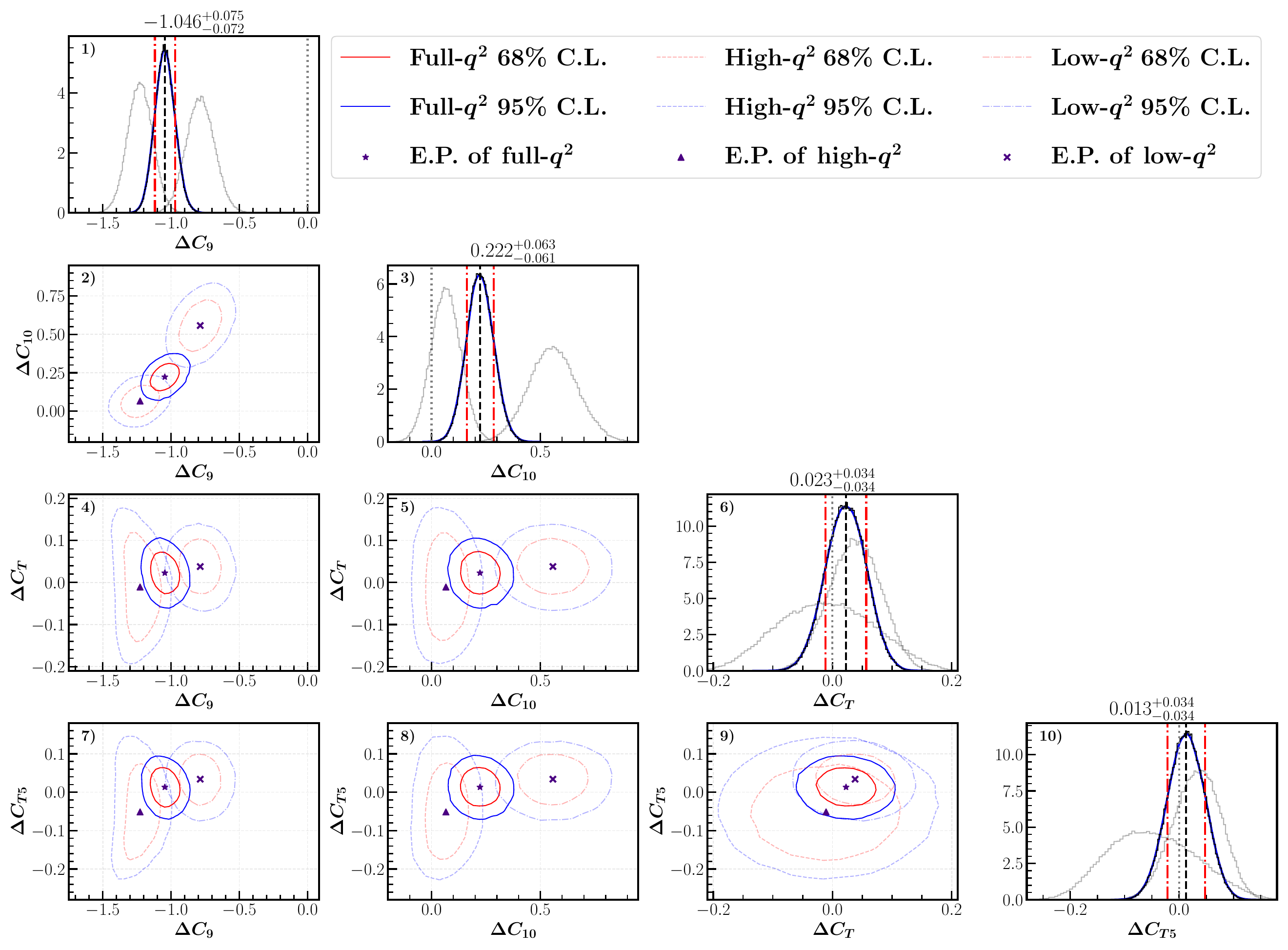}
	}
	\subfigure[S-IV]{
	\includegraphics[width=0.45\textwidth]{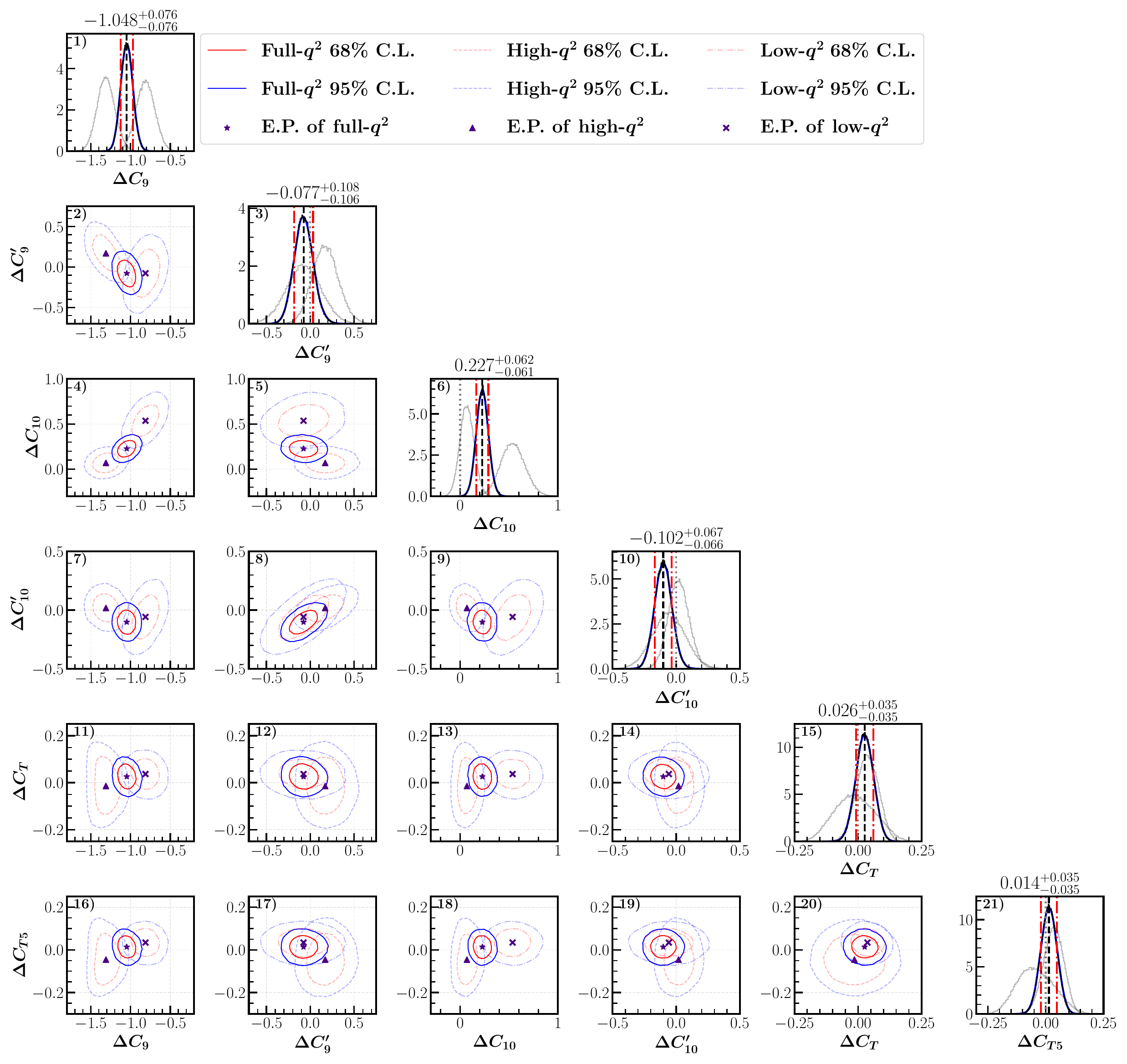}
	\label{fig:lfus4}
	}
	\subfigure[S-V]{
	\includegraphics[width=0.45\textwidth]{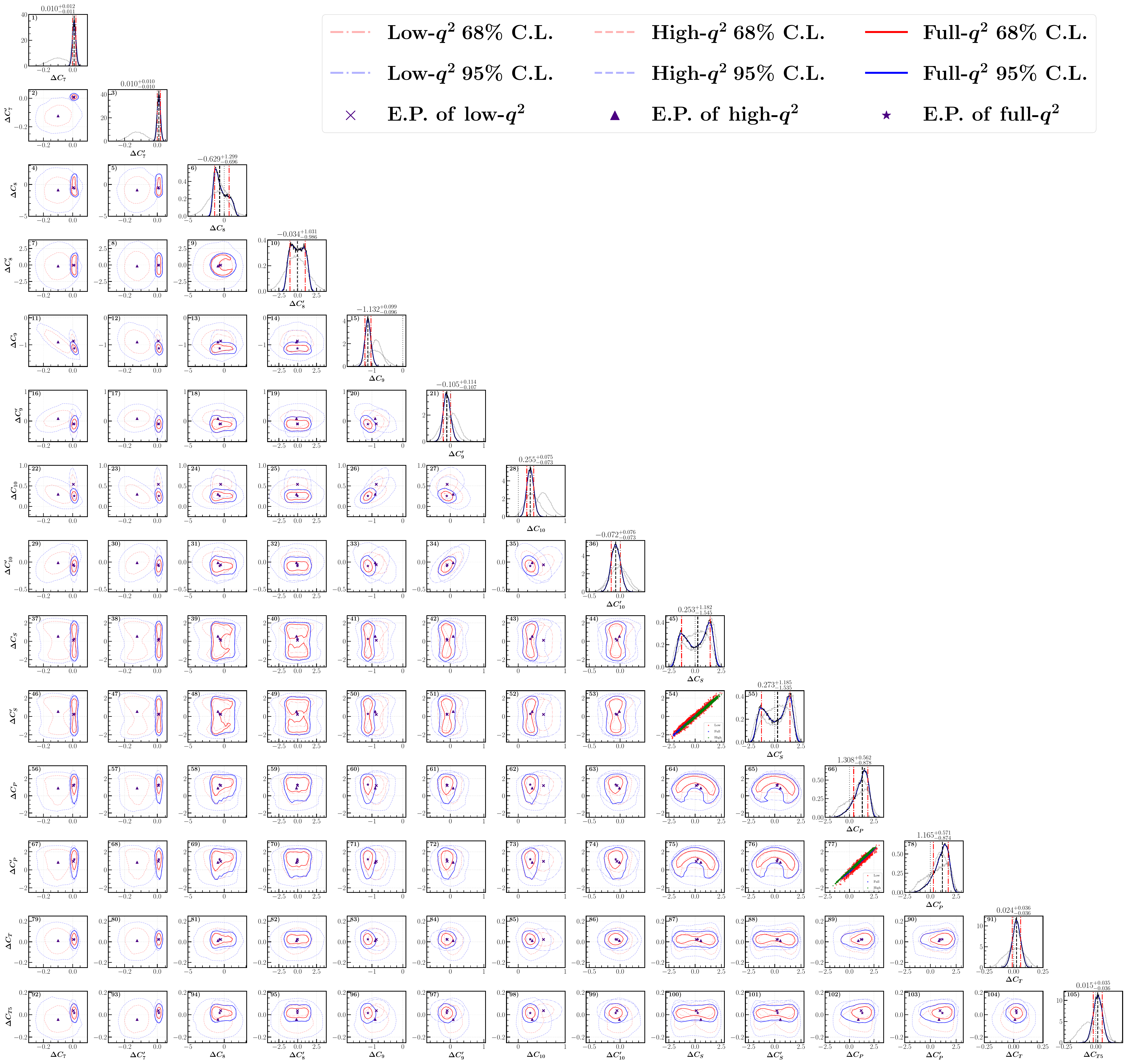}
	\label{fig:lfus5}
	}
	\caption{Complete plots of the confidence regions of various WCs under different scenarios settings (S-II to S-V), represented by subplots (a to d). 
		The diagonal panels in each subplot contain the density depicted by steps, KDE in a blue curve, and two red dot-dashed lines indicating the boundary of the region at 68\% C.L. The panels in the lower left corner of the diagonal represent the correlations and preferred regions, where those regions using the full data are highlighted in a darker color.
		The other plotting conventions are the same as the settings in \cref{fig:lfus1}.
		 }
\end{figure}

\normalem
\clearpage
\bibliographystyle{app/apsrev4-2}
\bibliography{app/reference}
\end{document}